\documentclass[12pt]{article}
\usepackage{iftex}
\ifPDFTeX\usepackage[T1]{fontenc}\fi
\usepackage{kpfonts}
\usepackage{amsmath,amsthm,mathtools}
\usepackage{enumitem}
\usepackage[margin=1.15in]{geometry}
\usepackage{setspace,microtype,xurl}
\usepackage{graphicx,booktabs,tabularx,array,xcolor,float}
\usepackage[round,authoryear]{natbib}
\usepackage[section]{placeins}
\usepackage[colorlinks=true,linkcolor=blue,citecolor=blue,urlcolor=blue]{hyperref}
\allowdisplaybreaks
\newtheorem{proposition}{Proposition}
\newtheorem{lemma}{Lemma}
\newtheorem{corollary}{Corollary}
\theoremstyle{remark}

\theoremstyle{plain}
\title{Who Pays for Product Returns?\\Advertising and Return Policies}
\author{\normalsize Sanxi Li\thanks{School of Economics, Renmin University of China, Beijing 100872, China. sanxi@ruc.edu.cn} \and Yufan Wu\thanks{Corresponding author. School of Economics, Capital University of Economics and Business, Beijing, China. wuyufan0103@126.com} \and Jun Yu\thanks{School of Economics, Shanghai University of Finance and Economics; Key Laboratory of Mathematical Economics (SUFE), Ministry of Education, Shanghai 200433, China. yu.jun@mail.shufe.edu.cn} \and Mingsheng Zhang\thanks{School of Economics, Renmin University of China, Beijing, China. mingsheng\_zhang@yeah.net}}
\date{}
\hypersetup{pdftitle={Who Pays for Product Returns? Advertising and Return Policies}}

\begin{document}
\maketitle
% \vspace{-1em}
% \begingroup
% \renewcommand{\thefootnote}{}
% Unnumbered author note; ordinary body footnotes keep their hyperlinks.
% \makeatletter\let\@footnotetext\H@@footnotetext\makeatother
% \footnotetext{%
% \textsuperscript{$a$}School of Economics, Renmin University of China, Beijing 100872, China.\par
% \textsuperscript{$b$}School of Economics, Capital University of Economics and Business, Beijing, China.\par
% \textsuperscript{$c$}School of Economics, Shanghai University of Finance and Economics; Key Laboratory of Mathematical Economics (SUFE), Ministry of Education, Shanghai 200433, China.\par
% \textsuperscript{$*$}Corresponding author. wuyufan0103@126.com}
% \textsuperscript{$\dagger$}\textit{Funding:} Sanxi Li acknowledges support from the National Natural Science Foundation of China (Grant No.~72192801), the Beijing Natural Science Foundation (Grant No.~Z220001), and the fund for building world-class universities and disciplines at Renmin University of China. Jun Yu acknowledges support from the National Natural Science Foundation of China (Grant No.~72273078) and the SUFE Theoretical Economics Gaofeng II Discipline Innovation Project (Grant No.~2018110721).}
% \endgroup
\begin{abstract}
An online marketplace decides how consumers, firms, and the platform share the cost of product returns. We study this decision when the platform earns commissions and sells prominent placement. Consumers inspect, try, and return products; firms bid for the first position and then set prices. Earlier display attracts more trials but can retain a smaller fraction of them. For small positive losses, assigning return costs to consumers raises the value of prominence relative to reimbursing firms by the same amount. Advertising therefore shifts the platform's preferred division of a given return bill toward consumers. We characterize the joint allocation for small positive return losses and give a cost condition under which the platform pays nothing. In markets with frequent returns, the platform can instead share the bill with consumers: a higher commission makes it pay more to encourage shopping and retained sales. Results at finite losses show that advertising can change the optimal policy from firm-paid free returns to full consumer payment, reducing both commissions and consumer surplus. This advertising-induced policy change can lower or raise total welfare, depending on whether savings from fewer trials and returns offset the loss from reduced participation.
\end{abstract}
\medskip
\noindent\textit{Keywords}: product returns; sponsored rankings; platform commissions; consumer search; cost sharing.
\clearpage
\label{main:start}
\section{Introduction}\label{sec:introduction}

A consumer buying shoes on an online marketplace first sees an ordered list of products. Listings reveal prices, materials, and features, while comfort becomes clear after delivery and trial. She may keep the first pair, return it, or try another before deciding. Browsing can continue between purchase and return: \citet{zhang2025browse} document this behavior using activity records from a Chinese apparel retailer. The order in which products are displayed helps determine which trials occur. The return rule determines what it costs to continue the comparison.

Free returns leave a bill to be paid. The National Retail Federation and Happy Returns estimated that 19.3 percent of U.S. online sales would be returned in 2025. A 2026 McKinsey estimate puts retailers' annual spending to recover value from returned goods at about \$200 billion.\footnote{National Retail Federation and Happy Returns, \href{https://nrf.com/media-center/press-releases/consumers-expected-to-return-nearly-850-billion-in-merchandise-in-2025}{2025 Retail Returns Landscape}, October 15, 2025; McKinsey, \href{https://www.mckinsey.com/industries/logistics/our-insights/from-cost-center-to-competitive-advantage-modernizing-reverse-logistics-with-ai}{Modernizing reverse logistics}, February 23, 2026. The return-rate estimate draws on a survey of 358 e-commerce professionals at large U.S. merchants; the spending figure is McKinsey's industry estimate. Accessed November 5, 2026.} A marketplace can require firms to bear shipping, handling, and recovery losses, allow a deduction from the consumer's refund, or reimburse firms itself. Firm-funded and platform-funded free returns both leave consumers with no return loss but create different pricing incentives. Walmart's seller-fulfilled rules prohibit return-shipping charges to consumers, while its fulfillment service charges firms for specified returns, including poor fit: the standard return-processing fee is \$4.70 for an item with a shipping weight of one pound or less. For eligible orders, Cainiao's return-shipping insurance can instead be bought by consumers or provided by firms or the platform.\footnote{Walmart, \href{https://marketplacelearn.walmart.com/guides/Policies\%20\%26\%20standards/Orders\%20\%26\%20returns/returns-policy}{seller-fulfilled returns policy}, August 31, 2026, and \href{https://marketplacelearn.walmart.com/guides/Walmart\%20Fulfillment\%20Services\%20\%28WFS\%29/WFS-fees}{WFS fees}, May 27, 2026; Cainiao, \href{https://page.cainiao.com/guoguo/statements/yfxinfo.html}{upgraded return-shipping insurance rules} (in Chinese). WFS uses the greater of unit and dimensional weight and applies additional fees to apparel and certain other items; the insurance provision concerns eligible insured orders and collections. The WFS fee schedule was accessed November 5, 2026; the other rules were accessed October 30, 2026.}

The same marketplace also sells visibility. It can earn a commission on a retained transaction and advertising revenue from a firm seeking earlier display. The relative importance of these revenues differs across platforms. In 2025, advertising generated \$1.99 billion, or about 18 percent of eBay's total revenue, with the rest classified as marketplace revenue. At PDD Holdings, which operates Pinduoduo and Temu, online marketing services and other revenues accounted for 50.4 percent of consolidated revenue, and transaction services for 49.6 percent.\footnote{eBay, \href{https://www.sec.gov/Archives/edgar/data/1065088/000106508826000027/ebay-20251231.htm}{2025 Form 10-K}, p.~40; PDD Holdings, \href{https://www.sec.gov/Archives/edgar/data/1737806/000110465926050727/pdd-20251231x20f.htm}{2025 Form 20-F}, revenue discussion and Note 13. These are shares of total revenue under each company's reporting categories. eBay's marketplace revenue includes commissions, listing and subscription fees, shipping fees, and other charges; PDD's transaction services cover transaction-related services. Accessed November 5, 2026.}

The allocation of return losses can change both sources of revenue. For example, eBay's refund accounting credits specified sales-based fees while retaining Priority advertising fees charged per click.\footnote{eBay, \href{https://www.ebay.com/help/selling/listings/promoted-listings/promoted-listings-standard?id=5295}{Promoted Listings} and \href{https://export.ebay.com/en/growth/seller-performance-guides/transactions/refunding-buyers/}{partial-refund fee credits}, accessed October 30, 2026. The stated credits apply to the sales-based fees and refund cases specified in the guidance.} A return rule changes the retained transactions that generate commissions and firms' willingness to pay for earlier display. The platform must choose how much of each returned order's loss to assign to consumers, how much to leave with firms, and how much to pay itself. How does selling prominence change this allocation? When should the platform fund free returns, and when should it leave their cost with firms?

We study these questions with two firms and consumers who acquire information through inspection and trial. Inspection establishes whether a product meets a basic requirement; a trial reveals its value in use. Consumers observe prices, decide whether to shop, and choose which products to inspect, try, return, and keep. The platform chooses a consumer-borne return cost and a reimbursement to firms. Firms bid for the first position in a second-price auction and then set prices. We compare a platform receiving commissions and auction payments with one receiving commissions alone. At each return rule, both use the same pricing equilibrium. The auction collects any positive profit advantage of appearing first. The comparison isolates how collecting that value changes the platform's allocation of return losses, at a given commission schedule.

The joint allocation has two margins. One determines how much return liability to remove from firms; the other determines whether consumers or the platform pay for that relief. We analyze these margins to solve the same policy problem. A first step is to distinguish trial purchases from retained sales. With firm-paid free returns, the first product is tried before the consumer knows the value of the alternative. By the time she considers the second product, she may already hold a satisfactory one. The second firm's orders pass an additional selection step. We show that the first firm receives more trials but retains a smaller fraction of them when inspection and return costs are low. The underlying selection mechanism also holds at positive costs when prices are equal. The equilibrium result establishes when firms' price adjustments preserve it.

The identity of the payer affects the value of prominence. Raising the consumer contribution while reducing reimbursement equally leaves firms' loss per return unchanged. Another trial and possible return now cost the consumer more. At given prices, this favors the first firm's retained sales relative to the second's. We show that, for small losses, the profit advantage of appearing first rises even after firms reprice and participation adjusts. This gives a policy comparison: at any given level of firm liability, adding advertising revenue weakly increases the optimal contribution from consumers. It describes the choice of payer for a given amount of relief; the joint optimum also determines that amount.

The amount of relief depends on pricing incentives and the platform's budget. Firms bearing return losses can cut prices to turn trials into retained purchases. Reimbursement weakens that reason to cut prices, but makes attracting consumers more profitable because their returns will be compensated. We characterize the resulting price changes. At the symmetric reference prices, the two products are equally likely to be kept, so consumers value their price reductions equally. The change in the sum of prices therefore determines whether reimbursement makes shopping more attractive. For a commission-only platform, reimbursement is profitable only if the additional commissions cover its bill. A platform selling prominence also counts the change in advertising receipts. Together with the conditional payer comparison, these incentives determine the joint allocation.

Combining the two margins gives a classification of the joint optimum. For fixed market conditions and sufficiently small losses, a strict comparison among the three full-payment policies determines the exact optimal payer over all feasible allocations. We also give a cost condition, valid at every suitability level, under which the platform optimally pays nothing. In a region where most orders are kept, the commission-only platform leaves the loss with firms, whereas advertising makes full consumer payment optimal. In markets where many trial orders are returned, the platform may instead share the bill with consumers. Consumer contributions raise advertising receipts, while platform payment can stimulate shopping and retained sales. Neither source of revenue need justify making one party pay the entire bill. In the sharing region, a higher commission makes encouraging retained sales more valuable and lowers the consumer contribution. We identify a positive interval of commission rates over which both parties pay, with full consumer payment below the interval and platform-paid free returns above it.

The finite-loss analysis establishes regions in which the commission-only platform chooses firm-paid free returns and the advertising platform assigns the loss to consumers. The policy change induced by advertising harms consumers in both regions. Total welfare falls when the saved resources do not offset the loss from reduced participation, but rises when return losses are large enough for those savings to dominate. The comparisons allow all feasible divisions of the return bill. We also vary consumers' responsiveness to shopping opportunities to identify its role in reimbursement and prices.

\paragraph{Related literature.}
Return policies let consumers defer commitment until uncertainty is resolved. \citet{anderson2009option} study this option value, and \citet{shulman2011competitive} analyze returns in competition. \citet{petrikaite2018returns} connects consumer search to the cost of physically returning products. \citet{janssen2024returns} identify the trade-off between return waste and the gains from learning product value after purchase. We examine how a platform's commissions and position revenue determine who pays for that learning, with participation and equilibrium prices linking the chosen payment rule to consumer and resource welfare.

\citet{wang2024freight} directly compare return freight paid by a platform, a retailer, or consumers, with prices and consumer welfare responding to the choice of payer. Our allocation problem additionally makes each firm's return exposure depend on the consumer's earlier trials. The platform chooses continuous contributions from all three parties, and their effect on competition for earlier display changes the return policy that maximizes platform revenue. This allows us to compare two payments that give firms the same relief per return but induce different trial choices and advertising bids.

The closest study of the information and return-policy decision is \citet{jerath2025returns}. A single-product firm chooses its price and return policy when consumers can resolve uncertainty about that product before or after buying. Inconclusive search allows search and returns to coexist, and the return fee balances incentives to acquire information and to purchase after an uninformative search. Here consumers compare products supplied by competing firms, and earlier display changes the selection of orders each firm receives. The platform values the resulting difference in position profits. This permits a comparison between consumer and platform payments that give firms the same relief per returned item but change willingness to pay for prominence.

The search and prominence literature establishes how being considered earlier affects competition \citep{armstrong2009prominence}. \citet{armstrong2011paying}, \citet{chen2011paid}, and \citet{athey2011position} study payments for favorable placement, while \citet{anderson2026direction} connect position-auction bids to incremental profits and position externalities. \citet{fishman2018prominence} study the cost of returning to a previously searched option. Our consumers physically return goods they have tried. The associated losses enter both firms' prices and their incremental profit from appearing first. The return-cost allocation is therefore an instrument that changes the platform's revenue from a given ranking technology.

Research on marketplace design also combines commissions and advertising. \citet{long2022designing} use auction information in marketplace design; \citet{saremi2026bidding} show how firms' price adjustments can move advertising and sales revenue in opposite directions. \citet{janssen2026platforms} examine informative sponsored positions on commission-earning platforms, and \citet{compiani2024rankings} compare rankings under consumer and platform objectives. Here the advertising mechanism is given, and the platform chooses who bears return losses. Reallocating the same relief of firm liability between consumer contributions and platform reimbursement changes trial selection and the relative profit of the first position. This identifies a return-policy channel through which position revenue and commissions can favor different payment rules.

Section~\ref{sec:model} defines the joint allocation problem and derives shopping and pricing behavior. Section~\ref{sec:rank} establishes how the payer affects advertising value and conditional optimal sharing. Section~\ref{sec:policy} combines this comparison with the return to relieving firms of liability to solve for the joint policy. Section~\ref{sec:welfare} evaluates welfare, finite return losses, and the role of consumer participation. Section~\ref{sec:implications} draws managerial implications.

\section{Model and equilibrium}\label{sec:model}

Two firms sell products through a marketplace. Consumers inspect listings, try products, and decide which purchase to keep. The platform allocates the loss on a returned order among the consumer, the firm, and itself. It earns commissions on retained sales and may also sell the prominent position.

\subsection{Products, information, and participation}
The firms are identical. Each product meets a consumer's basic requirements with probability $\alpha\in(0,1]$, independently across firms. We call this event suitability. Conditional on suitability, product $i$ gives match value $\varepsilon_i$, independently drawn from the uniform distribution on $[0,1]$. An unsuitable product has zero consumption value. Inspection reveals suitability; a trial purchase reveals match value. A listing can establish whether a shoe has the required features, while wearing it reveals comfort. The upper match value normalizes monetary units.

The prominent product, in position 1, can be inspected for free. Reaching and inspecting the second listing costs $s>0$. Consumers observe both posted prices before deciding whether to shop. Inspection is required before ordering: it establishes suitability and provides access to a trial purchase. Thus $s$ includes the attention and effort needed to consider and order the less prominent product; it is incurred even when suitability is certain. A trial then reveals match value. Consumers can inspect product 2 before buying product 1, hold one product while considering the other, recall products already tried, and keep at most one. They are risk neutral and face no liquidity constraint. A returned product physically goes back to its firm. The distinction between seeing a price and examining a product is consistent with evidence that rankings change which products consumers consider \citep{ursu2018rankings}.

There is a unit mass of potential consumers. Each has a private cost $e$, uniformly distributed on $[0,1]$, of undertaking the shopping occasion. The consumer incurs this cost before the free inspection of the first listing; inspecting the second requires the additional cost $s$. The participation cost is independent of product suitability and match values. Let $U$ denote expected utility from shopping before paying $e$, taking optimal inspection, trial, and retention decisions into account. We study active markets with $0<U<1$. We call consumers who enter participants. They enter when $e\le U$, so participation equals $U$ and consumer surplus is
\begin{equation}
 CS=\int_0^U(U-e)\,de=\frac{U^2}{2}.
 \label{eq:entry}
\end{equation}
The cost $e$ is sunk after entry and independent of product values, so all entrants face the same subsequent shopping problem. Participation and further inspection are separate decisions: a return policy changes both the value of starting a shopping occasion and the value of trying another product. Section~\ref{sec:entry} varies the responsiveness of participation.

\subsection{Return-cost allocation and firm profits}
A trial purchase commits the product itself and the resources needed to deliver it. Let $c$ be their combined cost, including production or procurement and outbound fulfillment. If the consumer keeps the product, these resources remain with that sale. If she returns it, return freight, inspection, handling, and any repacking use resources, while the recovered item can be reused or resold. Let $\eta$ be the value of that recovered item after deducting these return expenses, before allocating their cost among the consumer, the firm, and the platform. We consider $0\le\eta<c$. The net resource loss on a returned order is therefore
\begin{equation}
 \rho=c-\eta>0.\label{eq:loss}
\end{equation}
Equivalently, $\rho$ comprises outbound freight, return freight, return processing, and the loss in the product's resource value. A return consumes these resources even though the consumer ultimately keeps no product. The accounting follows the distinction between recoverable product value and the costs of recovering it in the reverse-logistics literature \citep{guide2006time}. In particular, $c$ includes the product's original cost, whereas $\eta$ is already net of return freight and processing. An observed return-processing fee measures only part of $\rho$.\footnote{For example, procurement costing 100 and outbound delivery costing 5 give $c=105$. If the returned item recovers 90 of product value, while return freight and handling cost 5 and 3, then $\eta=82$ and $\rho=23$. The loss consists of 10 in product value and 13 in logistics and handling. These monetary amounts illustrate the accounting before the model's normalization.}

Let $f$ denote the total loss borne by the consumer when returning an item, her share of the return loss $\rho$. Having paid the purchase price $p_i$, she obtains a payoff of $p_i-f$ upon return, net of the return costs she bears. Her net payoff from buying and then returning that item is therefore $-f$. The platform may reimburse the firm by $b$ per returned item. Its feasible policies are
\begin{equation}
 f\ge0,\qquad b\ge0,\qquad f+b\le\rho.\label{eq:feasible}
\end{equation}
The firm bears the remaining loss $\rho-f-b$. Thus $(0,0)$ gives firm-paid free returns, $(\rho,0)$ makes consumers pay the loss, and $(0,\rho)$ gives platform-paid free returns. The edge $f+b=\rho$ divides the bill between consumers and the platform. We study cost allocation without overcompensation: the consumer contribution and platform reimbursement together cover at most the loss on a returned item. The platform chooses this allocation holding $c$ and $\eta$ fixed. The loss per returned item is therefore fixed, but the total loss need not be: the policy changes how many consumers shop and how many products they return. The stated shares describe who bears the return bill directly; prices and shopping choices determine its economic incidence.

The platform collects a fixed commission rate $t\in[0,1)$ on the posted price of each retained sale. A returned order generates no commission. The commission schedule is given when the platform chooses its return policy. For firm $i$, let $S_i$, $D_i$, and $R_i=S_i-D_i$ denote shipments, retained purchases, and returns per participating consumer. The firm's operating profit per participant and its aggregate operating profit are
\begin{equation}
 \pi_i=[(1-t)p_i-c]D_i-(\rho-f-b)R_i,
 \qquad \Pi_i=U\pi_i.\label{eq:profit}
\end{equation}
The order-retention rate is $D_i/S_i$. It differs from suitability $\alpha$ and from the mass $U$ of consumers who shop. Both the consumer contribution and platform reimbursement reduce the firm's remaining return loss, but only the consumer contribution directly changes trial and retention choices.

\subsection{Timing, advertising, and the platform's objective}
The platform first commits to $(f,b)$. Firms then submit nonnegative bids for position 1 in a second-price auction without a reserve. The winner pays the other bid; the loser receives position 2 and pays nothing. Ties are resolved by a symmetric lottery. Firms next choose prices simultaneously from $[0,\infty)$. Consumers observe prices, decide whether to enter, and make their inspection and trial choices. We study subgame-perfect equilibria with weakly dominant auction bids and optimal consumer behavior at every price pair.

The auction sells a position for a period of exposure. Its payment is sunk when firms set prices. At each policy, a firm's value of winning is therefore the difference between the operating profits in positions 1 and 2, including the prices chosen in both positions. We consider markets in which appearing first is more profitable than appearing second under every feasible return policy. Both firms then bid a strictly positive amount. Lemma~\ref{prop:prices} gives sufficient conditions for the small-loss analysis, and Section~\ref{sec:finite} verifies this property at finite losses. Advertising revenue is
\begin{equation}
 A=\Pi_1-\Pi_2>0.
 \label{eq:accounts}
\end{equation}
The auction transfers the profit advantage of earlier display to the platform and leaves each firm with $\Pi_2$. Appendix~\ref{app:auction} proves the bidding rule.\footnote{Second-price bidding is also used in eBay's Priority advertising, although its actual payment is per click and depends on relevance, quality, and a reserve: \href{https://www.ebay.com/help/selling/listings/promoted-listings/promoted-listings-standard?id=5295}{eBay Promoted Listings guidance}, accessed October 30, 2026. The model uses a payment for the exposure period.}

Total returns are $\mathcal R=U(R_1+R_2)$, and commission revenue is $C=tU(p_1D_1+p_2D_2)$. The advertising-supported platform maximizes
\begin{equation}
 P^{\mathrm{ad}}=C+A-b\mathcal R.
 \label{eq:platform}
\end{equation}
The commission-only platform assigns positions by the same symmetric lottery and maximizes $P^0=C-b\mathcal R$. Products, consumers, commissions, and feasible policies are identical in the two cases, and both platforms use the same pricing continuation at each policy. At an unchanged policy, prices and purchases are therefore unchanged. Both objectives are maximized jointly over $(f,b)$ in~\eqref{eq:feasible}. The additional revenue $A$ changes the preferred allocation through the profit advantage of earlier display. Holding $f+b$ fixed compares who funds the same relief of firm liability. Varying $f+b$ determines how much liability firms retain. These are two margins of the joint choice, with no restriction on the order in which the platform chooses its instruments.

The auction collects this position value before prices are set. A policy-dependent listing fee $A(f,b)/2$ charged to each firm under the same lottery would produce the same platform revenue and expected net firm profits. The relevant incentive is how the revenue collected changes with the return policy; in the auction, firms' bids supply this dependence. The exposure-period payment makes that link explicit. With per-click billing, revenue would additionally depend on the volume of billed clicks.

\subsection{Consumer choices and demand}
A trial purchase provides information as well as the option to keep the product. After paying $p_i$, the consumer compares the consumption value $\varepsilon_i$ from keeping it with the payoff $p_i-f$ from returning it. The gain from keeping rather than returning is therefore $\varepsilon_i-p_i+f$. These choices have an equivalent search representation: assign a cost $f$ to each trial and a payoff $\varepsilon_i-p_i+f$ if that product is kept. For a single trial, keeping gives net utility $\varepsilon_i-p_i$, while returning gives $-f$, exactly as in the original decision. This representation assigns costs at trial while preserving the payoffs from keeping or returning the product. The consumer can retain the best product already tried and return the others, giving the independent sampling opportunities and recall property of the search problem in \citet{weitzman1979optimal}.

Let $u\ge0$ denote the best payoff currently available, measured relative to returning all products already tried. Before any trial, $u=0$. After trying product 1 alone, $u=\max\{\varepsilon_1-p_1+f,0\}$. In this equivalent representation, the cost assigned to each completed trial is sunk. Keeping product 1 yields its match value less the payoff forgone by not returning it. The continuation payoff $u$ differs from expected shopping utility $U$, which includes all trial and inspection costs.

Consider a suitable product $i$ whose match value is still unknown. A trial can improve the consumer's final choice by $(\varepsilon_i-p_i+f-u)_+$, where $(x)_+=\max\{x,0\}$. Before inspecting product 2, she must also allow for the possibility that it is unsuitable. If it is suitable, she may decline a trial that is no longer worthwhile. The trial and inspection decisions therefore satisfy, respectively,
\begin{align}
 \mathbb E[(\varepsilon_i-p_i+f-u)_+]&>f,\nonumber\\
 \alpha\big[\mathbb E[(\varepsilon_2-p_2+f-u)_+]-f\big]_+&>s.
 \label{eq:consumer-gain}
\end{align}
The first line compares the expected improvement from another trial with the cost $f$ in the equivalent search representation. The second compares the value of access to a suitable second product with the inspection cost. Once inspection has been worthwhile, a suitable product 2 is tried: the inspection cost is sunk, and the remaining trial has positive expected value.

With uniform match values, the expected improvement is $(1-p_i+f-u)^2/2$ when the cutoff lies inside $[0,1]$. The reservation utility is the available payoff at which the consumer is just willing to undertake the opportunity. For a suitable product 1 and an uninspected product 2, these reservation utilities are
\begin{equation}
 u_1=1-p_1+f-\sqrt{2f},\qquad
 u_2=1-p_2+f-\sqrt{2(f+s/\alpha)}.
 \label{eq:consumer-reservation}
\end{equation}
Inspection uses resources whether or not product 2 is suitable. In the equivalent search representation, the trial cost $f$ is incurred only when a suitable product is tried. The combined expected cost is $s+\alpha f$, which accounts for the term $s/\alpha$.

The consumer need not buy product 1 before inspecting product 2. Comparing these two choices gives priority to the opportunity with the higher reservation utility. The comparison requires accounting for the option to decline a trial after inspection; Appendix~\ref{app:consumer-choices} verifies it. A suitable product 1 is tried first when $u_1>u_2$, or
\begin{equation}
 p_1-p_2<\sqrt{2(f+s/\alpha)}-\sqrt{2f}.
 \label{eq:trialorder}
\end{equation}
A sufficiently low second price reverses trial order. At equal reservation utilities, consumers try product 2 first whenever both orders are feasible. This choice rule is separate from the auction's lottery over tied bids.

\begin{lemma}[Consumer choices]\label{lem:consumer-choices}
Suppose $s>0$, $f\ge0$, $p_i>f$, and $u_1>u_2>0$. An entering consumer inspects product 1 and tries it if suitable. If it is unsuitable, she inspects product 2 without buying product 1. After trying product 1, she inspects product 2 if $\max\{\varepsilon_1-p_1+f,0\}<u_2$; otherwise she stops. A suitable product 2 is tried following this inspection. Among the products tried, she keeps the one with the highest positive value of $\varepsilon_i-p_i+f$ and returns the others. If all these values are nonpositive, she returns every product tried.
\end{lemma}
A common consumer return cost changes the choice between keeping a product and returning all trials. It does not change the ranking of two products already tried, since $f$ enters both values equally. Reimbursement $b$ does not enter these choices at given prices.

This rule gives a simple construction of shipments. Product 1 is shipped whenever suitable. Product 2 is shipped either because product 1 was unsuitable or because the first trial did not provide enough value to stop:
\begin{equation}
 S_1=\alpha,\qquad
 S_2=\alpha\left[(1-\alpha)+\alpha\Pr\{\max(\varepsilon_1-p_1+f,0)<u_2\}\right].
 \label{eq:shipment-construction}
\end{equation}
The outer probability $\alpha$ requires product 2 itself to be suitable. Its orders are thus selected by what the consumer learned from product 1. Retained demand $D_i$ counts the trials that win the final comparison, and returns are $R_i=S_i-D_i$. Appendix~\ref{app:consumer-choices} integrates these events and obtains $U$, including inspection costs and consumer-borne return costs.

\subsection{Price equilibrium}
On a smooth consumer-choice region, the envelope relations are $U_{p_i}=-D_i$ and $U_f=-(R_1+R_2)$. A price increase affects the consumer only on a retained purchase; a higher consumer return cost affects her on each returned trial. Since participation is $U$, differentiating aggregate operating profit gives
\[
 \frac{\partial(U\pi_i)}{\partial p_i}
 =U\pi_{i,p_i}-D_i\pi_i.
\]
The first term is the price effect among participating consumers; the second is the loss from discouraging entry. Substituting per-participant profit from~\eqref{eq:profit}, an interior optimal price satisfies
\begin{equation}
 (1-t)D_i+[(1-t)p_i-c]D_{i,p_i}-(\rho-f-b)R_{i,p_i}
 =\frac{D_i}{U}\pi_i.
 \label{eq:firm-foc}
\end{equation}
A price must also be optimal against deviations that change trial order.

We first solve a reference market with free inspection and no net return loss, $s=\rho=0$. Consumers can then compare suitable products without wasting resources on trial, and the positions are symmetric. This gives a common price from which to trace how positive inspection and return costs change demand, pricing, and the value of earlier display. The reference market supplies the starting equilibrium for the small-loss results; the policies compared in those results have strictly positive losses and inspection costs. Denote the common reference price by $p_0$. A suitable product with value $x>p_0$ is kept unless the rival is suitable and has a higher value, an event with probability $\alpha(1-x)$. Integrating gives each firm's retained demand. Shopping utility is the expected highest positive value net of price among suitable products. These calculations give
\begin{align}
 D_0&=\alpha(1-p_0)\left[1-\frac{\alpha(1-p_0)}2\right],\nonumber\\
 U_0&=\alpha(1-p_0)^2\left[1-\frac{\alpha(1-p_0)}3\right].
 \label{eq:starting-quantities-main}
\end{align}
The price is the unique solution in $(\eta/(1-t),1)$ of
\begin{equation}
 [(1-t)p_0-\eta](D_0^2+\alpha U_0)=(1-t)U_0D_0.
 \label{eq:reference-price-main}
\end{equation}
The subscript 0 identifies this zero-cost comparison throughout the paper.

The next result allows positive costs and checks prices over the entire policy set. Write $\xi=s/(\alpha\rho)$ for the inspection cost relative to the return loss, adjusted for suitability. For the small-cost results, hold $\xi$ fixed and finite while setting
\begin{equation}
 c=\eta+\rho,\qquad s=\alpha\xi\rho.
 \label{eq:searchratio}
\end{equation}
\begin{lemma}[Price equilibrium]\label{prop:prices}
Fix $\alpha\in(0,1]$, $t\in[0,1)$, and $0\le\eta<1-t$. At $s=\rho=0$ there is a unique pure pricing equilibrium with positive firm profits. Suppose the fixed, finite inspection-cost ratio satisfies
\begin{equation}
 \xi\geq\max\left\{\frac94,\frac{1}{4(1-t)}\right\}.
 \label{eq:active-ratio}
\end{equation}
For sufficiently small positive $\rho$, there is a unique pure pricing equilibrium at every policy in~\eqref{eq:feasible}. Each price is a strict global best response, $p_i=p_0+O(\rho)$ uniformly over those policies, and the first position has strictly greater operating profit than the second.
\end{lemma}
Inspection costs make changing trial order through a price cut costly to the deviating firm. The first ratio bound ensures that the gain from becoming the first trial does not cover that cost at any feasible consumer return cost. The second ensures that the extra retained-sales value of earlier display exceeds its extra return exposure, so firms bid positively throughout the policy set. It adds no restriction to the first bound when the commission is at most $8/9$. We maintain~\eqref{eq:active-ratio} in the small-loss policy comparisons. Appendix~\ref{app:pricing} checks these deviations and excludes additional equilibria. The bound is sufficient rather than necessary. Section~\ref{sec:finite} separately evaluates positive losses away from the limiting comparison, reporting both the economic magnitudes and the pricing continuation used.

\section{Trial selection and the value of the first position}\label{sec:rank}

We begin with who pays for a given reduction in firms' return liability. Its advertising effect depends on the selection of orders into each position. Earlier display changes what a consumer knows when placing an order. She tries the first suitable product before knowing whether the alternative is better. Before trying the second, she may already hold a satisfactory product. An order from the second firm therefore passes an additional selection step. This difference matters because a trial uses resources even if it does not become a retained sale.

The selection effect can be established at positive inspection and return costs before considering price adjustments. The relevant interior has positive reservation utilities, prices above the consumer return cost, and product 1 tried first as in~\eqref{eq:trialorder}. These are the consumer choices used in the demand formulas in Appendix~\ref{app:consumer-choices}.

\begin{lemma}[Selection at given prices]\label{lem:selection}
In this interior, shifting a unit of return payment from the platform to consumers, while leaving the firm's loss per return unchanged, gives
\begin{equation}
 (\partial_f-\partial_b)(D_1-D_2)=\alpha^2(2+p_2-p_1)>0.
 \label{eq:selection-finite}
\end{equation}
With free returns and equal prices, the prominent firm receives more shipments but retains a smaller fraction of them, whenever both trial opportunities have strictly positive reservation utilities.
\end{lemma}

The consumer return cost makes a further trial less attractive when the consumer already holds a product. Stopping avoids a possible cost of another returned item. This favors the first firm's retained sales relative to the second's. Reimbursement to a firm leaves that consumer comparison unchanged. The result holds at finite costs on the stated choice region; equilibrium prices determine whether the region and the resulting profit advantage are maintained. Appendix~\ref{app:selection} gives the proof.

Under firm-paid free returns, the pricing result gives a particularly useful equilibrium implication.
\begin{proposition}[More trials with lower order retention]\label{prop:rank}
Under Lemma~\ref{prop:prices}, at $f=b=0$, sufficiently small positive $\rho$ under~\eqref{eq:searchratio} gives
\[
 S_1>S_2,\qquad D_1/S_1<D_2/S_2.
\]
Both firms earn positive operating profits. The retained-sales difference vanishes faster than the shipment difference as costs fall.
\end{proposition}

A good first draw prevents some second trials. The missing second orders are precisely those facing a strong alternative, so the orders that remain are more likely to be kept. With small costs, the reduction in second trials has order $\sqrt\rho$, whereas the equilibrium price and retained-demand differences have order $\rho$. Price adjustments therefore do not undo the selection ranking. This argument explains why the equal-price mechanism in Lemma~\ref{lem:selection} survives endogenous pricing in Proposition~\ref{prop:rank}; Appendix~\ref{app:rank} gives the proof.

Evidence on marketing-generated purchases also distinguishes orders from retained sales. \citet{elkihal2022keep} find that paid search and free-shipping campaigns increase the share of purchases returned at two online retailers. The model identifies a position-based source of such composition differences. It does not require the firms' products to have different value distributions: consumers reach the two firms with different information.

An advertising bid values the resulting profit difference, not the shipment difference alone. At every policy,
\begin{equation}
 \pi_1-\pi_2=[(1-t)p_1-c](D_1-D_2)
 +(1-t)(p_1-p_2)D_2
 -(\rho-f-b)(R_1-R_2).
 \label{eq:positionprofitdecomp}
\end{equation}
The three terms value relative retained demand, relative prices, and relative return exposure. Multiplication by participation gives the aggregate position value. The maintained ratio condition ensures that the retained-sales and price advantages together outweigh the extra return exposure. The policy comparisons below examine how the payer changes this positive value of prominence.

\subsection{Changing the payer and advertising receipts}\label{sec:funding}
Holding the firm's loss $\rho-f-b$ constant, the platform can replace reimbursement with an equal consumer contribution. Lemma~\ref{lem:selection} gives the relative-demand effect at fixed prices. Firms then reprice, and consumers revise their decision to enter. We first derive the change in advertising revenue and then use it to compare optimal ways to fund the same per-return relief.

\begin{proposition}[The payer and advertising revenue]\label{prop:ads}
Under Lemma~\ref{prop:prices}, for sufficiently small positive $\rho$,
\[
 A_f-A_b>0.
\]
Thus, whenever the platform pays part of the bill, replacing a small amount of reimbursement with the same consumer contribution raises advertising revenue. Starting from firm-paid free returns, an initial consumer contribution also raises advertising revenue more than an equal reimbursement. Reimbursement by itself can raise or lower advertising receipts. The comparison between payers also holds throughout the finite-loss regions in Section~\ref{sec:finite}.
\end{proposition}

The small-cost comparison separates the two effects. At the initial prices, each retained sale earns $m=(1-t)p_0-\eta>0$. The positions initially earn the same profit, so the direct change in participation multiplies a zero profit difference. Each firm's own price is optimal, so its price adjustment has no first-order effect on its own profit; the remaining price effect operates through the rival's profit. These observations give
\begin{align}
 (\partial_f-\partial_b)(\Pi_1-\Pi_2)\big|_0
 &=2U_0m\alpha^2\nonumber\\
 &\quad+m[D_0^2-U_0\alpha^2(1-p_0)]
 (\partial_f-\partial_b)(p_1-p_2)\big|_0.
 \label{eq:equalrelief-decomposition}
\end{align}
The first term is the gain in relative retention at unchanged prices. The second captures repricing. The first firm raises its price relative to the second. A higher first price hurts the second firm by discouraging participation, but helps it by diverting some remaining consumers toward its product. When the participation loss dominates, repricing increases the value of the first position further. When substitution dominates, repricing attenuates the direct effect. Appendix~\ref{app:advertising} proves that the attenuation does not reverse the total sign in this small-cost comparison.

At positive costs the initial position profits and margins differ. Changes in participation and return exposure therefore also enter the total effect. Section~\ref{sec:finite} evaluates these responses without taking costs to zero.

\begin{corollary}[Advertising and conditional optimal sharing]\label{cor:conditional-sharing}
Under the all-policy conditions of Lemma~\ref{prop:prices}, take sufficiently small positive losses and fix any feasible firm liability per returned item. Every optimal consumer contribution with advertising is weakly higher than every optimal consumer contribution without advertising. Advertising therefore shifts the conditional allocation toward consumer payment and away from platform payment. The result also holds on the common pricing branch in the finite-loss regions of Proposition~\ref{prop:finite}.
\end{corollary}

To see the decision implication, compare two ways to fund the same relief of firm liability. Advertising adds a revenue gain that rises with the consumer contribution. Adding this gain shifts the conditional optimum toward consumer payment. No concavity of platform profit is needed. The conclusion holds even if either platform has several optimal consumer contributions; Appendix~\ref{app:conditional-sharing} proves the comparison. The joint optimum also adjusts how much liability firms bear, so this conditional comparison is one step toward the full allocation rather than a ranking of consumer contributions across all markets.

\section{The optimal allocation of return costs}\label{sec:policy}

The platform maximizes revenue over the entire policy triangle in~\eqref{eq:feasible}. Corollary~\ref{cor:conditional-sharing} compares consumer and platform payments at a given firm liability. Completing the optimization requires determining that liability. Reimbursement provides the relevant comparison: it relieves firms without directly changing the consumer's trial cost. We derive its pricing and budget effects, combine the two margins to compare the three full-payment policies, and give conditions under which the platform pays nothing. We then identify when sharing the bill improves on full payment by a single party.

\subsection{How does reimbursement affect prices?}
Holding prices and the consumer contribution fixed, reimbursement changes neither shopping utility nor purchase choices. It raises the firm's profit by the number of returns it compensates. Differentiating the firm's aggregate profit gives
\begin{equation}
 \frac{\partial^2(U\pi_i)}{\partial p_i\partial b}
 =U R_{i,p_i}-D_iR_i.
 \label{eq:reimbursementpriceincentive}
\end{equation}
The first term weakens the incentive to cut prices to turn returns into retained purchases: the platform now pays more of the return loss. The second strengthens the incentive to attract consumers: a price increase loses consumers and the reimbursement associated with their returns. These are the incentives facing one firm at given prices. Solving both firms' price conditions jointly incorporates the rival's response and gives the equilibrium effect below. We report both prices and their sum: at the symmetric reference point, a consumer is equally likely to retain either product, so the sum determines the change in her expected price payment.

\begin{lemma}[Reimbursement and prices]\label{lem:funding-incidence}
Consider the limiting price response in Lemma~\ref{prop:prices}. There is a suitability threshold $\alpha_B^*\in(0,1)$. For $\alpha<\alpha_B^*$, there is a cost threshold $\vartheta_B(\alpha)\in(0,1/4)$, strictly decreasing in $\alpha$, such that reimbursement lowers the sum of prices precisely when
\begin{equation}
 \frac{\eta}{1-t}>\vartheta_B(\alpha).
 \label{eq:primitive-price-threshold}
\end{equation}
It raises the sum below the threshold. For $\alpha>\alpha_B^*$ it lowers the sum at every cost satisfying $0\le\eta<1-t$. At $\alpha=\alpha_B^*$ the reduction is strict for $\eta>0$ and the limiting response is zero at $\eta=0$. In particular, $\eta/(1-t)\ge1/4$ is sufficient for a strict reduction at every fixed $\alpha>0$. 

The first firm's price responds more positively than the second firm's price. The first price rises if $\eta/(1-t)\leq1/4$; the second falls if $\alpha\geq1/2$; and both fall if $\eta/(1-t)\geq3/10$. Each strict sign holds uniformly over feasible policies for sufficiently small positive losses.
\end{lemma}

The ratio $\eta/(1-t)$ is the cost of a retained product relative to the firm's revenue share when $c=\eta$. A higher value supports higher initial prices and more returns per participant. Losing participation then sacrifices reimbursement on more orders, strengthening the incentive to lower prices. At low product cost, the initial price is low and many suitable trial orders are kept. Reimbursement then adds relatively little income per participant, so attracting extra participants brings only a small compensation gain. Low suitability further limits this gain because many participants never order from the firm. By contrast, a price cut can still turn a marginal return into a retained purchase. Relieving the firm of that return loss weakens this retention incentive and can raise prices. Greater suitability lowers the cost threshold at which the participation effect prevails. Appendix~\ref{app:reimbursement-prices} derives the thresholds from the unique scalar price equation; $\alpha_B^*$ is approximately $0.675$.

The firms can therefore move their prices in opposite directions. A higher first price turns some initial trials into returns. A higher second price also discourages some consumers from trying that product at all. Reimbursement consequently removes more of the first firm's incentive to keep a trial from becoming a return. The participation incentive is initially the same for both firms. After strategic price responses, the first price still responds more positively. When suitability is at least one half and the cost ratio is at most one quarter, the first price rises while the second falls. Appendix~\ref{app:individual-prices} derives the individual responses.

The sum of prices matters because the consumer's envelope condition at the symmetric reference point is $dU=-D_0\,d(p_1+p_2)$ when only reimbursement changes. Each price is paid with the same probability $D_0$. A lower sum therefore makes shopping more valuable to consumers, even when the two prices move differently. Whether the additional sales cover the platform's reimbursement bill is a separate question.

\subsection{When do commissions cover the reimbursement?}
Reimbursement affects commissions through prices, retained purchases, and participation. At the symmetric initial prices, differentiating $C=tU(p_1D_1+p_2D_2)$ gives
\begin{equation}
 dC=t\{U_0D_0\,d(p_1+p_2)
       +p_0U_0\,d(D_1+D_2)+2p_0D_0\,dU\}.
 \label{eq:commission-accounting-main}
\end{equation}
The first term changes the revenue on existing retained purchases. The second changes retained purchases per participant, and the third changes the number of participants. A reduction in the sum of prices lowers the first term but increases the other two. In the two policy regions established below, the latter gains dominate, so a lower price sum raises commissions. Appendix~\ref{app:commission-budget} substitutes the demand and participation responses to obtain the exact coefficient in~\eqref{eq:funding-commission-slope}. Write $C_b^0$ for this limiting equilibrium response to reimbursement; on these response coefficients, a superscript 0 denotes the limiting derivative as costs fall along~\eqref{eq:searchratio}.

The initial reimbursement bill rises by $\mathcal R_0=2U_0(\alpha-D_0)$ per unit increase in $b$. Platform funding is locally profitable for the commission-only platform exactly when
\begin{equation}
 C_b^0>\mathcal R_0.
 \label{eq:commission-covers-returns}
\end{equation}
The platform collects only a share of sales revenue but compensates every returned order. A price reduction benefiting consumers need not generate enough additional commissions to meet that bill. Appendix~\ref{app:commission-budget} evaluates both sides from $(\alpha,\eta,t)$.

At finite costs, where advertising revenue is differentiable, the marginal profits are
\begin{align}
 P_f^{\mathrm{ad}}&=C_f+A_f-b\mathcal R_f,\nonumber\\
 P_b^{\mathrm{ad}}&=C_b+A_b-\mathcal R-b\mathcal R_b.
 \label{eq:platformmarginals}
\end{align}
The bill responds to both the reimbursement rate and the volume of returns. When firms bear none of the loss, $b=\rho-f$, shifting payment to consumers gives
\begin{equation}
 \frac{dP^{\mathrm{ad}}}{df}
 =\frac{dC}{df}+\frac{dA}{df}+\mathcal R
 -(\rho-f)\frac{d\mathcal R}{df}.
 \label{eq:edgeeffects}
\end{equation}
The terms are changes in commissions, advertising receipts, the reimbursement saved on existing returns, and the reimbursement effect of changed return volume. The commission-only platform omits the advertising term. These are the same comparisons used in the policy classification that follows.

% Follows reimbursement.tex; includes the two verified exact-policy regions.
\subsection{A common comparison of who pays}

The platform's choice depends on three revenue comparisons. A consumer contribution changes commissions and the value of prominence. Reimbursement changes both revenues and requires a payment on every returned order. Firm funding avoids that payment but can reduce the value of the prominent position. The following comparison combines these effects and solves for the joint allocation when one full-payment policy has a strict advantage in the small-loss revenue comparison.

Fix $\alpha\in(0,1]$, $t\in[0,1)$, $0\leq\eta<1-t$, and a finite $\xi=s/(\alpha\rho)$ satisfying~\eqref{eq:active-ratio}. Let $\rho$ fall with $c=\eta+\rho$ and $s=\alpha\xi\rho$, as in Lemma~\ref{prop:prices}. Write $C_f^0$ and $C_b^0$ for the limiting commission responses, and $\mathcal R_0$ for initial market returns. Let
\[
 B=\Gamma_b^0,\qquad S=\Gamma_f^0-\Gamma_b^0>0,
\]
where $\Gamma=\Pi_1-\Pi_2$ is the value of prominence. Thus $S$ measures the additional position value from assigning costs to consumers instead of reimbursing firms while leaving firms' return liability unchanged. The coefficient $B$ can have either sign. Appendix~\ref{app:policy-general} gives all four revenue coefficients from the initial price equation.

The position value has the expansion
\begin{equation}
 \Gamma=-(\rho-f-b)B+(f+s/\alpha)S+O(\rho^{3/2}),
 \label{eq:position-value-expansion-main}
\end{equation}
uniformly over feasible policies on this small-cost path. The first term values the firm's remaining return liability. The second captures the selection induced by consumer contributions and costly second inspection, including firms' price responses. Under firm, consumer, and platform funding, respectively, the first-order position values are $\rho(\xi S-B)$, $\rho(1+\xi)S$, and $\rho\xi S$. All three are positive under~\eqref{eq:active-ratio}. Adding the commission change and subtracting the reimbursement bill gives Table~\ref{tab:policy-gains}.

\begin{table}[htbp]
\centering
\caption{Revenue comparison among full-payment policies for small return losses}\label{tab:policy-gains}
\small
\renewcommand{\arraystretch}{1.15}
\begin{tabular}{@{}lcc@{}}
\toprule
Party paying the full loss & Commission only & Commission and advertising\\
\midrule
Firm & $0$ & $\xi S-B$\\
Consumer & $C_f^0$ & $C_f^0+(1+\xi)S$\\
Platform & $C_b^0-\mathcal R_0$ & $C_b^0-\mathcal R_0+\xi S$\\
\bottomrule
\end{tabular}
\par\smallskip
\begin{minipage}{\linewidth}\footnotesize
Each entry measures the first-order change in net platform revenue per unit of return loss, relative to commission income under firm-paid free returns at the same costs. Formally, revenue under a full-payment policy equals that common baseline plus $\rho$ times its table entry, with an error of order $\rho^{3/2}$. Thus the difference between two entries is the limiting derivative, with respect to $\rho$, of their revenue difference along~\eqref{eq:searchratio}. The entries include repricing and participation; they are not derivatives of revenue holding shipment and inspection costs fixed.
\end{minipage}
\end{table}

\begin{proposition}[Optimal assignment of return losses]\label{prop:policy-general}
Under the fixed-parameter small-cost path stated above, suppose one entry in a column of Table~\ref{tab:policy-gains} strictly exceeds the other two. For all sufficiently small positive $\rho$, the corresponding full-payment policy is the unique optimum over~\eqref{eq:feasible}: $(f,b)=(0,0)$ when the firm leads, $(\rho,0)$ when the consumer leads, and $(0,\rho)$ when the platform leads.
\end{proposition}

The comparison among full-payment policies also determines whether any shared allocation could do better. To first order, net commission revenue is linear in the two payments, and advertising adds another linear expression. Splitting the bill cannot exceed the best full-payment policy at this order. If one payer strictly leads, moving any part of its bill to another payer reduces revenue to first order. These strict marginal comparisons persist at small positive losses, selecting the exact full-payment policy. Appendix~\ref{app:policy-corners} establishes the uniform marginal comparisons. When the revenue difference between full-payment policies is small, the changing selection and volume of returns can instead make sharing profitable.

Advertising raises the gain from consumer funding relative to platform funding by $S$. The firm's return liability is zero under both arrangements, so this difference comes from who pays. The comparison with firm funding also depends on $B$, the effect of reimbursement on position value. Without advertising, consumer funding can itself lead when $C_f^0$ exceeds both zero and $C_b^0-\mathcal R_0$; Appendix~\ref{app:consumer-funding-example} gives an example. Firm-funded free returns follow under the market conditions in Proposition~\ref{prop:policy}(a).

\subsection{When does the platform pay nothing?}

Table~\ref{tab:policy-gains} compares all three payers. A simpler cost condition can rule out platform payment before making the remaining comparison between firms and consumers. The initial retained fraction is $D_0/\alpha$: each firm ships $\alpha$ orders per participant at the zero-cost reference point and keeps $D_0$ of them.

\begin{proposition}[No platform reimbursement]\label{prop:no-platform-payment}
Under the all-policy conditions of Lemma~\ref{prop:prices}, if $D_0/\alpha\geq2/5$, every optimal policy has $b=0$, for either revenue model and all sufficiently small positive losses. In particular, the cost condition
\begin{equation}
 \frac{\eta}{1-t}\leq\frac15
 \label{eq:no-platform-primitive}
\end{equation}
is sufficient at every fixed $\alpha\in(0,1]$ under the inspection-cost bound in~\eqref{eq:active-ratio}.
\end{proposition}

The comparison holds firms' liability constant. Replacing a unit of reimbursement with a consumer contribution saves payment on existing returns but also changes commissions and the number of compensated returns. In the stated region, the saving more than offsets the commission effect after prices and participation adjust. Advertising adds a further weak gain, by Corollary~\ref{cor:conditional-sharing}. Every allocation with positive reimbursement is therefore improved by moving payment toward consumers. The conclusion eliminates reimbursement throughout the feasible set, including shared-payment policies; it does not determine how the remaining bill is split between firms and consumers.

Condition~\eqref{eq:no-platform-primitive} uses the retained-product cost at the zero-loss reference point, relative to the firm's share of sales revenue. Low values support lower prices and a higher retained fraction. Appendix~\ref{app:obstruction} translates the retention condition into an exact suitability-dependent cost bound; one fifth is a convenient sufficient bound common to all suitability levels. The restriction is sufficient, and allows both firm-funded free returns and consumer contributions. Their comparison still depends on commissions and position value.

\subsection{Advertising, full payment, and shared payment}

Two regions show how the platform combines the choice of payer with the amount of relief given to firms. In the first, retained-product costs are low enough to exclude reimbursement. Advertising changes the preferred payer from firms to consumers. In the second, product costs support high prices and frequent returns. Reimbursement encourages enough additional shopping and retained sales to remove all firm liability. Consumer contributions can nevertheless remain attractive because they raise advertising receipts.

In this second region, the platform must balance the revenue from earlier display against the revenue from retained sales. Full consumer payment favors the former; platform-paid free returns favor the latter. At intermediate commission rates, neither extreme is optimal. As the consumer contribution rises, further reductions in trials and reimbursed returns change the gain from the next increase in $f$. This produces a range of commissions over which the platform and consumers share the bill.

\begin{proposition}[Advertising changes who pays for returns]\label{prop:policy}
\textnormal{(a) Firm-paid free returns versus consumer contributions.}
Fix
\[
 \alpha\in[1/10,3/10],\qquad t\in[1/10,1/5],\qquad
 \frac{\eta}{1-t}\in[1/25,1/10],
\]
and a finite $\xi\geq9/4$. For sufficiently small positive $\rho$, with $c=\eta+\rho$ and $s=\alpha\xi\rho$, the unique optimal policies are
\[
 (f^0,b^0)=(0,0),\qquad
 (f^{\mathrm{ad}},b^{\mathrm{ad}})=(\rho,0).
\]
Both firms earn positive profits after auction payments and retain more than half of their shipped orders. Section~\ref{sec:finite} gives parameter ranges with strictly positive return losses for which the same policy difference holds.

\textnormal{(b) Platform-paid free returns versus shared costs.}
For each $\alpha\in(0,1]$ and finite $\xi\geq9/4$, the following holds for a nonempty interval of sufficiently high recovery values $\eta<1$, specified in Appendix~\ref{app:interior}. For sufficiently small positive $\rho$, hold $c=\eta+\rho$ and $s=\alpha\xi\rho$ fixed. There are commission thresholds $t_C(\rho)<t_P(\rho)$ and an open interval containing them such that the advertising platform's unique policy on that interval is
\begin{equation}
 (f^{\mathrm{ad}},b^{\mathrm{ad}})=
 \begin{cases}
 (\rho,0),&t\leq t_C,\\
 (f^{\mathrm{ad}},\rho-f^{\mathrm{ad}}),\quad 0<f^{\mathrm{ad}}<\rho,
     &t_C<t<t_P,\\
 (0,\rho),&t\geq t_P.
 \end{cases}
 \label{eq:jointphase}
\end{equation}
The interior consumer contribution falls strictly as the commission rises. The commission-only platform uniquely chooses $(0,\rho)$ throughout the same neighborhood.
\end{proposition}

In part (a), eliminating reimbursement leaves the firm--consumer payment margin. Without advertising, raising the consumer contribution lowers commissions. With advertising, the added position value outweighs this loss, so the consumer contribution rises to the full return loss. The proof checks these signs over every feasible allocation. The finite-loss verification also checks each firm's price deviations, including those that reverse trial order. This yields a joint policy difference with positive profits and a majority of orders kept. Appendix~\ref{app:higher} proves part (a).

\paragraph{The range of shared payment.}
In Proposition~\ref{prop:policy}(b), raising reimbursement while firms still bear part of the loss increases platform profit. The optimum therefore lies on $f+b=\rho$. The remaining choice is how much of this bill to assign to consumers. Its marginal return is the sum in~\eqref{eq:edgeeffects}: changing the payer alters commissions, advertising receipts, and both the rate and volume of reimbursed returns.

A higher consumer contribution makes further trials less attractive. Fewer trials change retained sales and reduce the returns on which the platform still pays reimbursement. After firms reprice, the marginal gain from increasing the consumer contribution falls strictly in this region. The platform shares the bill when that gain is positive at a zero consumer contribution but negative at full consumer payment. These are inequalities, so sharing does not require an exact equality of revenues at the chosen positive loss.

A higher commission reduces the marginal gain from assigning costs to consumers: the extra retained sales encouraged by platform payment become more valuable relative to advertising receipts. The optimal consumer contribution consequently falls, reaching zero at $t_P$. The sharing interval has positive width for every sufficiently small positive loss. Appendix~\ref{app:interior} proves that its width is proportional to $\sqrt\rho$ as inspection and return costs fall together. Thus sharing occupies a narrower range of commissions when losses are very small. The commission comparison itself holds shipment cost, recovery value, and inspection cost fixed.

This construction has initial retention below one third. Proposition~\ref{prop:no-platform-payment} excludes reimbursement when initial retention is at least two fifths. The two results therefore describe distinct economic environments: frequent returns can make platform payment worth financing, while a sufficiently high retained fraction rules it out.

Suitability $\alpha$ is resolved before shipment: an unsuitable listing never becomes a trial order. Low order retention instead means frequent returns of products that passed this screen. The shared-payment construction describes this latter environment, with substantial recovery relative to the net return loss. It identifies a boundary of the funding problem where extensive trial and recovery make both consumer and platform payments relevant. The majority-retention and shared-payment regions represent different market conditions.

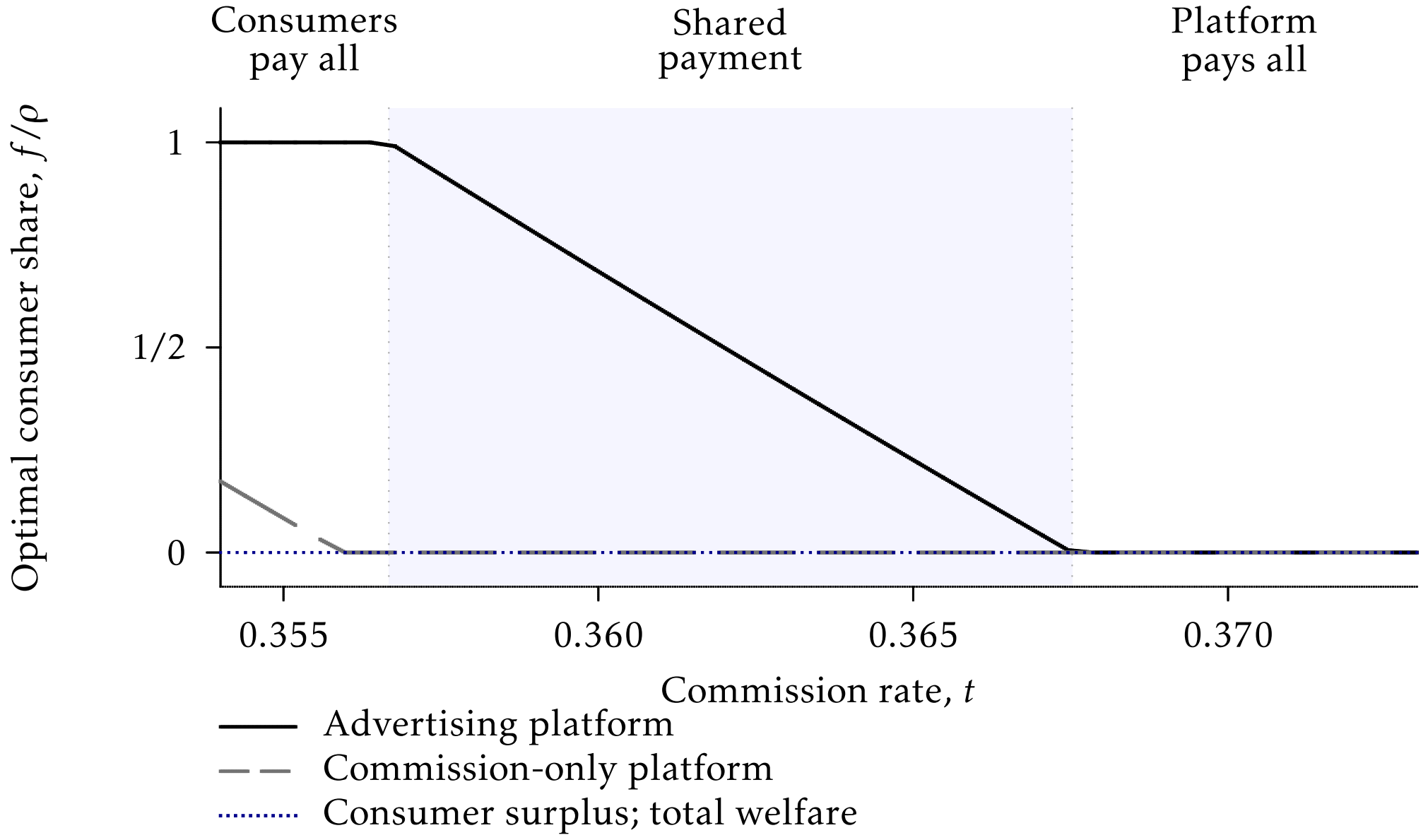
\begin{figure}[htbp]
\centering
\setstretch{1}
\caption{Return funding as the commission rate increases.}\label{fig:sharing-regions}
% Self-contained figure generated from numerical policy optimization.%
\begingroup%
\setlength{\unitlength}{0.32mm}%
\footnotesize%
\begin{picture}(450,250)(0,-22)%
\put(113.238,54.000){\color{blue!4}\rule{63.968mm}{44.800mm}}%
{\color{black!35}\linethickness{.4pt}\qbezier[45](113.238,54.000)(113.238,124.000)(113.238,194.000)}%
{\color{black!35}\linethickness{.4pt}\qbezier[45](313.139,54.000)(313.139,124.000)(313.139,194.000)}%
{\color{black}\linethickness{.55pt}\qbezier(64.000,54.000)(239.000,54.000)(414.000,54.000)}%
{\color{black}\linethickness{.55pt}\qbezier(64.000,54.000)(64.000,124.000)(64.000,194.000)}%
{\color{black}\linethickness{.55pt}\qbezier(82.421,54.000)(82.421,52.000)(82.421,50.000)}%
\put(82.421,40.000){\makebox(0,0)[c]{0.355}}%
{\color{black}\linethickness{.55pt}\qbezier(174.526,54.000)(174.526,52.000)(174.526,50.000)}%
\put(174.526,40.000){\makebox(0,0)[c]{0.360}}%
{\color{black}\linethickness{.55pt}\qbezier(266.632,54.000)(266.632,52.000)(266.632,50.000)}%
\put(266.632,40.000){\makebox(0,0)[c]{0.365}}%
{\color{black}\linethickness{.55pt}\qbezier(358.737,54.000)(358.737,52.000)(358.737,50.000)}%
\put(358.737,40.000){\makebox(0,0)[c]{0.370}}%
{\color{black}\linethickness{.55pt}\qbezier(60.000,64.000)(62.000,64.000)(64.000,64.000)}%
\put(54.000,64.000){\makebox(0,0)[r]{0}}%
{\color{black}\linethickness{.55pt}\qbezier(60.000,124.000)(62.000,124.000)(64.000,124.000)}%
\put(54.000,124.000){\makebox(0,0)[r]{$1/2$}}%
{\color{black}\linethickness{.55pt}\qbezier(60.000,184.000)(62.000,184.000)(64.000,184.000)}%
\put(54.000,184.000){\makebox(0,0)[r]{1}}%
\put(88.619,213.000){\makebox(0,0)[c]{\shortstack{Consumers\\pay all}}}%
\put(213.188,213.000){\makebox(0,0)[c]{\shortstack{Shared\\payment}}}%
\put(363.569,213.000){\makebox(0,0)[c]{\shortstack{Platform\\pays all}}}%
\put(239.000,23.000){\makebox(0,0)[c]{Commission rate, $t$}}%
\put(8.000,124.000){\makebox(0,0)[c]{\rotatebox{90}{Optimal consumer share, $f/\rho$}}}%
{\color{black}\linethickness{1pt}\qbezier(64.000,184.000)(67.646,184.000)(71.292,184.000)}%
{\color{black}\linethickness{1pt}\qbezier(71.292,184.000)(74.938,184.000)(78.583,184.000)}%
{\color{black}\linethickness{1pt}\qbezier(78.583,184.000)(82.229,184.000)(85.875,184.000)}%
{\color{black}\linethickness{1pt}\qbezier(85.875,184.000)(89.521,184.000)(93.167,184.000)}%
{\color{black}\linethickness{1pt}\qbezier(93.167,184.000)(96.812,184.000)(100.458,184.000)}%
{\color{black}\linethickness{1pt}\qbezier(100.458,184.000)(104.104,184.000)(107.750,184.000)}%
{\color{black}\linethickness{1pt}\qbezier(107.750,184.000)(111.396,183.438)(115.042,182.877)}%
{\color{black}\linethickness{1pt}\qbezier(115.042,182.877)(118.687,180.610)(122.333,178.343)}%
{\color{black}\linethickness{1pt}\qbezier(122.333,178.343)(125.979,176.081)(129.625,173.820)}%
{\color{black}\linethickness{1pt}\qbezier(129.625,173.820)(133.271,171.565)(136.917,169.309)}%
{\color{black}\linethickness{1pt}\qbezier(136.917,169.309)(140.562,167.060)(144.208,164.810)}%
{\color{black}\linethickness{1pt}\qbezier(144.208,164.810)(147.854,162.566)(151.500,160.322)}%
{\color{black}\linethickness{1pt}\qbezier(151.500,160.322)(155.146,158.084)(158.792,155.846)}%
{\color{black}\linethickness{1pt}\qbezier(158.792,155.846)(162.438,153.614)(166.083,151.382)}%
{\color{black}\linethickness{1pt}\qbezier(166.083,151.382)(169.729,149.156)(173.375,146.930)}%
{\color{black}\linethickness{1pt}\qbezier(173.375,146.930)(177.021,144.710)(180.667,142.489)}%
{\color{black}\linethickness{1pt}\qbezier(180.667,142.489)(184.313,140.275)(187.958,138.061)}%
{\color{black}\linethickness{1pt}\qbezier(187.958,138.061)(191.604,135.853)(195.250,133.645)}%
{\color{black}\linethickness{1pt}\qbezier(195.250,133.645)(198.896,131.443)(202.542,129.241)}%
{\color{black}\linethickness{1pt}\qbezier(202.542,129.241)(206.188,127.045)(209.833,124.849)}%
{\color{black}\linethickness{1pt}\qbezier(209.833,124.849)(213.479,122.659)(217.125,120.470)}%
{\color{black}\linethickness{1pt}\qbezier(217.125,120.470)(220.771,118.286)(224.417,116.103)}%
{\color{black}\linethickness{1pt}\qbezier(224.417,116.103)(228.062,113.926)(231.708,111.748)}%
{\color{black}\linethickness{1pt}\qbezier(231.708,111.748)(235.354,109.577)(239.000,107.407)}%
{\color{black}\linethickness{1pt}\qbezier(239.000,107.407)(242.646,105.242)(246.292,103.078)}%
{\color{black}\linethickness{1pt}\qbezier(246.292,103.078)(249.938,100.919)(253.583,98.761)}%
{\color{black}\linethickness{1pt}\qbezier(253.583,98.761)(257.229,96.610)(260.875,94.458)}%
{\color{black}\linethickness{1pt}\qbezier(260.875,94.458)(264.521,92.313)(268.167,90.168)}%
{\color{black}\linethickness{1pt}\qbezier(268.167,90.168)(271.812,88.029)(275.458,85.890)}%
{\color{black}\linethickness{1pt}\qbezier(275.458,85.890)(279.104,83.758)(282.750,81.626)}%
{\color{black}\linethickness{1pt}\qbezier(282.750,81.626)(286.396,79.501)(290.042,77.376)}%
{\color{black}\linethickness{1pt}\qbezier(290.042,77.376)(293.687,75.257)(297.333,73.138)}%
{\color{black}\linethickness{1pt}\qbezier(297.333,73.138)(300.979,71.026)(304.625,68.914)}%
{\color{black}\linethickness{1pt}\qbezier(304.625,68.914)(308.271,66.809)(311.917,64.704)}%
{\color{black}\linethickness{1pt}\qbezier(311.917,64.704)(315.562,64.352)(319.208,64.000)}%
{\color{black}\linethickness{1pt}\qbezier(319.208,64.000)(322.854,64.000)(326.500,64.000)}%
{\color{black}\linethickness{1pt}\qbezier(326.500,64.000)(330.146,64.000)(333.792,64.000)}%
{\color{black}\linethickness{1pt}\qbezier(333.792,64.000)(337.438,64.000)(341.083,64.000)}%
{\color{black}\linethickness{1pt}\qbezier(341.083,64.000)(344.729,64.000)(348.375,64.000)}%
{\color{black}\linethickness{1pt}\qbezier(348.375,64.000)(352.021,64.000)(355.667,64.000)}%
{\color{black}\linethickness{1pt}\qbezier(355.667,64.000)(359.313,64.000)(362.958,64.000)}%
{\color{black}\linethickness{1pt}\qbezier(362.958,64.000)(366.604,64.000)(370.250,64.000)}%
{\color{black}\linethickness{1pt}\qbezier(370.250,64.000)(373.896,64.000)(377.542,64.000)}%
{\color{black}\linethickness{1pt}\qbezier(377.542,64.000)(381.188,64.000)(384.833,64.000)}%
{\color{black}\linethickness{1pt}\qbezier(384.833,64.000)(388.479,64.000)(392.125,64.000)}%
{\color{black}\linethickness{1pt}\qbezier(392.125,64.000)(395.771,64.000)(399.417,64.000)}%
{\color{black}\linethickness{1pt}\qbezier(399.417,64.000)(403.062,64.000)(406.708,64.000)}%
{\color{black}\linethickness{1pt}\qbezier(406.708,64.000)(410.354,64.000)(414.000,64.000)}%
{\color{black!55}\linethickness{1pt}\qbezier(64.000,84.814)(67.646,82.678)(71.292,80.541)}%
{\color{black!55}\linethickness{1pt}\qbezier(71.292,80.541)(74.938,78.412)(78.583,76.283)}%
{\color{black!55}\linethickness{1pt}\qbezier(78.583,76.283)(82.229,74.161)(85.875,72.040)}%
{\color{black!55}\linethickness{1pt}\qbezier(93.167,67.812)(96.812,65.906)(100.458,64.000)}%
{\color{black!55}\linethickness{1pt}\qbezier(100.458,64.000)(104.104,64.000)(107.750,64.000)}%
{\color{black!55}\linethickness{1pt}\qbezier(107.750,64.000)(111.396,64.000)(115.042,64.000)}%
{\color{black!55}\linethickness{1pt}\qbezier(122.333,64.000)(125.979,64.000)(129.625,64.000)}%
{\color{black!55}\linethickness{1pt}\qbezier(129.625,64.000)(133.271,64.000)(136.917,64.000)}%
{\color{black!55}\linethickness{1pt}\qbezier(136.917,64.000)(140.562,64.000)(144.208,64.000)}%
{\color{black!55}\linethickness{1pt}\qbezier(151.500,64.000)(155.146,64.000)(158.792,64.000)}%
{\color{black!55}\linethickness{1pt}\qbezier(158.792,64.000)(162.438,64.000)(166.083,64.000)}%
{\color{black!55}\linethickness{1pt}\qbezier(166.083,64.000)(169.729,64.000)(173.375,64.000)}%
{\color{black!55}\linethickness{1pt}\qbezier(180.667,64.000)(184.313,64.000)(187.958,64.000)}%
{\color{black!55}\linethickness{1pt}\qbezier(187.958,64.000)(191.604,64.000)(195.250,64.000)}%
{\color{black!55}\linethickness{1pt}\qbezier(195.250,64.000)(198.896,64.000)(202.542,64.000)}%
{\color{black!55}\linethickness{1pt}\qbezier(209.833,64.000)(213.479,64.000)(217.125,64.000)}%
{\color{black!55}\linethickness{1pt}\qbezier(217.125,64.000)(220.771,64.000)(224.417,64.000)}%
{\color{black!55}\linethickness{1pt}\qbezier(224.417,64.000)(228.062,64.000)(231.708,64.000)}%
{\color{black!55}\linethickness{1pt}\qbezier(239.000,64.000)(242.646,64.000)(246.292,64.000)}%
{\color{black!55}\linethickness{1pt}\qbezier(246.292,64.000)(249.938,64.000)(253.583,64.000)}%
{\color{black!55}\linethickness{1pt}\qbezier(253.583,64.000)(257.229,64.000)(260.875,64.000)}%
{\color{black!55}\linethickness{1pt}\qbezier(268.167,64.000)(271.812,64.000)(275.458,64.000)}%
{\color{black!55}\linethickness{1pt}\qbezier(275.458,64.000)(279.104,64.000)(282.750,64.000)}%
{\color{black!55}\linethickness{1pt}\qbezier(282.750,64.000)(286.396,64.000)(290.042,64.000)}%
{\color{black!55}\linethickness{1pt}\qbezier(297.333,64.000)(300.979,64.000)(304.625,64.000)}%
{\color{black!55}\linethickness{1pt}\qbezier(304.625,64.000)(308.271,64.000)(311.917,64.000)}%
{\color{black!55}\linethickness{1pt}\qbezier(311.917,64.000)(315.562,64.000)(319.208,64.000)}%
{\color{black!55}\linethickness{1pt}\qbezier(326.500,64.000)(330.146,64.000)(333.792,64.000)}%
{\color{black!55}\linethickness{1pt}\qbezier(333.792,64.000)(337.438,64.000)(341.083,64.000)}%
{\color{black!55}\linethickness{1pt}\qbezier(341.083,64.000)(344.729,64.000)(348.375,64.000)}%
{\color{black!55}\linethickness{1pt}\qbezier(355.667,64.000)(359.313,64.000)(362.958,64.000)}%
{\color{black!55}\linethickness{1pt}\qbezier(362.958,64.000)(366.604,64.000)(370.250,64.000)}%
{\color{black!55}\linethickness{1pt}\qbezier(370.250,64.000)(373.896,64.000)(377.542,64.000)}%
{\color{black!55}\linethickness{1pt}\qbezier(384.833,64.000)(388.479,64.000)(392.125,64.000)}%
{\color{black!55}\linethickness{1pt}\qbezier(392.125,64.000)(395.771,64.000)(399.417,64.000)}%
{\color{black!55}\linethickness{1pt}\qbezier(399.417,64.000)(403.062,64.000)(406.708,64.000)}%
{\color{blue!55!black}\linethickness{.8pt}\qbezier[3](64.000,64.000)(67.646,64.000)(71.292,64.000)}%
{\color{blue!55!black}\linethickness{.8pt}\qbezier[3](71.292,64.000)(74.938,64.000)(78.583,64.000)}%
{\color{blue!55!black}\linethickness{.8pt}\qbezier[3](78.583,64.000)(82.229,64.000)(85.875,64.000)}%
{\color{blue!55!black}\linethickness{.8pt}\qbezier[3](85.875,64.000)(89.521,64.000)(93.167,64.000)}%
{\color{blue!55!black}\linethickness{.8pt}\qbezier[3](93.167,64.000)(96.812,64.000)(100.458,64.000)}%
{\color{blue!55!black}\linethickness{.8pt}\qbezier[3](100.458,64.000)(104.104,64.000)(107.750,64.000)}%
{\color{blue!55!black}\linethickness{.8pt}\qbezier[3](107.750,64.000)(111.396,64.000)(115.042,64.000)}%
{\color{blue!55!black}\linethickness{.8pt}\qbezier[3](115.042,64.000)(118.687,64.000)(122.333,64.000)}%
{\color{blue!55!black}\linethickness{.8pt}\qbezier[3](122.333,64.000)(125.979,64.000)(129.625,64.000)}%
{\color{blue!55!black}\linethickness{.8pt}\qbezier[3](129.625,64.000)(133.271,64.000)(136.917,64.000)}%
{\color{blue!55!black}\linethickness{.8pt}\qbezier[3](136.917,64.000)(140.562,64.000)(144.208,64.000)}%
{\color{blue!55!black}\linethickness{.8pt}\qbezier[3](144.208,64.000)(147.854,64.000)(151.500,64.000)}%
{\color{blue!55!black}\linethickness{.8pt}\qbezier[3](151.500,64.000)(155.146,64.000)(158.792,64.000)}%
{\color{blue!55!black}\linethickness{.8pt}\qbezier[3](158.792,64.000)(162.438,64.000)(166.083,64.000)}%
{\color{blue!55!black}\linethickness{.8pt}\qbezier[3](166.083,64.000)(169.729,64.000)(173.375,64.000)}%
{\color{blue!55!black}\linethickness{.8pt}\qbezier[3](173.375,64.000)(177.021,64.000)(180.667,64.000)}%
{\color{blue!55!black}\linethickness{.8pt}\qbezier[3](180.667,64.000)(184.313,64.000)(187.958,64.000)}%
{\color{blue!55!black}\linethickness{.8pt}\qbezier[3](187.958,64.000)(191.604,64.000)(195.250,64.000)}%
{\color{blue!55!black}\linethickness{.8pt}\qbezier[3](195.250,64.000)(198.896,64.000)(202.542,64.000)}%
{\color{blue!55!black}\linethickness{.8pt}\qbezier[3](202.542,64.000)(206.188,64.000)(209.833,64.000)}%
{\color{blue!55!black}\linethickness{.8pt}\qbezier[3](209.833,64.000)(213.479,64.000)(217.125,64.000)}%
{\color{blue!55!black}\linethickness{.8pt}\qbezier[3](217.125,64.000)(220.771,64.000)(224.417,64.000)}%
{\color{blue!55!black}\linethickness{.8pt}\qbezier[3](224.417,64.000)(228.062,64.000)(231.708,64.000)}%
{\color{blue!55!black}\linethickness{.8pt}\qbezier[3](231.708,64.000)(235.354,64.000)(239.000,64.000)}%
{\color{blue!55!black}\linethickness{.8pt}\qbezier[3](239.000,64.000)(242.646,64.000)(246.292,64.000)}%
{\color{blue!55!black}\linethickness{.8pt}\qbezier[3](246.292,64.000)(249.938,64.000)(253.583,64.000)}%
{\color{blue!55!black}\linethickness{.8pt}\qbezier[3](253.583,64.000)(257.229,64.000)(260.875,64.000)}%
{\color{blue!55!black}\linethickness{.8pt}\qbezier[3](260.875,64.000)(264.521,64.000)(268.167,64.000)}%
{\color{blue!55!black}\linethickness{.8pt}\qbezier[3](268.167,64.000)(271.812,64.000)(275.458,64.000)}%
{\color{blue!55!black}\linethickness{.8pt}\qbezier[3](275.458,64.000)(279.104,64.000)(282.750,64.000)}%
{\color{blue!55!black}\linethickness{.8pt}\qbezier[3](282.750,64.000)(286.396,64.000)(290.042,64.000)}%
{\color{blue!55!black}\linethickness{.8pt}\qbezier[3](290.042,64.000)(293.687,64.000)(297.333,64.000)}%
{\color{blue!55!black}\linethickness{.8pt}\qbezier[3](297.333,64.000)(300.979,64.000)(304.625,64.000)}%
{\color{blue!55!black}\linethickness{.8pt}\qbezier[3](304.625,64.000)(308.271,64.000)(311.917,64.000)}%
{\color{blue!55!black}\linethickness{.8pt}\qbezier[3](311.917,64.000)(315.562,64.000)(319.208,64.000)}%
{\color{blue!55!black}\linethickness{.8pt}\qbezier[3](319.208,64.000)(322.854,64.000)(326.500,64.000)}%
{\color{blue!55!black}\linethickness{.8pt}\qbezier[3](326.500,64.000)(330.146,64.000)(333.792,64.000)}%
{\color{blue!55!black}\linethickness{.8pt}\qbezier[3](333.792,64.000)(337.438,64.000)(341.083,64.000)}%
{\color{blue!55!black}\linethickness{.8pt}\qbezier[3](341.083,64.000)(344.729,64.000)(348.375,64.000)}%
{\color{blue!55!black}\linethickness{.8pt}\qbezier[3](348.375,64.000)(352.021,64.000)(355.667,64.000)}%
{\color{blue!55!black}\linethickness{.8pt}\qbezier[3](355.667,64.000)(359.313,64.000)(362.958,64.000)}%
{\color{blue!55!black}\linethickness{.8pt}\qbezier[3](362.958,64.000)(366.604,64.000)(370.250,64.000)}%
{\color{blue!55!black}\linethickness{.8pt}\qbezier[3](370.250,64.000)(373.896,64.000)(377.542,64.000)}%
{\color{blue!55!black}\linethickness{.8pt}\qbezier[3](377.542,64.000)(381.188,64.000)(384.833,64.000)}%
{\color{blue!55!black}\linethickness{.8pt}\qbezier[3](384.833,64.000)(388.479,64.000)(392.125,64.000)}%
{\color{blue!55!black}\linethickness{.8pt}\qbezier[3](392.125,64.000)(395.771,64.000)(399.417,64.000)}%
{\color{blue!55!black}\linethickness{.8pt}\qbezier[3](399.417,64.000)(403.062,64.000)(406.708,64.000)}%
{\color{blue!55!black}\linethickness{.8pt}\qbezier[3](406.708,64.000)(410.354,64.000)(414.000,64.000)}%
{\color{black}\linethickness{.9pt}\qbezier(64.000,13.000)(75.000,13.000)(86.000,13.000)}%
\put(94.000,13.000){\makebox(0,0)[l]{Advertising platform}}%
{\color{black!55}\linethickness{.9pt}\qbezier(64.000,0.000)(68.000,0.000)(72.000,0.000)}%
{\color{black!55}\linethickness{.9pt}\qbezier(76.000,0.000)(80.000,0.000)(84.000,0.000)}%
\put(94.000,0.000){\makebox(0,0)[l]{Commission-only platform}}%
{\color{blue!55!black}\linethickness{.9pt}\qbezier[10](64.000,-13.000)(75.000,-13.000)(86.000,-13.000)}%
\put(94.000,-13.000){\makebox(0,0)[l]{Consumer surplus; total welfare}}%
\end{picture}%
\endgroup%

\par\smallskip
\begin{minipage}{.98\textwidth}\footnotesize
Numerical illustration with $\alpha=1$, $\eta=0.4$, $\rho=0.002$, and $s=0.006$. Each curve optimizes over all feasible payments. At all plotted optima firms bear none of the loss, so the platform pays the share $1-f/\rho$. The upper labels identify the advertising platform's choice. Its thresholds are approximately $t_C=0.3567$ and $t_P=0.3675$. The commission-only platform also shares payment at the lowest displayed rates, but assigns consumers a smaller contribution. Consumer surplus and total welfare favor full platform payment. The curves are numerical comparisons on the selected price equilibria, including rates outside the local interval in Proposition~\ref{prop:policy}(b).
\end{minipage}
\end{figure}

Figure~\ref{fig:sharing-regions} illustrates the commission comparison at positive costs. Advertising increases the consumer contribution relative to a commission-only platform. As commissions rise, the platform gradually substitutes its own payment for consumer contributions. In this example consumers would prefer full platform payment throughout, including the range where sharing maximizes platform profit.

\section{Welfare and the size of policy effects}\label{sec:welfare}

The joint optimum identifies who pays; it need not identify the allocation preferred by consumers or society. A returned product uses resources regardless of who bears their cost. The return policy allocates this loss among the three parties. We first evaluate the small-loss funding policies using consumer surplus and total welfare. We then study finite losses to measure the policy effects and determine whether the welfare ranking changes.

Let $M$ denote expected retained product value per participant, and let $q$ be the probability of inspecting product 2. Resource surplus per participant, before the private participation cost, and total welfare are
\begin{equation}
 V=M-c(S_1+S_2)+\eta(R_1+R_2)-sq,
 \qquad W=UV-\frac{U^2}{2}.
 \label{eq:welfare}
\end{equation}
The shipment and recovery terms count the resources used by kept and returned trials. These terms already include the resource loss on each returned item. Allocating that loss through $f$ and $b$ changes who bears it; commissions and auction payments are transfers. Consumer surplus is $U^2/2$, so consumer protection depends on the private value of shopping, whereas total welfare also values the resources and products generated by participants.

\begin{proposition}[Who should fund free returns?]\label{prop:welfare}
Under the all-policy conditions of Lemma~\ref{prop:prices}, fix the primitives away from the zero price-response threshold in Lemma~\ref{lem:funding-incidence}. For sufficiently small positive losses, consumer surplus and total welfare have the same unique optimal policy over~\eqref{eq:feasible}:
\begin{enumerate}[label=\textnormal{(\alph*)}]
\item If reimbursement raises the limiting sum of prices, both are maximized by firm-paid free returns, $(0,0)$.
\item If reimbursement lowers that sum, both are maximized by platform-paid free returns, $(0,\rho)$.
\end{enumerate}
Thus the suitability and cost thresholds in Lemma~\ref{lem:funding-incidence} also classify the welfare-maximizing payer in the small-cost region.
\end{proposition}

The proposition applies across the parameter range of the pricing lemma, rather than only to the two policy examples. When return liability encourages firms to keep prices low, removing it harms consumers; leaving firms to pay preserves both free returns and the incentive to retain purchases. When reimbursement instead lowers expected price payments through participation, platform funding improves shopping opportunities. Consumer contributions add a direct burden that does not improve either welfare criterion enough to be optimal in these small-cost comparisons.

These claims follow from the same price response used for platform revenue. At the symmetric starting point,
\begin{align}
 CS_b^0&=-U_0D_0\,\partial_b(p_1+p_2)\big|_0,\nonumber\\
 W_b^0&=-(p_0-\eta)
 [U_0\alpha(1-\alpha(1-p_0))+2D_0^2]
 \partial_b(p_1+p_2)\big|_0.
 \label{eq:reference-welfare-slopes}
\end{align}
Both factors preceding the price response are positive. The remaining step is to compare the two ways of relieving firms' return liability. Appendix~\ref{app:welfareproof} shows that, when reimbursement improves welfare, replacing a consumer contribution with reimbursement improves both welfare measures as well. Starting from any policy, one can therefore raise reimbursement until firms bear no return loss, then replace the remaining consumer contribution along $f+b=\rho$. These changes lead to platform-paid free returns. When reimbursement reduces welfare, the same calculations show that consumer contributions also reduce both measures. Lowering both payments then leads to firm-paid free returns. The strict signs persist over the entire feasible policy set for sufficiently small losses, giving the unique optima in the proposition.

The platform's preferred payer need not coincide with this welfare choice. In the firm-funding region of Proposition~\ref{prop:policy}, advertising moves the platform from the common welfare optimum to consumer funding. In the sharing region, platform funding is the common welfare optimum, and advertising reduces both welfare measures whenever it induces a positive consumer contribution. The difference comes from the revenue attached to the relative profit of the prominent firm.

\subsection{Why the welfare ranking can change at larger losses}
The link between private utility and resource surplus is useful beyond the small-cost comparison. Consumers receive the value of the product they keep, pay its price, and bear their allocated return costs and inspection costs. Hence
\begin{align}
 U&=M-\sum_i p_iD_i-f(R_1+R_2)-sq,\nonumber\\
 V&=U+\sum_i(p_i-c)D_i-(\rho-f)(R_1+R_2).
 \label{eq:private-resource-accounting}
\end{align}
The second line follows from $S_i=D_i+R_i$ and $\rho=c-\eta$. It adds the retained-sale surplus to shopping utility and subtracts the part of the return loss not already borne by consumers. For a policy change holding product and inspection costs fixed, differentiate this identity and use the consumer envelope relations:

\begin{align}
 dU&=-D_1\,dp_1-D_2\,dp_2-(R_1+R_2)\,df,\nonumber\\
 dV&=\sum_i(p_i-c)\,dD_i-(\rho-f)\,d(R_1+R_2),\nonumber\\
 dW&=U\,dV+(V-U)\,dU.
 \label{eq:welfareidentities}
\end{align}
In the second line, the price and consumer-cost terms cancel against $dU$. Match values and inspection costs enter through consumers' optimal choices, leaving the marginal resource effects of retained purchases and returns. The last line separates the change in resource surplus per participant from the change in participation. The marginal entrant has private participation cost $U$, so $V-U$ is the social surplus from that additional participant.

For a finite comparison, let $\Delta$ denote consumer funding minus firm funding at unchanged primitives, and put a bar over the arithmetic mean across the two policies. Then
\begin{equation}
 \Delta W=\bar U\,\Delta V+(\bar V-\bar U)\,\Delta U.
 \label{eq:welfare-endpoint-decomposition}
\end{equation}
A consumer return cost can lower the private value of shopping while improving the allocation among those who enter. When the resource gain exceeds the loss from reduced participation, total welfare rises even though consumer surplus falls. Section~\ref{sec:finite} establishes such a comparison. Participation costs are independent of product preferences, so the change in resource surplus per participant comes from shopping decisions rather than a different distribution of match draws among entrants.

Each firm retains the operating profit of position 2 after paying for advertising. Aggregate accounting therefore also gives
\begin{equation}
 P^{\mathrm{ad}}=W-CS-2U\pi_2.
 \label{eq:welfare-platform-accounting}
\end{equation}
Higher platform profit can reflect an improved allocation or a transfer from consumers and firms. The finite-cost comparisons distinguish these possibilities rather than inferring efficiency from either platform revenue or the fraction of orders kept.

\subsection{Finite return losses}\label{sec:finite}

At positive return losses, consumer contributions affect the cutoff for another trial, and the two firms' prices respond differently. The price equations and platform revenues therefore retain terms that disappear from the small-loss comparison. We establish a result on intervals over which suitability, commission, recovery value, return loss, and the inspection-cost ratio can vary independently. The return-loss interval has a strictly positive lower bound. We then hold the other market conditions fixed and compare two loss intervals to distinguish the policy disagreement from its welfare consequences.

Both platforms face the same price equilibria at each payment rule. We follow these equilibria continuously across policies and check every unilateral price deviation, including deviations that change trial order.{\interfootnotelinepenalty=10000\footnote{At finite losses, the proof establishes this continuous selection of global price equilibria over the full policy set. It does not exclude other, disconnected equilibria. The policy comparisons use the same selection for both revenue models.}}

\begin{proposition}[Policy choice at positive return losses]\label{prop:finite}
\textnormal{(a) Variation in market conditions.}
There exist intervals of positive length within the model's parameter domain,
\begin{equation}
\begin{gathered}
 \alpha\in[\underline\alpha,\overline\alpha],\quad
 t\in[\underline t,\overline t],\quad
 \eta\in[\underline\eta,\overline\eta],\\
 \rho\in[\underline\rho,\overline\rho],\quad
 \xi\in[\underline\xi,\overline\xi],\qquad
 0<\underline\rho<\overline\rho,
\end{gathered}
\label{eq:finite-parameter-intervals}
\end{equation}
such that the following holds for every independent choice of these five parameters, with $c=\eta+\rho$ and $s=\alpha\xi\rho$. Commission-only profit, consumer surplus, and total welfare are uniquely maximized by firm-paid free returns, $(f,b)=(0,0)$. The advertising platform uniquely chooses full consumer payment, $(\rho,0)$.

\textnormal{(b) The welfare comparison.}
There also exist fixed values of $(\alpha,t,\eta,\xi)$ and two return-loss intervals, a lower interval $[\underline\rho_L,\overline\rho_L]$ and a higher interval $[\underline\rho_H,\overline\rho_H]$, with
\[
 0<\underline\rho_L<\overline\rho_L<\underline\rho_H<\overline\rho_H,
\]
such that, with $c=\eta+\rho$ and $s=\alpha\xi\rho$, the unique platform and welfare optima are
\[
\begin{array}{c|cccc}
 \text{Return loss} & P^0 & P^{\mathrm{ad}} & CS & W\\ \hline
 {[\underline\rho_L,\overline\rho_L]} &(0,0)&(\rho,0)&(0,0)&(0,0)\\
 {[\underline\rho_H,\overline\rho_H]} &(0,0)&(\rho,0)&(0,0)&(\rho,0).
\end{array}
\]
In both parts, the continuous price equilibria described above cover every feasible payment rule. Each price is a strict global best response, advertising bids and profits after auction payments are positive, and both firms retain more than half their shipments. All stated optima are over the entire feasible policy set.
\end{proposition}

Part (a) shows that the policy difference persists as several market conditions vary independently, with return losses bounded away from zero. Appendix~\ref{app:finite-box} constructs the intervals and verifies the comparison for every parameter combination. On those intervals, the change from firm-paid free returns to full consumer payment lowers consumer surplus by between $1.2$ and $2.7$ percent. Part (b) holds suitability, commission, recovery value, and the inspection-cost ratio fixed. The same change in the payer harms consumers in both loss intervals, while its effect on total welfare differs. Appendix~\ref{app:finite-policy} proves these comparisons.

The construction for part (b) also gives a scale for the losses. In the lower interval, return losses are about $6.35$--$9.24$ percent of shipment cost; advertising changes the preferred payer from firms to consumers and reduces both consumer surplus and total welfare. In the higher interval, losses are about one quarter of shipment cost. Consumers are still worse off, but full consumer payment now uniquely maximizes total welfare over all feasible sharing arrangements. Thus the same policy disagreement can have different efficiency consequences.

At the representative loss $\rho=0.02$ in the higher interval of this construction, the consumer contribution raises the value of products retained per participant. The additional retained purchases cost more to supply, but their value exceeds that added cost. Fewer returned trials and less second-product inspection save further resources. These gains among participants outweigh the welfare loss from reduced participation. Appendix~\ref{app:resource-decomposition} verifies each component on the same equilibrium branch. Their signs describe this comparison; the policy ranking in the proposition covers the stated interval.

The proof checks all unilateral price deviations, including changes in trial order, and compares policies over the whole feasible triangle. For the second interval, commission-only profit need not fall with the consumer contribution at every policy. Reducing reimbursement always raises that profit; once reimbursement is zero, reducing the consumer contribution raises it along $b=0$. This path establishes the commission-only optimum. The advertising and welfare optima follow from their policy-derivative signs.

A complete classification at arbitrary positive losses requires additional comparisons. Within the current shopping order, the trial cutoff depends on $\sqrt{2(f+s/\alpha)}$, and the two price equations are coupled polynomial equations of degree up to five. A change in the consumer contribution or a price deviation can also change which product is tried first or whether a further trial is worthwhile. The global pricing problem must compare these regions before the platform can choose its payment rule. Moreover, the linear revenue comparison in Table~\ref{tab:policy-gains} no longer applies: return volume and participation can make partial payment optimal. The finite-loss proof therefore verifies equilibrium and policy inequalities over stated parameter ranges, rather than extrapolating the small-loss classification to every market.

To assess magnitudes and examine losses between and beyond the two proved intervals, Table~\ref{tab:finite-magnitudes} reports numerical comparisons for the fixed market parameters used to construct part (b), stated in the table note. Each entry compares full consumer
funding with firm-paid free returns in the same market, with prices and
participation adjusting. Each percentage uses the value under firm-paid free returns as its denominator.
These calculated magnitudes complement the interval-certified statements in
Proposition~\ref{prop:finite}.

\begin{table}[htbp]
\centering
\caption{Consumer funding relative to firm-paid free returns: numerical magnitudes}
\label{tab:finite-magnitudes}
\small
\renewcommand{\arraystretch}{1.15}
\begin{tabular}{rrrrr}
\toprule
$\rho$ & $\rho/c$ & $\Delta CS/CS$ & $\Delta W/W$ & $\Delta P^{\mathrm{ad}}/P^{\mathrm{ad}}$\\
\midrule
$0.005$ & $7.82\%$ & $-1.81\%$ & $-0.051\%$ & $0.83\%$\\
$0.010$ & $14.50\%$ & $-3.50\%$ & $-0.041\%$ & $1.58\%$\\
$0.020$ & $25.33\%$ & $-6.51\%$ & $0.182\%$ & $2.66\%$\\
$0.040$ & $40.42\%$ & $-10.87\%$ & $1.623\%$ & $3.06\%$\\
\bottomrule
\end{tabular}
\par\smallskip
\begin{minipage}{0.96\linewidth}\footnotesize
Parameters are $\alpha=1/5$, $t=3/20$, $\eta=783/13280$, $c=\eta+\rho$, and $s=3\rho/5$. Each change compares full consumer funding, $(f,b)=(\rho,0)$, with firm-paid free returns, $(0,0)$, on the common price branch. Percentage changes use the firm-funded level at the same $\rho$ as the denominator. These are numerical magnitudes; Proposition~\ref{prop:finite} specifies the interval-certified policy regions.
\end{minipage}
\end{table}

At $\rho=0.005$, moving from firm funding to consumer funding lowers commission revenue by $2.2502\times10^{-6}$ per potential consumer and raises advertising revenue by $1.03135\times10^{-5}$. The latter more than covers the former, producing the platform gain of $0.83$ percent. Consumer surplus falls by $1.81$ percent, while total welfare falls by $0.051$ percent. At $\rho=0.02$, consumers again lose, although the balance of resource gains and losses turns positive. Consumer harm is common to the two proved regions; the efficiency ranking depends on resource use and participation.

Under firm funding at $\rho=0.005$, the two firms retain approximately $59.0$ and $60.4$ percent of their shipments. This example describes frequent trial returns. The aggregate return rate in the introduction establishes the scale of returns across online retail; the calculations here quantify the mechanism in a higher-return market.

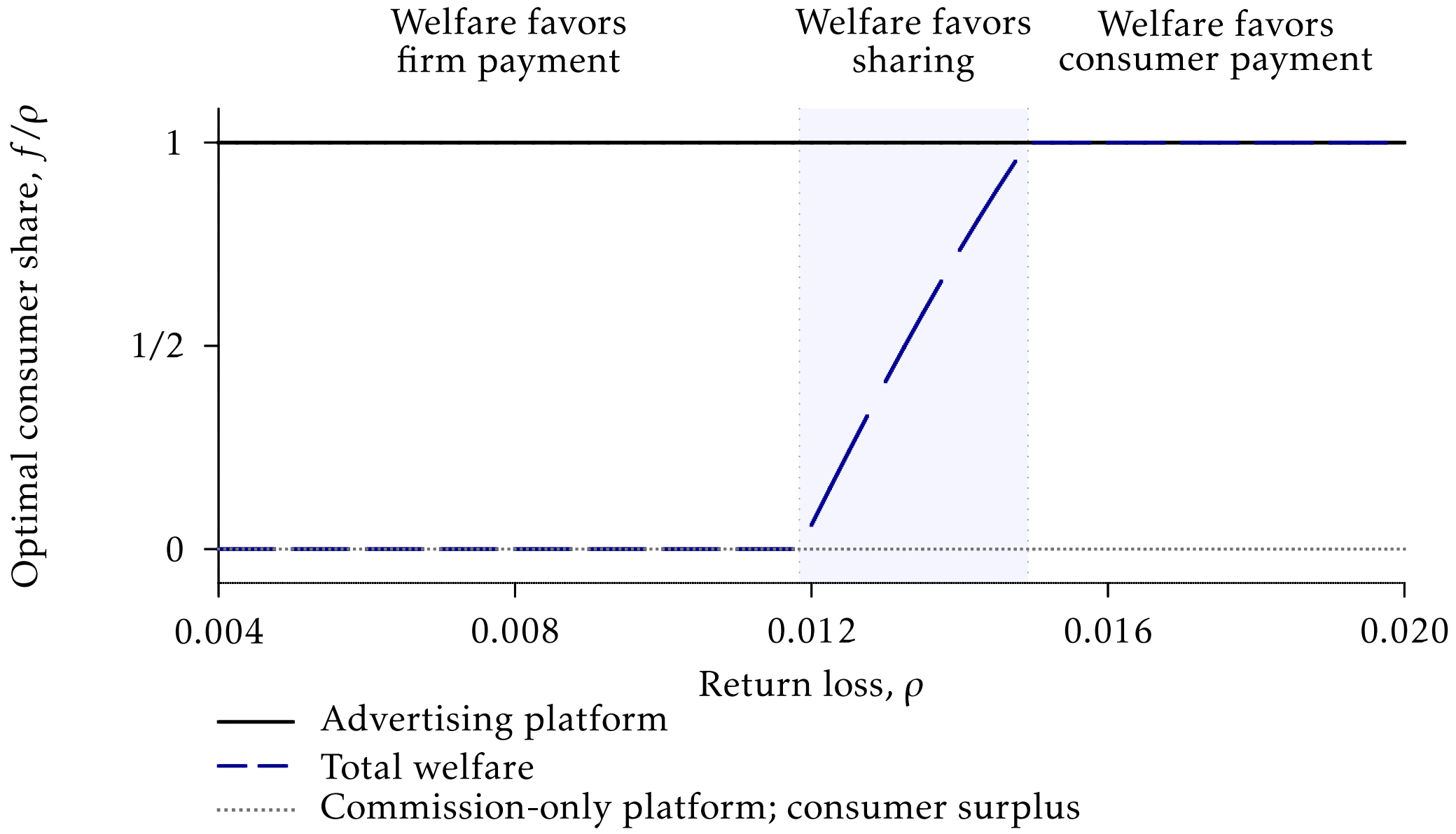
\begin{figure}[htbp]
\centering
\setstretch{1}
\caption{Return funding and welfare at positive losses.}\label{fig:finite-regions}
% Self-contained figure generated from numerical policy optimization.%
\begingroup%
\setlength{\unitlength}{0.32mm}%
\footnotesize%
\begin{picture}(450,250)(0,-22)%
\put(235.443,54.000){\color{blue!4}\rule{21.593mm}{44.800mm}}%
{\color{black!35}\linethickness{.4pt}\qbezier[45](235.443,54.000)(235.443,124.000)(235.443,194.000)}%
{\color{black!35}\linethickness{.4pt}\qbezier[45](302.920,54.000)(302.920,124.000)(302.920,194.000)}%
{\color{black}\linethickness{.55pt}\qbezier(64.000,54.000)(239.000,54.000)(414.000,54.000)}%
{\color{black}\linethickness{.55pt}\qbezier(64.000,54.000)(64.000,124.000)(64.000,194.000)}%
{\color{black}\linethickness{.55pt}\qbezier(64.000,54.000)(64.000,52.000)(64.000,50.000)}%
\put(64.000,40.000){\makebox(0,0)[c]{0.004}}%
{\color{black}\linethickness{.55pt}\qbezier(151.500,54.000)(151.500,52.000)(151.500,50.000)}%
\put(151.500,40.000){\makebox(0,0)[c]{0.008}}%
{\color{black}\linethickness{.55pt}\qbezier(239.000,54.000)(239.000,52.000)(239.000,50.000)}%
\put(239.000,40.000){\makebox(0,0)[c]{0.012}}%
{\color{black}\linethickness{.55pt}\qbezier(326.500,54.000)(326.500,52.000)(326.500,50.000)}%
\put(326.500,40.000){\makebox(0,0)[c]{0.016}}%
{\color{black}\linethickness{.55pt}\qbezier(414.000,54.000)(414.000,52.000)(414.000,50.000)}%
\put(414.000,40.000){\makebox(0,0)[c]{0.020}}%
{\color{black}\linethickness{.55pt}\qbezier(60.000,64.000)(62.000,64.000)(64.000,64.000)}%
\put(54.000,64.000){\makebox(0,0)[r]{0}}%
{\color{black}\linethickness{.55pt}\qbezier(60.000,124.000)(62.000,124.000)(64.000,124.000)}%
\put(54.000,124.000){\makebox(0,0)[r]{$1/2$}}%
{\color{black}\linethickness{.55pt}\qbezier(60.000,184.000)(62.000,184.000)(64.000,184.000)}%
\put(54.000,184.000){\makebox(0,0)[r]{1}}%
\put(149.721,213.000){\makebox(0,0)[c]{\shortstack{Welfare favors\\firm payment}}}%
\put(269.181,213.000){\makebox(0,0)[c]{\shortstack{Welfare favors\\sharing}}}%
\put(358.460,213.000){\makebox(0,0)[c]{\shortstack{Welfare favors\\consumer payment}}}%
\put(239.000,23.000){\makebox(0,0)[c]{Return loss, $\rho$}}%
\put(8.000,124.000){\makebox(0,0)[c]{\rotatebox{90}{Optimal consumer share, $f/\rho$}}}%
{\color{black}\linethickness{1pt}\qbezier(64.000,184.000)(66.734,184.000)(69.469,184.000)}%
{\color{black}\linethickness{1pt}\qbezier(69.469,184.000)(72.203,184.000)(74.938,184.000)}%
{\color{black}\linethickness{1pt}\qbezier(74.938,184.000)(77.672,184.000)(80.406,184.000)}%
{\color{black}\linethickness{1pt}\qbezier(80.406,184.000)(83.141,184.000)(85.875,184.000)}%
{\color{black}\linethickness{1pt}\qbezier(85.875,184.000)(88.609,184.000)(91.344,184.000)}%
{\color{black}\linethickness{1pt}\qbezier(91.344,184.000)(94.078,184.000)(96.812,184.000)}%
{\color{black}\linethickness{1pt}\qbezier(96.812,184.000)(99.547,184.000)(102.281,184.000)}%
{\color{black}\linethickness{1pt}\qbezier(102.281,184.000)(105.016,184.000)(107.750,184.000)}%
{\color{black}\linethickness{1pt}\qbezier(107.750,184.000)(110.484,184.000)(113.219,184.000)}%
{\color{black}\linethickness{1pt}\qbezier(113.219,184.000)(115.953,184.000)(118.688,184.000)}%
{\color{black}\linethickness{1pt}\qbezier(118.688,184.000)(121.422,184.000)(124.156,184.000)}%
{\color{black}\linethickness{1pt}\qbezier(124.156,184.000)(126.891,184.000)(129.625,184.000)}%
{\color{black}\linethickness{1pt}\qbezier(129.625,184.000)(132.359,184.000)(135.094,184.000)}%
{\color{black}\linethickness{1pt}\qbezier(135.094,184.000)(137.828,184.000)(140.562,184.000)}%
{\color{black}\linethickness{1pt}\qbezier(140.562,184.000)(143.297,184.000)(146.031,184.000)}%
{\color{black}\linethickness{1pt}\qbezier(146.031,184.000)(148.766,184.000)(151.500,184.000)}%
{\color{black}\linethickness{1pt}\qbezier(151.500,184.000)(154.234,184.000)(156.969,184.000)}%
{\color{black}\linethickness{1pt}\qbezier(156.969,184.000)(159.703,184.000)(162.438,184.000)}%
{\color{black}\linethickness{1pt}\qbezier(162.438,184.000)(165.172,184.000)(167.906,184.000)}%
{\color{black}\linethickness{1pt}\qbezier(167.906,184.000)(170.641,184.000)(173.375,184.000)}%
{\color{black}\linethickness{1pt}\qbezier(173.375,184.000)(176.109,184.000)(178.844,184.000)}%
{\color{black}\linethickness{1pt}\qbezier(178.844,184.000)(181.578,184.000)(184.312,184.000)}%
{\color{black}\linethickness{1pt}\qbezier(184.312,184.000)(187.047,184.000)(189.781,184.000)}%
{\color{black}\linethickness{1pt}\qbezier(189.781,184.000)(192.516,184.000)(195.250,184.000)}%
{\color{black}\linethickness{1pt}\qbezier(195.250,184.000)(197.984,184.000)(200.719,184.000)}%
{\color{black}\linethickness{1pt}\qbezier(200.719,184.000)(203.453,184.000)(206.188,184.000)}%
{\color{black}\linethickness{1pt}\qbezier(206.188,184.000)(208.922,184.000)(211.656,184.000)}%
{\color{black}\linethickness{1pt}\qbezier(211.656,184.000)(214.391,184.000)(217.125,184.000)}%
{\color{black}\linethickness{1pt}\qbezier(217.125,184.000)(219.859,184.000)(222.594,184.000)}%
{\color{black}\linethickness{1pt}\qbezier(222.594,184.000)(225.328,184.000)(228.062,184.000)}%
{\color{black}\linethickness{1pt}\qbezier(228.062,184.000)(230.797,184.000)(233.531,184.000)}%
{\color{black}\linethickness{1pt}\qbezier(233.531,184.000)(236.266,184.000)(239.000,184.000)}%
{\color{black}\linethickness{1pt}\qbezier(239.000,184.000)(241.734,184.000)(244.469,184.000)}%
{\color{black}\linethickness{1pt}\qbezier(244.469,184.000)(247.203,184.000)(249.938,184.000)}%
{\color{black}\linethickness{1pt}\qbezier(249.938,184.000)(252.672,184.000)(255.406,184.000)}%
{\color{black}\linethickness{1pt}\qbezier(255.406,184.000)(258.141,184.000)(260.875,184.000)}%
{\color{black}\linethickness{1pt}\qbezier(260.875,184.000)(263.609,184.000)(266.344,184.000)}%
{\color{black}\linethickness{1pt}\qbezier(266.344,184.000)(269.078,184.000)(271.812,184.000)}%
{\color{black}\linethickness{1pt}\qbezier(271.812,184.000)(274.547,184.000)(277.281,184.000)}%
{\color{black}\linethickness{1pt}\qbezier(277.281,184.000)(280.016,184.000)(282.750,184.000)}%
{\color{black}\linethickness{1pt}\qbezier(282.750,184.000)(285.484,184.000)(288.219,184.000)}%
{\color{black}\linethickness{1pt}\qbezier(288.219,184.000)(290.953,184.000)(293.688,184.000)}%
{\color{black}\linethickness{1pt}\qbezier(293.688,184.000)(296.422,184.000)(299.156,184.000)}%
{\color{black}\linethickness{1pt}\qbezier(299.156,184.000)(301.891,184.000)(304.625,184.000)}%
{\color{black}\linethickness{1pt}\qbezier(304.625,184.000)(307.359,184.000)(310.094,184.000)}%
{\color{black}\linethickness{1pt}\qbezier(310.094,184.000)(312.828,184.000)(315.562,184.000)}%
{\color{black}\linethickness{1pt}\qbezier(315.562,184.000)(318.297,184.000)(321.031,184.000)}%
{\color{black}\linethickness{1pt}\qbezier(321.031,184.000)(323.766,184.000)(326.500,184.000)}%
{\color{black}\linethickness{1pt}\qbezier(326.500,184.000)(329.234,184.000)(331.969,184.000)}%
{\color{black}\linethickness{1pt}\qbezier(331.969,184.000)(334.703,184.000)(337.438,184.000)}%
{\color{black}\linethickness{1pt}\qbezier(337.438,184.000)(340.172,184.000)(342.906,184.000)}%
{\color{black}\linethickness{1pt}\qbezier(342.906,184.000)(345.641,184.000)(348.375,184.000)}%
{\color{black}\linethickness{1pt}\qbezier(348.375,184.000)(351.109,184.000)(353.844,184.000)}%
{\color{black}\linethickness{1pt}\qbezier(353.844,184.000)(356.578,184.000)(359.312,184.000)}%
{\color{black}\linethickness{1pt}\qbezier(359.312,184.000)(362.047,184.000)(364.781,184.000)}%
{\color{black}\linethickness{1pt}\qbezier(364.781,184.000)(367.516,184.000)(370.250,184.000)}%
{\color{black}\linethickness{1pt}\qbezier(370.250,184.000)(372.984,184.000)(375.719,184.000)}%
{\color{black}\linethickness{1pt}\qbezier(375.719,184.000)(378.453,184.000)(381.188,184.000)}%
{\color{black}\linethickness{1pt}\qbezier(381.188,184.000)(383.922,184.000)(386.656,184.000)}%
{\color{black}\linethickness{1pt}\qbezier(386.656,184.000)(389.391,184.000)(392.125,184.000)}%
{\color{black}\linethickness{1pt}\qbezier(392.125,184.000)(394.859,184.000)(397.594,184.000)}%
{\color{black}\linethickness{1pt}\qbezier(397.594,184.000)(400.328,184.000)(403.062,184.000)}%
{\color{black}\linethickness{1pt}\qbezier(403.062,184.000)(405.797,184.000)(408.531,184.000)}%
{\color{black}\linethickness{1pt}\qbezier(408.531,184.000)(411.266,184.000)(414.000,184.000)}%
{\color{blue!55!black}\linethickness{1pt}\qbezier(64.000,64.000)(66.734,64.000)(69.469,64.000)}%
{\color{blue!55!black}\linethickness{1pt}\qbezier(69.469,64.000)(72.203,64.000)(74.938,64.000)}%
{\color{blue!55!black}\linethickness{1pt}\qbezier(74.938,64.000)(77.672,64.000)(80.406,64.000)}%
{\color{blue!55!black}\linethickness{1pt}\qbezier(85.875,64.000)(88.609,64.000)(91.344,64.000)}%
{\color{blue!55!black}\linethickness{1pt}\qbezier(91.344,64.000)(94.078,64.000)(96.812,64.000)}%
{\color{blue!55!black}\linethickness{1pt}\qbezier(96.812,64.000)(99.547,64.000)(102.281,64.000)}%
{\color{blue!55!black}\linethickness{1pt}\qbezier(107.750,64.000)(110.484,64.000)(113.219,64.000)}%
{\color{blue!55!black}\linethickness{1pt}\qbezier(113.219,64.000)(115.953,64.000)(118.688,64.000)}%
{\color{blue!55!black}\linethickness{1pt}\qbezier(118.688,64.000)(121.422,64.000)(124.156,64.000)}%
{\color{blue!55!black}\linethickness{1pt}\qbezier(129.625,64.000)(132.359,64.000)(135.094,64.000)}%
{\color{blue!55!black}\linethickness{1pt}\qbezier(135.094,64.000)(137.828,64.000)(140.562,64.000)}%
{\color{blue!55!black}\linethickness{1pt}\qbezier(140.562,64.000)(143.297,64.000)(146.031,64.000)}%
{\color{blue!55!black}\linethickness{1pt}\qbezier(151.500,64.000)(154.234,64.000)(156.969,64.000)}%
{\color{blue!55!black}\linethickness{1pt}\qbezier(156.969,64.000)(159.703,64.000)(162.438,64.000)}%
{\color{blue!55!black}\linethickness{1pt}\qbezier(162.438,64.000)(165.172,64.000)(167.906,64.000)}%
{\color{blue!55!black}\linethickness{1pt}\qbezier(173.375,64.000)(176.109,64.000)(178.844,64.000)}%
{\color{blue!55!black}\linethickness{1pt}\qbezier(178.844,64.000)(181.578,64.000)(184.312,64.000)}%
{\color{blue!55!black}\linethickness{1pt}\qbezier(184.312,64.000)(187.047,64.000)(189.781,64.000)}%
{\color{blue!55!black}\linethickness{1pt}\qbezier(195.250,64.000)(197.984,64.000)(200.719,64.000)}%
{\color{blue!55!black}\linethickness{1pt}\qbezier(200.719,64.000)(203.453,64.000)(206.188,64.000)}%
{\color{blue!55!black}\linethickness{1pt}\qbezier(206.188,64.000)(208.922,64.000)(211.656,64.000)}%
{\color{blue!55!black}\linethickness{1pt}\qbezier(217.125,64.000)(219.859,64.000)(222.594,64.000)}%
{\color{blue!55!black}\linethickness{1pt}\qbezier(222.594,64.000)(225.328,64.000)(228.062,64.000)}%
{\color{blue!55!black}\linethickness{1pt}\qbezier(228.062,64.000)(230.797,64.000)(233.531,64.000)}%
{\color{blue!55!black}\linethickness{1pt}\qbezier(239.000,71.133)(241.734,76.561)(244.469,81.990)}%
{\color{blue!55!black}\linethickness{1pt}\qbezier(244.469,81.990)(247.203,87.340)(249.938,92.691)}%
{\color{blue!55!black}\linethickness{1pt}\qbezier(249.938,92.691)(252.672,97.950)(255.406,103.210)}%
{\color{blue!55!black}\linethickness{1pt}\qbezier(260.875,113.523)(263.609,118.565)(266.344,123.607)}%
{\color{blue!55!black}\linethickness{1pt}\qbezier(266.344,123.607)(269.078,128.525)(271.812,133.442)}%
{\color{blue!55!black}\linethickness{1pt}\qbezier(271.812,133.442)(274.547,138.227)(277.281,143.013)}%
{\color{blue!55!black}\linethickness{1pt}\qbezier(282.750,152.303)(285.484,156.804)(288.219,161.304)}%
{\color{blue!55!black}\linethickness{1pt}\qbezier(288.219,161.304)(290.953,165.654)(293.688,170.005)}%
{\color{blue!55!black}\linethickness{1pt}\qbezier(293.688,170.005)(296.422,174.203)(299.156,178.401)}%
{\color{blue!55!black}\linethickness{1pt}\qbezier(304.625,184.000)(307.359,184.000)(310.094,184.000)}%
{\color{blue!55!black}\linethickness{1pt}\qbezier(310.094,184.000)(312.828,184.000)(315.562,184.000)}%
{\color{blue!55!black}\linethickness{1pt}\qbezier(315.562,184.000)(318.297,184.000)(321.031,184.000)}%
{\color{blue!55!black}\linethickness{1pt}\qbezier(326.500,184.000)(329.234,184.000)(331.969,184.000)}%
{\color{blue!55!black}\linethickness{1pt}\qbezier(331.969,184.000)(334.703,184.000)(337.438,184.000)}%
{\color{blue!55!black}\linethickness{1pt}\qbezier(337.438,184.000)(340.172,184.000)(342.906,184.000)}%
{\color{blue!55!black}\linethickness{1pt}\qbezier(348.375,184.000)(351.109,184.000)(353.844,184.000)}%
{\color{blue!55!black}\linethickness{1pt}\qbezier(353.844,184.000)(356.578,184.000)(359.312,184.000)}%
{\color{blue!55!black}\linethickness{1pt}\qbezier(359.312,184.000)(362.047,184.000)(364.781,184.000)}%
{\color{blue!55!black}\linethickness{1pt}\qbezier(370.250,184.000)(372.984,184.000)(375.719,184.000)}%
{\color{blue!55!black}\linethickness{1pt}\qbezier(375.719,184.000)(378.453,184.000)(381.188,184.000)}%
{\color{blue!55!black}\linethickness{1pt}\qbezier(381.188,184.000)(383.922,184.000)(386.656,184.000)}%
{\color{blue!55!black}\linethickness{1pt}\qbezier(392.125,184.000)(394.859,184.000)(397.594,184.000)}%
{\color{blue!55!black}\linethickness{1pt}\qbezier(397.594,184.000)(400.328,184.000)(403.062,184.000)}%
{\color{blue!55!black}\linethickness{1pt}\qbezier(403.062,184.000)(405.797,184.000)(408.531,184.000)}%
{\color{black!55}\linethickness{.8pt}\qbezier[3](64.000,64.000)(66.734,64.000)(69.469,64.000)}%
{\color{black!55}\linethickness{.8pt}\qbezier[3](69.469,64.000)(72.203,64.000)(74.938,64.000)}%
{\color{black!55}\linethickness{.8pt}\qbezier[3](74.938,64.000)(77.672,64.000)(80.406,64.000)}%
{\color{black!55}\linethickness{.8pt}\qbezier[3](80.406,64.000)(83.141,64.000)(85.875,64.000)}%
{\color{black!55}\linethickness{.8pt}\qbezier[3](85.875,64.000)(88.609,64.000)(91.344,64.000)}%
{\color{black!55}\linethickness{.8pt}\qbezier[3](91.344,64.000)(94.078,64.000)(96.812,64.000)}%
{\color{black!55}\linethickness{.8pt}\qbezier[3](96.812,64.000)(99.547,64.000)(102.281,64.000)}%
{\color{black!55}\linethickness{.8pt}\qbezier[3](102.281,64.000)(105.016,64.000)(107.750,64.000)}%
{\color{black!55}\linethickness{.8pt}\qbezier[3](107.750,64.000)(110.484,64.000)(113.219,64.000)}%
{\color{black!55}\linethickness{.8pt}\qbezier[3](113.219,64.000)(115.953,64.000)(118.688,64.000)}%
{\color{black!55}\linethickness{.8pt}\qbezier[3](118.688,64.000)(121.422,64.000)(124.156,64.000)}%
{\color{black!55}\linethickness{.8pt}\qbezier[3](124.156,64.000)(126.891,64.000)(129.625,64.000)}%
{\color{black!55}\linethickness{.8pt}\qbezier[3](129.625,64.000)(132.359,64.000)(135.094,64.000)}%
{\color{black!55}\linethickness{.8pt}\qbezier[3](135.094,64.000)(137.828,64.000)(140.562,64.000)}%
{\color{black!55}\linethickness{.8pt}\qbezier[3](140.562,64.000)(143.297,64.000)(146.031,64.000)}%
{\color{black!55}\linethickness{.8pt}\qbezier[3](146.031,64.000)(148.766,64.000)(151.500,64.000)}%
{\color{black!55}\linethickness{.8pt}\qbezier[3](151.500,64.000)(154.234,64.000)(156.969,64.000)}%
{\color{black!55}\linethickness{.8pt}\qbezier[3](156.969,64.000)(159.703,64.000)(162.438,64.000)}%
{\color{black!55}\linethickness{.8pt}\qbezier[3](162.438,64.000)(165.172,64.000)(167.906,64.000)}%
{\color{black!55}\linethickness{.8pt}\qbezier[3](167.906,64.000)(170.641,64.000)(173.375,64.000)}%
{\color{black!55}\linethickness{.8pt}\qbezier[3](173.375,64.000)(176.109,64.000)(178.844,64.000)}%
{\color{black!55}\linethickness{.8pt}\qbezier[3](178.844,64.000)(181.578,64.000)(184.312,64.000)}%
{\color{black!55}\linethickness{.8pt}\qbezier[3](184.312,64.000)(187.047,64.000)(189.781,64.000)}%
{\color{black!55}\linethickness{.8pt}\qbezier[3](189.781,64.000)(192.516,64.000)(195.250,64.000)}%
{\color{black!55}\linethickness{.8pt}\qbezier[3](195.250,64.000)(197.984,64.000)(200.719,64.000)}%
{\color{black!55}\linethickness{.8pt}\qbezier[3](200.719,64.000)(203.453,64.000)(206.188,64.000)}%
{\color{black!55}\linethickness{.8pt}\qbezier[3](206.188,64.000)(208.922,64.000)(211.656,64.000)}%
{\color{black!55}\linethickness{.8pt}\qbezier[3](211.656,64.000)(214.391,64.000)(217.125,64.000)}%
{\color{black!55}\linethickness{.8pt}\qbezier[3](217.125,64.000)(219.859,64.000)(222.594,64.000)}%
{\color{black!55}\linethickness{.8pt}\qbezier[3](222.594,64.000)(225.328,64.000)(228.062,64.000)}%
{\color{black!55}\linethickness{.8pt}\qbezier[3](228.062,64.000)(230.797,64.000)(233.531,64.000)}%
{\color{black!55}\linethickness{.8pt}\qbezier[3](233.531,64.000)(236.266,64.000)(239.000,64.000)}%
{\color{black!55}\linethickness{.8pt}\qbezier[3](239.000,64.000)(241.734,64.000)(244.469,64.000)}%
{\color{black!55}\linethickness{.8pt}\qbezier[3](244.469,64.000)(247.203,64.000)(249.938,64.000)}%
{\color{black!55}\linethickness{.8pt}\qbezier[3](249.938,64.000)(252.672,64.000)(255.406,64.000)}%
{\color{black!55}\linethickness{.8pt}\qbezier[3](255.406,64.000)(258.141,64.000)(260.875,64.000)}%
{\color{black!55}\linethickness{.8pt}\qbezier[3](260.875,64.000)(263.609,64.000)(266.344,64.000)}%
{\color{black!55}\linethickness{.8pt}\qbezier[3](266.344,64.000)(269.078,64.000)(271.812,64.000)}%
{\color{black!55}\linethickness{.8pt}\qbezier[3](271.812,64.000)(274.547,64.000)(277.281,64.000)}%
{\color{black!55}\linethickness{.8pt}\qbezier[3](277.281,64.000)(280.016,64.000)(282.750,64.000)}%
{\color{black!55}\linethickness{.8pt}\qbezier[3](282.750,64.000)(285.484,64.000)(288.219,64.000)}%
{\color{black!55}\linethickness{.8pt}\qbezier[3](288.219,64.000)(290.953,64.000)(293.688,64.000)}%
{\color{black!55}\linethickness{.8pt}\qbezier[3](293.688,64.000)(296.422,64.000)(299.156,64.000)}%
{\color{black!55}\linethickness{.8pt}\qbezier[3](299.156,64.000)(301.891,64.000)(304.625,64.000)}%
{\color{black!55}\linethickness{.8pt}\qbezier[3](304.625,64.000)(307.359,64.000)(310.094,64.000)}%
{\color{black!55}\linethickness{.8pt}\qbezier[3](310.094,64.000)(312.828,64.000)(315.562,64.000)}%
{\color{black!55}\linethickness{.8pt}\qbezier[3](315.562,64.000)(318.297,64.000)(321.031,64.000)}%
{\color{black!55}\linethickness{.8pt}\qbezier[3](321.031,64.000)(323.766,64.000)(326.500,64.000)}%
{\color{black!55}\linethickness{.8pt}\qbezier[3](326.500,64.000)(329.234,64.000)(331.969,64.000)}%
{\color{black!55}\linethickness{.8pt}\qbezier[3](331.969,64.000)(334.703,64.000)(337.438,64.000)}%
{\color{black!55}\linethickness{.8pt}\qbezier[3](337.438,64.000)(340.172,64.000)(342.906,64.000)}%
{\color{black!55}\linethickness{.8pt}\qbezier[3](342.906,64.000)(345.641,64.000)(348.375,64.000)}%
{\color{black!55}\linethickness{.8pt}\qbezier[3](348.375,64.000)(351.109,64.000)(353.844,64.000)}%
{\color{black!55}\linethickness{.8pt}\qbezier[3](353.844,64.000)(356.578,64.000)(359.312,64.000)}%
{\color{black!55}\linethickness{.8pt}\qbezier[3](359.312,64.000)(362.047,64.000)(364.781,64.000)}%
{\color{black!55}\linethickness{.8pt}\qbezier[3](364.781,64.000)(367.516,64.000)(370.250,64.000)}%
{\color{black!55}\linethickness{.8pt}\qbezier[3](370.250,64.000)(372.984,64.000)(375.719,64.000)}%
{\color{black!55}\linethickness{.8pt}\qbezier[3](375.719,64.000)(378.453,64.000)(381.188,64.000)}%
{\color{black!55}\linethickness{.8pt}\qbezier[3](381.188,64.000)(383.922,64.000)(386.656,64.000)}%
{\color{black!55}\linethickness{.8pt}\qbezier[3](386.656,64.000)(389.391,64.000)(392.125,64.000)}%
{\color{black!55}\linethickness{.8pt}\qbezier[3](392.125,64.000)(394.859,64.000)(397.594,64.000)}%
{\color{black!55}\linethickness{.8pt}\qbezier[3](397.594,64.000)(400.328,64.000)(403.062,64.000)}%
{\color{black!55}\linethickness{.8pt}\qbezier[3](403.062,64.000)(405.797,64.000)(408.531,64.000)}%
{\color{black!55}\linethickness{.8pt}\qbezier[3](408.531,64.000)(411.266,64.000)(414.000,64.000)}%
{\color{black}\linethickness{.9pt}\qbezier(64.000,13.000)(75.000,13.000)(86.000,13.000)}%
\put(94.000,13.000){\makebox(0,0)[l]{Advertising platform}}%
{\color{blue!55!black}\linethickness{.9pt}\qbezier(64.000,0.000)(68.000,0.000)(72.000,0.000)}%
{\color{blue!55!black}\linethickness{.9pt}\qbezier(76.000,0.000)(80.000,0.000)(84.000,0.000)}%
\put(94.000,0.000){\makebox(0,0)[l]{Total welfare}}%
{\color{black!55}\linethickness{.9pt}\qbezier[10](64.000,-13.000)(75.000,-13.000)(86.000,-13.000)}%
\put(94.000,-13.000){\makebox(0,0)[l]{Commission-only platform; consumer surplus}}%
\end{picture}%
\endgroup%

\par\smallskip
\begin{minipage}{.98\textwidth}\footnotesize
Numerical illustration with $\alpha=1/5$, $t=3/20$, $\eta=783/13280$, $c=\eta+\rho$, and $s=3\rho/5$. All plotted optima have $b=0$: a consumer share of zero means firms pay, and a share of one means consumers pay. The advertising platform chooses full consumer payment, while the commission-only platform and consumers prefer firm-paid free returns. Total welfare favors a partial consumer contribution between losses of approximately $0.01184$ and $0.01492$. Curves optimize over the full policy set on the selected price equilibria; they are numerical. Proposition~\ref{prop:finite} gives the proved parameter regions.
\end{minipage}
\end{figure}

Figure~\ref{fig:finite-regions} compares the optimal policies under all four objectives as the return loss rises along the stated cost path. At small losses, advertising induces a consumer contribution that is too high from both consumer and total-welfare perspectives. At intermediate losses, a partial consumer contribution saves return resources while preserving more participation than full consumer payment. At larger losses, the advertising platform's choice coincides with the total-welfare optimum, although consumers still prefer free returns. Sharing in this figure is between firms and consumers; the platform pays nothing.

\subsection{How participation affects reimbursement}\label{sec:entry}\label{sec:entry-extension}
Reimbursement can encourage price cuts because firms gain compensation on the returns generated by an additional consumer. To examine this incentive, let participation costs have distribution $H(e)=e^\gamma$ on $[0,1]$, where $\gamma>0$. Participation is now $U^\gamma$, and $\gamma$ is its elasticity with respect to shopping utility. Uniform participation costs correspond to $\gamma=1$. A higher $\gamma$ makes participation more responsive to utility; at a given $U\in(0,1)$, it also lowers the number of participants.

The shopping decisions of those who enter are unchanged. Aggregate firm profit becomes $U^\gamma\pi_i$, while consumer surplus and total welfare are
\begin{equation}
 CS=\frac{U^{\gamma+1}}{\gamma+1},\qquad
 W=U^\gamma V-\frac{\gamma U^{\gamma+1}}{\gamma+1}.
 \label{eq:general-entry-welfare}
\end{equation}
At given prices and consumer contributions, reimbursement raises profit by the number of returned orders. Its effect on the incentive to raise price is
\begin{equation}
 \frac{\partial^2(U^\gamma\pi_i)}{\partial p_i\partial b}
 =U^\gamma R_{i,p_i}-\gamma U^{\gamma-1}D_iR_i.
 \label{eq:general-entry-incentive}
\end{equation}
The first term is the compensation gained when a higher price turns existing trials into returns. The second is compensation lost when the higher price discourages participation. Greater participation responsiveness strengthens this second reason to cut prices. Equilibrium effects also incorporate the rival's price response.

\begin{proposition}[Participation and reimbursement]\label{prop:entry-incidence}
Fix $\gamma>0$, $t\in[0,1)$, and $0\le\eta<1-t$, with $\eta/(1-t)\ne(1+\gamma)^{-2}$. For sufficiently low positive suitability, the limiting price response at the symmetric zero-cost prices satisfies
\[
 \partial_b(p_1+p_2)<0\quad\Longleftrightarrow\quad
 \frac{\eta}{1-t}>\frac{1}{(1+\gamma)^2}.
\]
The response is taken from the smooth solution of the price conditions with positive costs and fixed $s/(\alpha\rho)$; it is an equilibrium response wherever the global best-response conditions hold. The strict finite-loss policy and welfare comparisons in Proposition~\ref{prop:finite} remain valid for $\gamma$ sufficiently close to one.
\end{proposition}

Higher participation responsiveness lowers the cost threshold at which reimbursement reduces the sum of prices. Firms with many returned orders have more compensation to lose when a consumer stays away, so attracting consumers can outweigh the weaker incentive to retain existing purchases. At $\gamma=1$, the threshold approaches one quarter as suitability falls, as in Lemma~\ref{lem:funding-incidence}. This is a condition for the price response; whether the platform should reimburse firms still depends on whether its added revenues cover the bill. Appendix~\ref{app:entry-extension} derives the response and proves the persistence of the finite-loss comparisons.

\section{Managerial implications}\label{sec:implications}

\paragraph{Measure retained transactions by position.}
Prominence attracts trials made before consumers have a satisfactory alternative. A bid should value retained margins and return losses relative to the position the firm would occupy if it lost the auction. The same increase in orders can therefore have different advertising value depending on which orders are kept. Platforms already distinguish promotion from retained transactions: in January 2025, Taobao and Tmall announced fee refunds for specified abnormal orders and plans to optimize advertising for net transactions.\footnote{Xinhua, \href{https://www.xinhuanet.com/tech/20250120/df83cf747351430ba5f09df78202d550/c.html}{report on Taobao and Tmall's twelve firm-support measures}, January 20, 2025, measure 6 (in Chinese). The announced fee refunds concern orders identified as abnormal.} The model shows why return terms also change what firms will pay for earlier display.

\paragraph{Choose the consumer contribution and reimbursement jointly.}
The platform allocates a loss among three payers with two instruments. An experiment that raises the consumer contribution while reducing reimbursement equally holds firm liability constant and identifies the effect of changing the payer. Under the conditions of Corollary~\ref{cor:conditional-sharing}, advertising shifts the optimal split toward consumers. A second comparison varies firms' remaining liability. Both are needed to evaluate the final allocation. Record shipments and retained sales by position, prices, participation, advertising payments, and the total reimbursement bill. A change in the reimbursement rate alone does not measure the platform's expense because the policy also changes return volume. For per-click advertising, bids, payments per click, and click volume are separate outcomes.

\paragraph{Evaluate reimbursement against its bill.}
Keeping consumer returns free while varying reimbursement provides a comparison of how firm liability affects prices and sales. The price response measures the benefit to participating consumers; additional commission revenue relative to compensated returns determines the platform's financial return. These are different tests. Alibaba's Tuikobao announcement, for example, reported a reduction in participating firms' return-shipping costs, illustrating the practical relevance of changing firm expenses while preserving a consumer-facing return service.\footnote{Alibaba Group, \href{https://www.alibabagroup.com/zh-HK/document-1784016094848090112}{Double 11 announcement}, October 14, 2024 (in Chinese), reported a 23 percent decline during the preceding week among subscribers to the recently launched service.} The model holds physical shipment and recovery costs fixed when comparing payers. A service that reduces those costs offers an additional benefit. Proposition~\ref{prop:no-platform-payment} also gives a screening condition: for small losses and a sufficiently low retained-product cost relative to the firm's revenue share, no positive reimbursement is optimal. Outside that region, the commission and advertising responses must be assessed together. When the platform and consumers share the bill, the relevant question is whether increasing the consumer contribution raises advertising and commission income enough to justify the associated loss of shopping activity.

\paragraph{Measure participation and resource use alongside retention.}
A higher retained-order fraction can accompany lower consumer surplus. In two longitudinal field studies, \citet{bower2012return} find lower subsequent spending among customers who paid return shipping, illustrating why participation and demand belong in the evaluation. Fewer consumers may shop, and those who enter may try fewer products. Total returns should therefore be separated into the number of participants and returns per participant. Consumer protection calls for prices, consumer return costs, and the private value of shopping. Efficiency additionally requires shipment costs, recoveries, inspection costs, and the value of kept products. The finite-cost analysis shows why a platform's policy can harm consumers whether its resource effect is negative or positive.

\section{Conclusion}\label{sec:conclusion}

A platform allocates return losses among consumers, firms, and itself. Advertising changes this choice because consumer contributions alter the selection of trials into retained purchases and the profit advantage of earlier display. For a given amount of relief to firms, collecting position revenue shifts payment toward consumers. The joint optimum also depends on how relief changes prices, retained sales, and the reimbursement bill.

The small-loss results identify conditions for full payment by one party and for sharing between consumers and the platform. Reimbursement can move the two firms' prices in opposite directions, and a higher commission can increase the platform's contribution. The finite-loss results establish that advertising can induce consumer contributions that reduce commissions, consumer surplus, and total welfare over an explicit range of market conditions. At larger losses, the same policy disagreement can instead improve total welfare while still harming consumers. Evaluating return funding therefore requires distinguishing revenue from retained transactions, payments for prominence, and the resources used by trial and return. Heterogeneous firms and alternative advertising payment rules are natural extensions.
\label{main:end}

\clearpage
\begingroup
\raggedright

\endgroup
\clearpage
\appendix

% Number each type of appendix object from A.1, continuously across appendices.
\setcounter{equation}{0}
\setcounter{figure}{0}
\setcounter{table}{0}
\setcounter{proposition}{0}
\setcounter{lemma}{0}
\setcounter{corollary}{0}
\setcounter{remark}{0}
\renewcommand{\theequation}{A.\arabic{equation}}
\renewcommand{\thefigure}{A.\arabic{figure}}
\renewcommand{\thetable}{A.\arabic{table}}
\renewcommand{\theproposition}{A.\arabic{proposition}}
\renewcommand{\thelemma}{A.\arabic{lemma}}
\renewcommand{\thecorollary}{A.\arabic{corollary}}
\renewcommand{\theremark}{A.\arabic{remark}}
% Keep hyperlink destinations distinct from those in the main text.
\def\theHequation{appendix.\arabic{equation}}
\def\theHfigure{appendix.\arabic{figure}}
\def\theHtable{appendix.\arabic{table}}
\def\theHproposition{appendix.\arabic{proposition}}
\def\theHlemma{appendix.\arabic{lemma}}
\def\theHcorollary{appendix.\arabic{corollary}}
\def\theHremark{appendix.\arabic{remark}}

\section*{Online Appendix}
\addcontentsline{toc}{section}{Online Appendix}
Appendix~\ref{app:consumer} derives the consumer quantities and defines the notation reused in the proofs. The table below links the results in the text to their proofs. Appendix~\ref{app:verification} explains the exact sign and interval arguments used in those proofs; the accompanying reproducibility files contain the calculation records and instructions.

\begin{center}\small
\begin{tabular}{@{}>{\raggedright\arraybackslash}p{.72\textwidth}l@{}}
\toprule
Result in the text & Proof\\
\midrule
Lemma~\ref{lem:consumer-choices}: consumer choices and demand & \ref{app:consumer-choices}\\
Lemma~\ref{prop:prices}: global price equilibrium & \ref{app:pricing}\\
Lemma~\ref{lem:selection}; Proposition~\ref{prop:rank}: trial selection & \ref{app:selection}--\ref{app:rank}\\
Proposition~\ref{prop:ads}; Corollary~\ref{cor:conditional-sharing}: the payer and advertising & \ref{app:advertising}--\ref{app:conditional-sharing}\\
Lemma~\ref{lem:funding-incidence}: reimbursement and prices & \ref{app:reimbursement-prices}--\ref{app:individual-prices}\\
Table~\ref{tab:policy-gains}; Proposition~\ref{prop:policy-general}: joint allocation & \ref{app:policy-general}--\ref{app:policy-corners}\\
Proposition~\ref{prop:no-platform-payment}: no platform payment & \ref{app:obstruction}\\
Proposition~\ref{prop:policy}(a): firms versus consumers & \ref{app:higher}\\
Proposition~\ref{prop:policy}(b): platform and consumer sharing & \ref{app:interior}\\
Proposition~\ref{prop:welfare}: welfare-maximizing allocation & \ref{app:welfareproof}\\
Proposition~\ref{prop:finite}(a): independently varying parameters & \ref{app:finite-box}\\
Proposition~\ref{prop:finite}(b): finite-loss intervals & \ref{app:finite-prices}--\ref{app:finite-policy}\\
Proposition~\ref{prop:entry-incidence}: participation and reimbursement & \ref{app:entry-extension}\\
\bottomrule
\end{tabular}
\end{center}

\section{Consumer choices and common calculations}\label{app:statements}\label{app:consumer}
This appendix derives the consumer rule in Lemma~\ref{lem:consumer-choices}, the quantities used in the price condition~\eqref{eq:firm-foc}, and the common starting point for the small-loss proofs. All quantities are per participating consumer unless multiplied by participation $U$.

\subsection{Consumer choices and quantities}\label{app:consumer-choices}
\begin{proof}[Proof of Lemma~\ref{lem:consumer-choices}]
Use the equivalent payoff representation in~\eqref{eq:consumer-gain}, and write
\[
 w_i=1-p_i+f,\qquad g=\sqrt{2f},\qquad h=\sqrt{2(f+s/\alpha)}.
\]
Here $w_i$ is the upper bound on a suitable product's value relative to the payoff from returning it. The lemma's conditions imply $g<w_1<1$, $h<w_2<1$, and $w_1-g>w_2-h>0$.

For an available payoff $u$, the expected improvement from a suitable trial is
\[
 \int_0^1(\varepsilon-p_i+f-u)_+\,d\varepsilon
 =\frac{(w_i-u)^2}{2}
 \quad\text{if }0\le p_i-f+u\le1.
\]
Subtracting $f$ gives the postinspection trial value. Before inspection of product 2, the consumer obtains this value only with probability $\alpha$ and can decline the trial if it is negative; inspection itself costs $s$. The zero-gain equations are therefore
\[
 (w_1-u)^2/2=f,\qquad
 \alpha[(w_2-u)^2/2-f]=s.
\]
The relevant roots are $u_1=w_1-g$ and $u_2=w_2-h$, as in~\eqref{eq:consumer-reservation}; the other quadratic roots exceed the largest attainable payoff. The inequalities $w_i<1$ and $u_2>0$ ensure that all cutoffs used below lie inside the uniform support. The initial choice, after product 1 is found suitable, is between trying it and paying to inspect product 2. Inspection need not be followed by trial, so the order comparison warrants a separate check.

Suppose first that $p_1\le p_2$. A suitable product 2 then has trial reservation utility $w_2-g\le w_1-g=u_1$. If the consumer were to inspect product 2 before any trial, she would still try product 1 first when product 2 proved suitable: the transformed problem then consists of two known-suitable products with the same trial cost, and product 1 has the higher reservation utility. If product 2 proved unsuitable, she would again try product 1. Moving that inspection after the first trial preserves the available decisions and may avoid the inspection cost when the first trial is satisfactory. Trying product 1 first is therefore weakly better.

If $p_2<p_1$, a suitable product 2 has trial reservation utility $w_2-g>u_1$. A strategy that inspects it before any trial will therefore try it first whenever it is suitable. For this strategy, inspection and the conditional trial form a single opportunity: the payoff is zero when the product is unsuitable and $\varepsilon_2-p_2+f$ when suitable, and its expected cost is $s+\alpha f$. These payoffs are independent of product 1. Replacing the contingent trial cost by its expectation preserves expected utility because the decision to undertake the opportunity precedes the suitability draw. The reservation equation is
\[
 \alpha\mathbb E[(\varepsilon_2-p_2+f-u)_+]=s+\alpha f,
\]
which gives $u_2$. The independent-opportunity ordering result of \citet{weitzman1979optimal}, with recall and outside option zero, gives product 1 priority when $u_1>u_2$. Its assumptions hold here: match and suitability draws are independent, a previous trial remains available, and each opportunity has a fixed expected cost before its draw. When product 2 is inspected after a first trial, every inspection that is worthwhile is followed by a trial if suitable: its trial reservation utility $w_2-g$ exceeds its inspection reservation utility $u_2=w_2-h$. Thus the combined-opportunity calculation also respects the option to decline a trial and establishes~\eqref{eq:trialorder} on the lemma's region.

If product 1 is unsuitable, $u_2>0$ makes inspection of product 2 worthwhile. If it is suitable, $u_1>0$ makes the first trial worthwhile, and the realized available payoff determines continuation. The postinspection trial cutoff for product 2 is $w_2-g>w_2-h=u_2$, so every suitable product 2 reached by a strictly optimal inspection is tried. Final retention maximizes the gain from keeping rather than returning a product over the tried products and zero. The trial costs assigned in the equivalent search representation and the inspection costs are sunk and cancel from this final comparison.
\end{proof}

To obtain quantities, the second trial occurs after a suitable first product exactly when
\[
 \varepsilon_1<1-w_1+w_2-h.
\]
The cutoff lies strictly between $1-w_1$ and $1$ because $0<w_2-h<w_1$. Combining this event with an unsuitable first product gives the shipment formula~\eqref{eq:shipment-construction}. Conditional on both products being suitable, let $x=\varepsilon_1-p_1+f$. Product 1 is retained without a second trial for $x\ge w_2-h$ and wins a subsequent comparison with probability $1-w_2+x$ for $0\le x<w_2-h$. Its conditional retained demand is consequently
\[
 w_1-(w_2-h)+\int_0^{w_2-h}(1-w_2+x)\,dx
 =w_1-\frac{w_2^2-h^2}{2}.
\]
For product 2, the first term below covers $x\le0$, whose probability is $1-w_1$; the integral covers a positive but insufficient first draw:
\[
 (1-w_1)w_2+\int_0^{w_2-h}(w_2-x)\,dx
 =w_2(1-w_1)+\frac{w_2^2-h^2}{2}.
\]
If only product $i$ is suitable, it is retained with probability $w_i$. Weighting the single-suitable states by $\alpha(1-\alpha)$ and the both-suitable state by $\alpha^2$ gives
\begin{align}
 S_1&=\alpha,\qquad S_2=\alpha[1+\alpha(w_2-w_1-h)],\nonumber\\
 D_1&=\alpha\{w_1-\alpha(w_2^2-h^2)/2\},\nonumber\\
 D_2&=\alpha\{w_2(1-\alpha w_1)+\alpha(w_2^2-h^2)/2\},\nonumber\\
 U&=\alpha\{(w_1^2+w_2^2)/2-2f-s/\alpha\}\nonumber\\
 &\quad+\alpha^2\{-w_1(w_2^2-h^2)/2+w_2^3/6-h^2w_2/2+h^3/3\}.
 \label{eq:demands}
\end{align}
For the last line, the expected value from product 1 alone is $\alpha(w_1^2/2-f)$. The probability that it leaves no positive available payoff is $1-\alpha w_1$; from that state product 2 is worth $\alpha(w_2^2/2-f)-s$. Positive first-product payoffs below $w_2-h$ permit a further improvement. Thus
\begin{align*}
 U={}&\alpha(w_1^2/2-f)
 +(1-\alpha w_1)[\alpha(w_2^2/2-f)-s]\\
 &+\alpha^2\int_0^{w_2-h}\frac{(w_2-x)^2-h^2}{2}\,dx,
\end{align*}
which gives~\eqref{eq:demands}. Returns are $R_i=S_i-D_i$.

Two envelope identities simplify the pricing and welfare calculations:
\[
 U_{p_i}=-D_i,\qquad U_f=-(R_1+R_2).
\]
A price increase affects the consumer only on a kept purchase; a higher consumer return cost affects her on every returned trial. These derivatives hold all other prices and policies fixed. The envelope theorem applies on this smooth choice region \citep{milgrom2002envelope}; differentiating~\eqref{eq:demands} gives the same identities.

\subsection{The initial price and quantities}\label{app:reference-quantities}
The proofs for small positive costs begin at the reference market in~\eqref{eq:starting-quantities-main}--\eqref{eq:reference-price-main}, where $s=\rho=f=b=0$. Let $p_0$ be the common initial price and write
\[
 k=1-t,\qquad \ell=1-p_0,\qquad z=\alpha\ell.
\]
The firm keeps revenue share $k$. Conditional on suitability, $\ell$ is the probability that match value exceeds $p_0$; $z$ includes suitability as well. Actual retention also depends on the alternative product. Substitution of~\eqref{eq:starting-quantities-main} into~\eqref{eq:reference-price-main} gives the scalar price equation
\begin{equation}
 \Phi(z):=\frac{z(5z^2-26z+36)}{3z^2-16z+24}
 =\alpha\left(1-\frac\eta k\right).
 \label{eq:primitiveroot}
\end{equation}
Appendix~\ref{app:pricing} proves that there is one root, with $0<z<5/7$. Expected utility $U_0$ and each firm's retained demand $D_0$ are given in~\eqref{eq:starting-quantities-main}. Total returns and the retained-sale margin are
\begin{equation}
 \mathcal R_0=2U_0(\alpha-D_0),\qquad m=kp_0-\eta>0.
 \label{eq:basequantities}
\end{equation}
These quantities have the same economic meaning throughout the proofs.

\subsection{How a policy change affects prices and revenue}\label{app:policy-derivatives}
The same two pricing equations~\eqref{eq:firm-foc} determine all policy responses. Write $r=\rho-f-b$ for the firm's remaining loss per return. Using $c=\eta+\rho$ and $R_i=S_i-D_i$ in~\eqref{eq:profit}, and substituting $p_i=1+f-w_i$, gives
\[
 \pi_i=(kp_i-\eta-f-b)D_i-rS_i,\qquad
 F_i=\frac{\partial(U\pi_i)}{\partial w_i}=0.
\]
Let $J$ be the derivative matrix of $(F_1,F_2)$ with respect to $(w_1,w_2)$ at the initial price. For $j\in\{f,b,\rho,s\}$, the column $F_j$ contains the partial derivatives with respect to $j$, holding $w_1,w_2$ fixed. Implicit differentiation then yields
\begin{equation}
 \binom{w_{1j}}{w_{2j}}=-J^{-1}F_j,\qquad
 p_{ij}=\mathbf 1_{\{j=f\}}-w_{ij}.
 \label{eq:referencecoeff}
\end{equation}
Appendix~\ref{app:pricing} establishes that $J$ is nonsingular. Thus each policy first changes firms' pricing incentives, and~\eqref{eq:referencecoeff} gives the resulting price adjustments. Substitution into the profit, commission, and welfare expressions gives their total policy effects.

For the small-loss comparisons, hold $\xi=s/(\alpha\rho)>0$ fixed as $\rho$ falls, as in~\eqref{eq:searchratio}. The global-equilibrium claims impose~\eqref{eq:active-ratio}. On response coefficients, a superscript 0 denotes the limiting derivative along these positive-loss equilibria, rather than a derivative that changes trial order at the zero-cost point. In~\eqref{eq:demands}, $h$ enters retained demand through $h^2$; the terms $h^3$ and $rh$ have first derivatives of order $\sqrt\rho$. Indeed, $h$ is bounded below by $\sqrt{2\xi\rho}$ and $0\le r\le\rho$, so the potentially singular term $r/h$ also tends to zero uniformly over feasible policies. These limits yield the derivatives used below. In particular, $\Gamma_j^0$ denotes the total derivative of the signed position value $\Gamma=\Pi_1-\Pi_2$, and $C_j^0$ that of commissions. Under Lemma~\ref{prop:prices}, position value is positive and~\eqref{eq:accounts} gives the same derivatives for advertising receipts.

Three polynomials recur in the denominators and sign calculations:
\begin{align*}
 P_2(z)&=3z^2-16z+24,\\
 P_3(z)&=15z^3-130z^2+408z-432,\\
 P_4(z)&=29z^4-216z^3+596z^2-816z+576.
\end{align*}
On $0<z<5/7$, $P_2,P_4>0$ and $P_3<0$. For $P_2$ and $P_3$, substitution of $z=1-u$ makes the signs immediate; for the fourth-degree polynomial, $P_4(1-u)=29u^4+100u^3+122u^2+156u+169>0$. This notation keeps the subsequent expressions compact. Appendix~\ref{app:verification} explains the exact sign tests used below.

\section{Global pricing: proof of Lemma~\ref{prop:prices}}\label{app:pricing}
We prove the pricing claims in Lemma~\ref{prop:prices} from the operating profit~\eqref{eq:profit} and price condition~\eqref{eq:firm-foc}. First, the zero-cost market has a unique equilibrium. Second, its nearby price solution is a global equilibrium at every feasible policy. Third, no further equilibrium occurs, including at an order tie. The lemma's positive-position-value conclusion is linked to its separate calculation at the end. Throughout, use $k,\ell,z$, the consumer thresholds $g,h$, and the polynomials $P_2,P_3,P_4$ defined in Appendix~\ref{app:consumer}.

\paragraph{Step 1: Each firm has a unique best response at zero costs.}
At $s=\rho=0$, a retained sale has margin $kp_i-\eta$ and a return causes no loss. The price $\bar p=(1+\eta/k)/2$ guarantees a positive margin and positive demand against every nonnegative rival price. In particular,
\begin{gather*}
 U(\bar p,p_j)\ge\frac{\alpha(1-\bar p)^2}{2},\qquad
 k\bar p-\eta=\frac{k-\eta}{2}>0,\\
 D_i(\bar p,p_j)\ge\alpha(1-\alpha)(1-\bar p)+\frac{\alpha^2(1-\bar p)^2}{2}>0.
\end{gather*}
The second demand term remains positive when both products are always suitable, $\alpha=1$: a sufficiently good trial can still beat the alternative. Prices at or below $\eta/k$ have nonpositive margins, and prices at or above one attract no retained purchases. Every equilibrium price therefore lies in $(\eta/k,1)$.

To examine the curvature of a firm's profit, fix a rival price in this interval and write $L=1-p_j$, $x=1-p_i$, and $d(x)=D_i/\alpha$. These local coordinates make the two demand regions easy to compare:
\[
 d(x)=\begin{cases}
 x-\alpha L^2/2,&x\ge L,\\
 (1-\alpha L)x+\alpha x^2/2,&x\le L.
 \end{cases}
\]
On each region, $U_x=\alpha d>0$ and $d'>0$. For $x\le L$, $(d')^2-2dd''=(1-\alpha L)^2\ge0$; for $x\ge L$, $d''=0$. Thus
\[
 (\log(Ud))''=\frac{\alpha d'}{U}-\frac{\alpha^2d^2}{U^2}
 +\frac{d''}{d}-\frac{(d')^2}{d^2}\le-\frac{(d')^2}{4d^2}<0.
\]
For the inequality, complete the square in $\alpha d/U$ to bound the first two terms by $(d')^2/(4d^2)$; the last two are at most $-(d')^2/(2d^2)$. Adding the log margin, $\log(1-\eta/k-x)$, preserves strict concavity on $0<x<1-\eta/k$. Profit and its first derivative agree at $x=L$, so the decreasing log-profit derivative joins continuously across the two regions. Profit vanishes at the endpoints and is positive inside. It therefore has exactly one maximizer, characterized by its first-order condition.

\paragraph{Step 2: The unique zero-cost equilibrium is symmetric.}
At a symmetric price, the quantities in~\eqref{eq:starting-quantities-main} and $D_{i,p_i}=-\alpha$ give~\eqref{eq:reference-price-main}. With $\ell=1-p_0$ and $z=\alpha\ell$, this is
\[
 (1-\eta/k-\ell)\{1-z/3+(1-z/2)^2\}=\ell(1-z/2)(1-z/3).
\]
This yields the price equation~\eqref{eq:primitiveroot}. Its left-hand function $\Phi$ is strictly increasing on $[0,1]$: the numerator of its derivative is $(z-2)P_3(z)>0$ there. Also, $\Phi(0)=0$, $\Phi(5/7)>1$, and $\Phi(z)>z$ for $z>0$ in this interval. Since the right side of~\eqref{eq:primitiveroot} lies in $(0,1]$, there is exactly one symmetric root, with $0<z<5/7$. Moreover, $z<\alpha(1-\eta/k)$, so its price is in $(\eta/k,1)$. Step 1 makes this stationary price a best response. The later price and policy comparisons also use
\[
 \left(\frac{\Phi(z)}z\right)'
 =-\frac{2(z^2-12z+24)}{P_2(z)^2}<0,\qquad
 \frac32-\frac{\Phi(z)}z=\frac{z(4-z)}{2P_2(z)}>0.
\]
Both signs hold for $0<z<5/7$; the ratio tends to $3/2$ as $z$ falls to zero.

It remains to exclude unequal prices. A firm's first-order condition equates its markup to an expression involving demand:
\[
 p_i-\eta/k=\frac{UD_i}{D_i^2-UD_{i,p_i}}.
\]
Suppose $\eta/k<p_1<p_2<1$. Subtracting these two conditions would require the difference on the right to equal $p_2-p_1$. The demand formulas instead give the strict inequality
\begin{equation}
 \frac{UD_2}{D_2^2-UD_{2,p_2}}-\frac{UD_1}{D_1^2-UD_{1,p_1}}<p_2-p_1,
 \qquad 0<\alpha\le1.
 \label{eq:finite-alpha-asymmetry-certificate}
\end{equation}
Both denominators are positive. To verify the strict inequality, put $x=1-p_1>y=1-p_2>0$ and define normalized demands and utility, only for this calculation, by
\begin{align*}
 d_1&=x-\alpha y^2/2, & d_2&=y-\alpha xy+\alpha y^2/2,\\
 \mathcal U&=(x^2+y^2)/2-\alpha xy^2/2+\alpha y^3/6,
 &q_2&=1-\alpha(x-y).
\end{align*}
Thus $D_i=\alpha d_i$, $U=\alpha\mathcal U$, $D_{1,p_1}=-\alpha$, and $D_{2,p_2}=-\alpha q_2$. The difference between the right and left sides of~\eqref{eq:finite-alpha-asymmetry-certificate} is
\[
 \frac{(x-y)N}{(\mathcal U+d_1^2)(\mathcal Uq_2+d_2^2)},
\]
where
\[
 \begin{aligned}
 N={}&(\mathcal U+d_1^2)(\mathcal Uq_2+d_2^2)\\
 &+\mathcal U^2(1+\alpha y-\alpha d_1)
 -\mathcal U d_1d_2(1+\alpha y).
 \end{aligned}
\]
To obtain this polynomial, use $d_1-d_2=(x-y)(1+\alpha y)$ and $d_2-d_1q_2=(x-y)(\alpha d_1-1-\alpha y)$ when combining the two fractions. For its sign, set $E=\alpha(x-y)$, $Z=\alpha y$, and $H=1-\alpha x$ in this calculation only. These coordinates are nonnegative and sum to one. After homogenizing with $E+Z+H=1$, the polynomial $144\alpha^4N-169Z^8$ has nonnegative coefficients, so
\[
 144\alpha^4N\ge169Z^8>0.
\]
The full expansion is in \path{pricing_certificate_details.txt}, and \path{original_verification/src/verify_reference_uniqueness.py} checks the identity. Hence the two price conditions cannot hold at unequal prices. Together with Step 1, this establishes the unique pure equilibrium at zero costs.

\paragraph{Step 3: Equilibrium prices continue to small positive costs.}
Fix the primitives in the lemma and the finite cost ratio $\xi=s/(\alpha\rho)\ge9/4$. Define the logarithmic price conditions as $\partial_{p_i}\log(U\pi_i)=0$, which are equivalent to~\eqref{eq:firm-foc} where profits are positive. At $(p_0,p_0)$, the two eigenvalues of their derivative matrix, using the smooth continuation of~\eqref{eq:demands} with product 1 tried first, are
\[
 \frac{P_3(z)}{4\ell^2(3-z)^2(2-z)},\qquad
 -\frac{P_4(z)}{4\ell^2(3-z)^2(2-z)^2}.
\]
Both are strictly negative by the signs established in Appendix~\ref{app:policy-derivatives}, so the matrix is nonsingular. Since $\partial_{w_i}=-\partial_{p_i}$, differentiating $F_i=-(U\pi_i)\partial_{p_i}\log(U\pi_i)$ with respect to $w_j$ at the stationary point shows that $J$ in~\eqref{eq:referencecoeff} is $U_0mD_0>0$ times this matrix. Hence $J$ is nonsingular as well. This argument concerns the fixed-order continuation; it does not require the zero-cost payoff to be twice differentiable across equal prices.

The policy shares $(f/\rho,b/\rho)$ range over the compact triangle implied by~\eqref{eq:feasible}. Uniformly over these shares, $0\le g\le\sqrt{2\rho}$ and
\[
 \sqrt{2\xi\rho}\le h\le\sqrt{2(1+\xi)\rho},\qquad
 h-g=\frac{2\xi\rho}{h+g}\ge
 \frac{\sqrt2\,\xi}{\sqrt{1+\xi}+1}\sqrt\rho.
\]
Regard $h$ as a parameter in the polynomial expressions~\eqref{eq:demands} before imposing $h^2=2(f+s/\alpha)$. This makes the continuation smooth at zero without differentiating a square root there. In the pricing equations, terms linear in $h$ multiply the remaining return loss $r=\rho-f-b$ and have order $\rho^{3/2}$; the $h^3$ terms have the same order, and all other perturbations have order $\rho$. The implicit function theorem and compactness of the policy shares give a common neighborhood containing a unique solution with
\[
 p_i=p_0+O(\rho),
\]
positive participation, and positive firm profits. Since $p_1-p_2=O(\rho)$ while the displayed lower bound for $h-g$ has order $\sqrt\rho$, the solution satisfies $p_1-p_2<h-g$. Thus it lies strictly in the consumer-choice region used to construct it.

\paragraph{Step 4: Changing the trial order is not profitable.}
We now check all unilateral price deviations, as required after~\eqref{eq:firm-foc}. The zero-cost best response is uniquely $p_0$ against $p_0$. Hence continuity on a bounded price set gives a strictly positive payoff gap outside any fixed neighborhood of $p_0$. Payoffs converge uniformly to their zero-cost values, so this gap excludes such deviations for small costs. Prices that yield no positive retained demand give nonpositive profit and are excluded as well.

Within that neighborhood, own-price curvature stays negative in each fixed-order region. The constructed stationary price therefore maximizes profit within the first-product-first region. At the order boundary, a reversal requires firm 1 to raise its price, or firm 2 to cut its price, by $h-g+O(\rho)$ from $p_0$. The reverse-order derivative there has the sign of the negative initial curvature times this displacement, with an $O(\rho)$ error. It points back toward the boundary: it is negative for firm 1 and positive for firm 2. Negative curvature preserves that direction farther into the nearby reversed region. Thus its best possible deviation is the boundary payoff on the reversed side.

For completeness, the quantities on that side can be obtained by the same integrals as in Appendix~\ref{app:consumer-choices}. Superscript $R$ denotes the reversed trial order, with the original position labels retained:
\begin{align*}
 S_2^R&=\alpha,\qquad S_1^R=\alpha[1+\alpha(w_1-w_2-g)],\\
 D_2^R&=\alpha\{w_2-\alpha(w_1^2-g^2)/2\},\\
 D_1^R&=\alpha\{w_1(1-\alpha w_2)+\alpha(w_1^2-g^2)/2\},\\
 U^R&=\alpha\{(w_1^2+w_2^2)/2-2f\}-s\\
 &\quad+\alpha^2\{-w_2(w_1^2-g^2)/2+w_1^3/6-g^2w_1/2+g^3/3\}.
\end{align*}
Product 1's suitability is still learned first and for free. Product 2 is then inspected in either suitability state, so its inspection cost is paid with probability one. If product 2 is suitable, its trial precedes any trial of product 1; the latter uses the cutoff $w_1-g$. These observations explain both the unconditional $-s$ term and the use of $g$, rather than $h$, in the continuation terms.

The following two polynomials describe that payoff comparison:
\begin{align*}
 K_1(z)&=13z^4-110z^3+358z^2-552z+360,\\
 K_2(z)&=9z^4-78z^3+274z^2-480z+360.
\end{align*}
Substitute these quantities into~\eqref{eq:profit}, set $p_1=p_2+h-g$ for firm 1's deviation or $p_2=p_1-h+g$ for firm 2's, and expand about $(p_0,p_0)$. The constant terms cancel, and the terms linear in the price displacement vanish by~\eqref{eq:reference-price-main}. The resulting gains relative to equilibrium profit are
\begin{equation}
 \begin{aligned}
 \Delta_1^{\rm rev}&=-\frac{\alpha^2k\ell^2}{12P_2(z)}\{K_2(h^2+g^2)-2K_1hg\}+O(\rho^{3/2}),\\
 \Delta_2^{\rm rev}&=-\frac{\alpha^2k\ell^2}{12P_2(z)}\{K_1(h^2+g^2)-2K_2hg\}+O(\rho^{3/2}).
 \end{aligned}
 \label{eq:finite-alpha-switch-gains}
\end{equation}
The own equilibrium-price correction is $O(\rho)$ and has no first-order payoff effect. The rival correction is also $O(\rho)$ and has the same first-order effect on equilibrium and deviation profit at the initial point, so it cancels. A change in shipments at the boundary is $O(\sqrt\rho)$ and is multiplied by $r\le\rho$. These observations leave the displayed quadratic terms and a uniform $O(\rho^{3/2})$ remainder.

Both braces are uniformly positive after division by $h^2$. Positivity of $K_1$ follows by substituting $z=1-v$:
\[
 K_1(1-v)=13v^4+58v^3+106v^2+114v+69>0.
\]
Also $K_2-K_1=4z(2-z)(3-z)^2>0$. The first brace is
$K_1(h-g)^2+(K_2-K_1)(h^2+g^2)>0$.
For the second, feasibility and $\xi\ge9/4$ give
\[
 \frac gh=\sqrt{\frac{f}{f+\xi\rho}}\le\frac2{\sqrt{13}}.
\]
The expression $K_1[1+(g/h)^2]-2K_2g/h$ decreases as $g/h$ increases over this range. At the upper endpoint, positivity is equivalent to $17\sqrt{13}K_1>52K_2$. Both sides are positive, and squaring reduces this inequality to
\[
 3757K_1(z)^2-2704K_2(z)^2\ge\frac{3136426116525}{5764801}>0,\qquad 0\le z\le5/7.
\]
This lower bound is the smallest coefficient in the degree-eight Bernstein representation after substituting $z=5x/7$, $x\in[0,1]$. The nine exact coefficients are supplied in \path{pricing_certificate_details.txt}; \path{original_verification/src/verify_search_threshold.py} checks them. Since $h^2\ge2\xi\rho$, both leading losses dominate their $O(\rho^{3/2})$ remainders. Each constructed price is therefore a strict global best response. 

\paragraph{Step 5: No additional equilibrium occurs near an order tie.}
The fixed price $\bar p$ in Step 1 continues to guarantee positive profit against every rival price when costs are small. Every equilibrium price therefore lies in $(c/k,1+f)$: lower prices have nonpositive retained and returned margins, and higher prices generate no positive retained demand. Uniform payoff convergence on this bounded range implies that every convergent subsequence of equilibria is an equilibrium of the zero-cost game. Its uniqueness in Step 2 therefore makes every sequence of pure equilibria converge to $(p_0,p_0)$.

The implicit function theorem excludes additional roots on the first-product-first region. Relabelling the firms in the reverse-order formulas gives the same nonsingular initial derivative matrix with its rows and columns interchanged. The reverse-order conditions therefore also have only one nearby root, with $p_1-p_2=O(\rho)$, which cannot satisfy $p_1-p_2\ge h-g$. The only remaining possibility is the boundary itself.

At $p_1-p_2=h-g$, consumers try the second product first by the tie rule following~\eqref{eq:trialorder}. Write $\Pi_1^F$ and $\Pi_1^R$ for firm 1's aggregate profits under the two orders. Substitution of $w_2=w_1+h-g$ and $h^2-g^2=2s/\alpha$ gives equal shopping utility under both orders, but
\[
 D_1^F-D_1^R=\alpha^2hg,\qquad S_1^F-S_1^R=\alpha^2h.
\]
The profit difference is consequently
\[
 \Pi_1^F-\Pi_1^R=U\alpha^2h\{(kp_1-\eta-f-b)g-r\}.
\]
Since prices converge to $p_0$, the coefficient $kp_1-\eta-f-b$ is at least $m/2$ for small costs, where $m=kp_0-\eta>0$. If $g>2\rho/m$, then $(kp_1-\eta-f-b)g-r>0$, and an arbitrarily small price cut raises profit by changing the trial order. Otherwise $g\le2\rho/m$, so any adverse payoff jump is only $O(\rho^{3/2})$. Differentiating the two order-specific quantities then gives the difference between right and left own-price derivatives:
\[
 (\Pi_1^R)_{p_1}-(\Pi_1^F)_{p_1}=U_0\alpha^2(kp_0-\eta)h+o(\sqrt\rho)>0.
\]
If the right derivative is positive, raising the price is profitable. Otherwise the left derivative is negative by a multiple of $\sqrt\rho$. A price cut of $\rho^{3/4}$ then yields a gain of order $\rho^{5/4}$, which exceeds both the adverse jump and the second-order error, of order $\rho^{3/2}$. An order tie is therefore not an equilibrium, including when $f=0$.

These steps establish the unique global pricing equilibrium uniformly over~\eqref{eq:feasible}. To complete Lemma~\ref{prop:prices}, its remaining condition $\xi\ge1/(4k)$ ensures positive position value. Appendix~\ref{app:positive-position-value} calculates this value on the price continuation just constructed, before using any auction receipts, and proves it is strictly positive throughout the policy triangle. The auction formula~\eqref{eq:accounts} then applies.

All bounds can also be chosen uniformly on compact sets of $(\alpha,k,\eta,\xi)$ bounded away from $\alpha=0$, $k=0$, and $\eta=k$, with $\xi$ bounded above. The endpoint $\alpha=1$ is included.\hfill$\square$

\section{Positions and advertising}\label{app:coefficients}
\subsection{The auction payment}\label{app:auction}
Fix the pricing continuation for each position. Write $\Gamma=\Pi_1-\Pi_2>0$ for the profit advantage of winning. Since auction payments are sunk before pricing, winning against a bid $y$ changes a firm's profit relative to losing by $\Gamma-y$. Bidding $\Gamma$ wins when this gain is positive and loses when it is negative, with indifference at equality. The standard second-price argument of \citet{vickrey1961counterspeculation} therefore makes this bid weakly dominant. Identical firms bid $\Gamma$, so the auction collects $\Gamma$ and each firm retains $\Pi_2$ after payment. The same price continuation applies to the commission-only platform at a given return rule.

\subsection{Proof of Lemma~\ref{lem:selection}}\label{app:selection}
Use the consumer quantities in Appendix~\ref{app:consumer}. At fixed prices, $w_{1f}=w_{2f}=1$, $\partial_fh^2=2$, and reimbursement does not change consumer choices. Subtracting retained demands gives
\[
 D_1-D_2=\alpha\{(w_1-w_2)(1+\alpha w_2)+\alpha h^2\}.
\]
The difference $w_1-w_2$ does not change with $f$, so
\[
 (\partial_f-\partial_b)(D_1-D_2)
 =\alpha^2\{w_1-w_2+2\}=\alpha^2(2+p_2-p_1).
\]
Since $0<w_i<1$, $w_1-w_2>-1$, establishing the strict sign in~\eqref{eq:selection-finite}.

For free returns at equal prices, let $w=1-p$ and $h=\sqrt{2s/\alpha}$. The positive-reservation-utility interior gives $0<h<w<1$. Direct substitution gives
\[
 S_1-S_2=\alpha^2h>0,\qquad
 \frac{D_1}{S_1}-\frac{D_2}{S_2}
 =-\frac{\alpha h(w-h)[1-\alpha(w+h)/2]}{1-\alpha h}<0.
\]
Indeed, $w-h>0$, $\alpha(w+h)/2<1$, and $\alpha h<1$.\hfill$\square$

\subsection{Proof of Proposition~\ref{prop:rank}}\label{app:rank}
Set $f=b=0$ and $s/\alpha=\xi\rho$, maintaining the ratio bound in Lemma~\ref{prop:prices}. Appendix~\ref{app:pricing} gives $p_i=p_0+O(\rho)$, hence $w_i=\ell+O(\rho)$ and $h=\sqrt{2\xi\rho}$. The shipment formulas give $S_1=\alpha$ and $S_2=\alpha-\alpha^2h+O(\rho)$. Retained demands depend on $h^2$, so both equal $D_0+O(\rho)$. Dividing by shipments then yields
\begin{equation}
 \begin{aligned}
 S_1-S_2&=\alpha^2\sqrt{2\xi\rho}+O(\rho),\\
 D_1-D_2&=O(\rho),\\
 \frac{D_1}{S_1}-\frac{D_2}{S_2}&=-D_0\sqrt{2\xi\rho}+O(\rho).
 \end{aligned}
 \label{eq:trialquality}
\end{equation}
The shipment difference has a positive leading term of order $\sqrt\rho$, while retained-demand and price differences have order $\rho$. Since $D_0>0$, the prominent firm receives more trials and retains a smaller share of them for sufficiently small positive $\rho$. Positive operating profits follow from Lemma~\ref{prop:prices}.\hfill$\square$

\subsection{Proof of Proposition~\ref{prop:ads}}\label{app:advertising}
The comparison raises the consumer contribution and lowers platform reimbursement by the same amount. The firm's loss on each returned order stays fixed. We first hold prices fixed, then incorporate firms' price responses. All reference responses below use the limiting derivatives defined in Appendix~\ref{app:consumer}.

\paragraph{Step 1: Consumer choices at fixed prices.}
At the initial prices, $\partial_fw_i=1$ and $\partial_fh^2=2$, while $b$ does not affect choices. Differentiating~\eqref{eq:demands} gives
\[
 (\partial_f-\partial_b)D_1\big|_p=\alpha(1+\alpha-\alpha\ell),\qquad
 (\partial_f-\partial_b)D_2\big|_p=\alpha(1-\alpha-\alpha\ell).
\]
The difference is $2\alpha^2$. Multiplying by participation and the initial retained-sale margin gives the direct effect $2U_0m\alpha^2$ on $\Gamma$. The participation response multiplies an initially zero profit difference and contributes nothing at this step.

\paragraph{Step 2: Relative prices.}
Each firm's own price is optimal. A price change therefore affects the profit difference through the other firm's payoff:
\[
 \Gamma_{p_1}=-(U\pi_2)_{p_1}=m(D_0^2-U_0\alpha^2\ell),\qquad
 \Gamma_{p_2}=-\Gamma_{p_1}.
\]
Here $U_{p_1}=-D_0$ and $D_{2,p_1}=\alpha^2\ell$ at the reference point. Thus $mD_0^2$ is the rival's loss from reduced participation, and $mU_0\alpha^2\ell$ is its gain from substitution. Together with Step 1, these terms give~\eqref{eq:equalrelief-decomposition}.

Differentiate the pricing conditions $F_i=0$ from Appendix~\ref{app:consumer}. Their derivative rows and equal-exchange forcing satisfy
\[
 J_{1\cdot}-J_{2\cdot}=-\frac{kz^2P_4(z)}{12P_2(z)}(1,-1),\qquad
 (\partial_f-\partial_b)(F_1-F_2)\big|_w
 =-\frac{2kz^4(3-z)(8-3z)}{3\ell P_2(z)}.
\]
Subtract the differentiated equations and use $p_1-p_2=w_2-w_1$. Division by the nonzero row coefficient gives
\begin{equation}
 (\partial_f-\partial_b)(p_1-p_2)\big|_0
 =\frac{8z^2(3-z)(8-3z)}{\ell P_4(z)}>0.
 \label{eq:equalrelief-pricegap}
\end{equation}
Thus the prominent firm raises its price relative to the second. Whether this reinforces the direct consumer effect depends on which of the two effects on the rival is larger.

\paragraph{Step 3: The total advertising effect.}
Write $S=\Gamma_f^0-\Gamma_b^0$. Substitute~\eqref{eq:equalrelief-pricegap} into~\eqref{eq:equalrelief-decomposition}, using $D_0=z(1-z/2)$, $U_0=z\ell(1-z/3)$, and $m=2k\ell(3-z)(2-z)/P_2(z)$. This gives
\begin{align}
 \frac{S}{2U_0m\alpha^2}&=\frac{8(4-z)(3-z)(z^2-4z+6)}{P_4(z)}>0,\label{eq:equalrelief-multiplier}\\
 \frac{S}{2U_0m\alpha^2}-1&=\frac{z(8-3z)(7z^2-24z+12)}{P_4(z)}.\label{eq:equalrelief-feedback}
\end{align}
For $0<z<5/7$, the first numerator is positive, since $z^2-4z+6=(z-2)^2+2$, and $P_4(z)>0$ by Appendix~\ref{app:consumer}. Hence the price response cannot overturn the direct effect. The second identity states when repricing reinforces or attenuates it. With positive bids, $A=\Gamma$, so $A_f-A_b=S+O(\sqrt\rho)>0$ uniformly over feasible policies for sufficiently small positive losses.

\paragraph{Step 4: Reimbursement and positive position value.}
Write $B=\Gamma_b^0$. At fixed prices the direct reimbursement effect is $U_0(R_1-R_2)=0$, since both firms initially return $\alpha-D_0$ orders per participant. The differenced pricing equations give
\[
 (F_1-F_2)_b\big|_w=-\frac{z^3(3-z)}{3\ell},\qquad
 p_{1b}^0-p_{2b}^0=\frac{4z(3-z)P_2(z)}{\ell kP_4(z)}.
\]
Consequently $B=m(D_0^2-U_0\alpha^2\ell)(p_{1b}^0-p_{2b}^0)$. Together with Step 3, this yields
\begin{equation}
 \begin{aligned}
 B(z)&=-\frac{2z^3(z-3)^2(z-2)(7z^2-24z+12)}{3P_4(z)},\\
 S(z,k)&=-\frac{32kz^3(z-4)(z-3)^3(z-2)(z^2-4z+6)}{3P_2(z)P_4(z)}>0.
 \end{aligned}
 \label{eq:position-response-coefficients}
\end{equation}
For completeness, differentiating the same relative-price equation with respect to $\rho$ and $s$ gives
\[
 p_{1\rho}^0-p_{2\rho}^0=-(p_{1b}^0-p_{2b}^0),\qquad
 p_{1s}^0-p_{2s}^0=\frac1\alpha(\partial_f-\partial_b)(p_1-p_2)\big|_0.
\]
The direct effect of $\rho$ on relative profit is zero: it changes each firm's profit by $-U_0S_i$, and initially $S_1=S_2=\alpha$. The direct effect of $s$ is $2U_0m\alpha$, or $1/\alpha$ times Step 1's direct effect. Thus $\Gamma_f^0=B+S$, $\Gamma_\rho^0=-B$, and $\Gamma_s^0=S/\alpha$. The uniform expansion consequently gives
\begin{equation}
 \Gamma=-rB+(f+s/\alpha)S+O(\rho^{3/2}).\label{eq:adexpansion}
\end{equation}
Both $B$ and the price feedback in~\eqref{eq:equalrelief-feedback} have the sign of $7z^2-24z+12$. Its only root in $(0,5/7)$ is $(12-2\sqrt{15})/7$. Since $\Phi$ is increasing, they are positive precisely when
\begin{equation}
 \alpha\left(1-\frac\eta{1-t}\right)<\frac{510-64\sqrt{15}}{301}.\label{eq:adprimitivecutoff}
\end{equation}
Both sides of this cutoff are feasible, for example with $\eta=0$ and suitability varied. Reimbursement alone can therefore either increase or decrease advertising receipts. Each strict leading sign persists at sufficiently small positive losses.

\paragraph{Positive bids throughout the policy triangle.}\label{app:positive-position-value}
Use~\eqref{eq:position-response-coefficients} to obtain
\[
 \frac14-\frac{kB}{S}
 =\frac{z(408-404z+140z^2-17z^3)}{16(4-z)(3-z)(z^2-4z+6)}>0.
\]
For $z=1-u$, the cubic in the numerator is $127+175u+89u^2+17u^3>0$. Hence $B/S<1/(4k)$. Under $\xi\ge1/(4k)$, the three vertex values of the leading position advantage are strictly positive:
$\rho(\xi S-B)$, $\rho(1+\xi)S$, and $\rho\xi S$.
Every feasible allocation is a convex combination of these vertices, so its leading advantage is bounded below by $\rho$ times their smallest positive coefficient. The uniform $O(\rho^{3/2})$ remainder cannot overturn that bound for sufficiently small positive losses. Thus equilibrium bids are positive throughout the triangle, as required in Step 3 and Lemma~\ref{prop:prices}.

\paragraph{Step 5: The finite-loss comparison.}
On the common pricing branch in Proposition~\ref{prop:finite}, the auction is active and $A=P^{\mathrm{ad}}-P^0$. Subtracting the certified marginal platform revenues therefore bounds
$A_f-A_b=(P_f^{\mathrm{ad}}-P_f^0)-(P_b^{\mathrm{ad}}-P_b^0)$.
Appendices~\ref{app:finite-policy} and~\ref{app:finite-box}, specifically Table~\ref{tab:finite-box-bounds} and~\eqref{eq:finite-equal-relief}, give strictly positive lower bounds throughout the feasible policy set in every finite-loss region asserted in the proposition. At policy boundaries, derivatives are interpreted from the feasible side. At $(f,b)=(0,0)$, the comparison is between the initial revenue changes under $(f,b)=(\varepsilon,0)$ and $(0,\varepsilon)$ as $\varepsilon\downarrow0$.\hfill$\square$

\subsection{Proof of Corollary~\ref{cor:conditional-sharing}}\label{app:conditional-sharing}
Fix firm liability $r\in[0,\rho]$. Policies with this liability have $(f,b)=(f,\rho-r-f)$ for $0\le f\le\rho-r$. Both revenue functions are continuous on this compact segment, so each has an optimum. By Proposition~\ref{prop:ads}, $A(f,\rho-r-f)$ is strictly increasing in $f$ whenever the segment has positive length.

Take any optimum $f^{\mathrm{ad}}$ with advertising and any optimum $f^0$ without it. Their optimality inequalities are
\begin{align*}
 P^0(f^0,\rho-r-f^0)&\ge P^0(f^{\mathrm{ad}},\rho-r-f^{\mathrm{ad}}),\\
 P^0(f^{\mathrm{ad}},\rho-r-f^{\mathrm{ad}})+A(f^{\mathrm{ad}},\rho-r-f^{\mathrm{ad}})
 &\ge P^0(f^0,\rho-r-f^0)+A(f^0,\rho-r-f^0).
\end{align*}
Adding gives $A(f^{\mathrm{ad}},\rho-r-f^{\mathrm{ad}})\ge A(f^0,\rho-r-f^0)$. Strict monotonicity implies $f^{\mathrm{ad}}\ge f^0$. Since the two optimizers were arbitrary, the conclusion applies to every pair of optima, without concavity or uniqueness. If $r=\rho$, the segment consists of $f=0$ and the conclusion is immediate. The finite-loss result uses the same argument and Step 5's all-policy bounds.\hfill$\square$

\section{Reimbursement, prices, and commission income}\label{app:welfare}
This section proves Lemma~\ref{lem:funding-incidence} and derives the commission comparison used in Section~\ref{sec:policy}. A superscript 0 denotes a limiting equilibrium derivative along the small-loss path, including both firms' price responses. It does not denote a derivative holding prices fixed. We use $k=1-t$, $\ell=1-p_0$, and $z=\alpha\ell$ from Appendix~\ref{app:consumer}.

\subsection{The price sum: proof of Lemma~\ref{lem:funding-incidence}}\label{app:reimbursement-prices}
\paragraph{Step 1: Differentiate both pricing conditions.}
At fixed prices, reimbursement changes a firm's pricing incentive by $UR_{i,p_i}-D_iR_i$, as in~\eqref{eq:reimbursementpriceincentive}. To incorporate both firms' responses, differentiate $F_i=0$ in~\eqref{eq:referencecoeff}. Since $p_{ib}^0=-w_{ib}^0$, the equations are $J(p_{1b}^0,p_{2b}^0)'=F_b$. Subtracting and adding their rows gives, respectively,
\begin{align}
 p_{1b}^0-p_{2b}^0
 &=\frac{4z(3-z)P_2(z)}{\ell kP_4(z)}>0,\label{eq:reimbursement-price-gap}\\
 (z-2)P_3(z)(p_{1b}^0+p_{2b}^0)
 &-8z(3-z)^2(z-2)(p_{1b}^0-p_{2b}^0)
 =\frac{2P_2(z)}{\ell k}\{\ell P_2(z)+2z^2-12\}.\nonumber
\end{align}
The first identity also follows from the differenced equations displayed in Appendix~\ref{app:advertising}. Substitute it into the second equation and collect terms in $\ell$ to obtain
\begin{equation}
 \left.\frac{\partial(p_1+p_2)}{\partial b}\right|_0
 =\frac{2P_2(z)^2}{\ell k(z-2)P_3(z)}[\ell-\ell_B(z)],\label{eq:reimbursement-price-sum}
\end{equation}
where
\begin{equation}
 \ell_B(z)=-\frac{42z^6-256z^5+124z^4+2256z^3-6864z^2+9792z-6912}{P_2(z)P_4(z)}.\label{eq:reimbursement-threshold}
\end{equation}
The coefficient outside the brackets is positive: $P_2>0$, $P_3<0$, and $z<5/7<2$. Thus reimbursement lowers the price sum exactly when $\ell<\ell_B(z)$, raises it when $\ell>\ell_B(z)$, and has zero limiting effect at equality.

\paragraph{Step 2: Convert the price threshold into primitive conditions.}
First, $\ell_B(z)>0$ on the relevant interval. To check its numerator, put $u=1-z\in(0,1)$. The numerator of~\eqref{eq:reimbursement-threshold}, including its minus sign, becomes
\[
 1818+2300u+1282u^2+1032u^3+u^4(526-4u-42u^2)>0,
\]
since the last bracket exceeds $480$. The denominator is positive.

The sign certificates below establish that $\ell_B(z)$, $z/\ell_B(z)$, and $\ell_B(z)\Phi(z)/z$ are strictly increasing. The first property is also used in Appendix~\ref{app:welfareproof}. Appendix~\ref{app:pricing} establishes the same property for $\Phi$. Direct substitution gives
\[
 \left.\left(\frac{\ell_B(z)\Phi(z)}z-1\right)\right|_{z=2/5}
 =-\frac{1621474}{41384329}<0,\qquad
 \left.\left(\frac{\ell_B(z)\Phi(z)}z-1\right)\right|_{z=1/2}
 =\frac{455937}{20959141}>0.
\]
There is therefore exactly one $z_B^*\in(2/5,1/2)$ satisfying $\ell_B(z_B^*)\Phi(z_B^*)/z_B^*=1$. Define $\alpha_B^*=\Phi(z_B^*)$. Since $\Phi$ is increasing and $\Phi(1/2)<1$, this threshold belongs to $(0,1)$.

For fixed $\alpha$, the equilibrium relation $\Phi(z)=\alpha(1-\eta/k)$ makes $z$ strictly decrease with $\eta/k$. Its largest value is the root of $\Phi(z)=\alpha$, attained at zero cost. Also,
\[
 \ell<\ell_B(z)\quad\Longleftrightarrow\quad \frac{z}{\ell_B(z)}<\alpha.
\]
If $\alpha<\alpha_B^*$, then at that largest $z$ we have $\ell_B(z)\Phi(z)/z<1$, hence $z/\ell_B(z)>\alpha$. Because $z/\ell_B(z)$ increases continuously from zero, it equals $\alpha$ at exactly one smaller value $z_B(\alpha)$. Define
\[
 \vartheta_B(\alpha)=1-\frac{\Phi(z_B(\alpha))}{\alpha}.
\]
Reimbursement lowers the sum precisely when $z<z_B(\alpha)$, equivalently when $\eta/k>\vartheta_B(\alpha)$.

As $\alpha$ increases, $z_B(\alpha)$ increases. Using $\alpha=z_B/\ell_B(z_B)$ in this definition shows that $\vartheta_B=1-\ell_B(z_B)\Phi(z_B)/z_B$ strictly decreases. At $z=0$, $\ell_B=1/2$ and $\Phi(z)/z\to3/2$, so its limit is $1/4$. At $\alpha_B^*$ it reaches zero. For $\alpha>\alpha_B^*$, even the largest equilibrium $z$ satisfies $z/\ell_B(z)<\alpha$; the strict reduction therefore holds at every cost $0\le\eta<k$. At $\alpha=\alpha_B^*$, equality occurs only at $\eta=0$. These observations prove every price-sum claim in the lemma, including the sufficient condition $\eta/k\ge1/4$.

\subsection{The individual prices}\label{app:individual-prices}
The positive difference in~\eqref{eq:reimbursement-price-gap} establishes the ordering of the two responses. Its underlying direct incentives differ because $R_{1p_1}=\alpha$ and $R_{2p_2}=\alpha(1-\alpha)$ at the reference point, while $D_iR_i=D_0(\alpha-D_0)$ is the same for both firms. Equation~\eqref{eq:reimbursement-price-gap} includes the strategic responses as well.

For the signs of the individual derivatives, define their zero-response thresholds by
\[
 \ell_1(z)=\ell_B(z)-\frac{2z(3-z)(z-2)P_3(z)}{P_2(z)P_4(z)},\qquad
 \ell_2(z)=\ell_B(z)+\frac{2z(3-z)(z-2)P_3(z)}{P_2(z)P_4(z)}.
\]
The added term is positive, so $\ell_1<\ell_B<\ell_2$. Taking half the sum plus or minus half the difference in~\eqref{eq:reimbursement-price-sum} and~\eqref{eq:reimbursement-price-gap} gives
\[
 p_{ib}^0=\frac{P_2(z)^2}{\ell k(z-2)P_3(z)}[\ell-\ell_i(z)],\qquad i=1,2.
\]
Again the multiplier is positive. The exact inequalities needed below are
\begin{align}
 \ell_1(z)\Phi(z)/z&<3/4 &&(0<z\le4/7),\nonumber\\
 \ell_1(z)&<z &&(4/7\le z\le5/7),\nonumber\\
 \ell_2(z)&>2z &&(0\le z\le1/4),\label{eq:individual-price-signs}\\
 \ell_2(z)\Phi(z)/z&>1 &&(1/4\le z\le5/7),\nonumber\\
 \ell_1(z)\Phi(z)/z&>7/10 &&(0\le z\le5/7).\nonumber
\end{align}
Ratios involving $\Phi(z)/z$ at zero mean their continuous limits. Here is how these inequalities imply the primitive conditions in the lemma.

If $\eta/k\le1/4$, the price equation gives $\ell=(1-\eta/k)z/\Phi(z)\ge3z/[4\Phi(z)]$. Moreover $\ell=z/\alpha\ge z$. The first inequality in~\eqref{eq:individual-price-signs} gives $\ell>\ell_1$ for $z\le4/7$, and the second gives it for larger $z$. Thus $p_{1b}^0>0$.

If $\alpha\ge1/2$, then $\ell=z/\alpha\le2z$. The price equation and $\eta\ge0$ also give $\ell\le z/\Phi(z)$. The third inequality gives $\ell<\ell_2$ when $z\le1/4$, and the fourth does so when $z\ge1/4$. Thus $p_{2b}^0<0$.

Finally, if $\eta/k\ge3/10$, then $\ell\le7z/[10\Phi(z)]<\ell_1<\ell_2$ by the last inequality. Both prices fall. The response expansion in Appendix~\ref{app:consumer} has error $O(\sqrt\rho)$ uniformly over feasible policies at fixed primitives. It preserves each strict limiting sign for sufficiently small positive losses. This completes the proof of Lemma~\ref{lem:funding-incidence}.\hfill$\square$

\paragraph{Exact sign certificates for the threshold comparisons.}
For completeness, Table~\ref{tab:price-sign-certificates} supplies finite algebraic certificates for the three monotonicity claims and the five inequalities just used. Every entry in its first column is a polynomial after cancellation; all removed factors are positive on the specified interval. No numerical root approximation is used for these signs.

To verify the table, set $z=a+(b-a)x$ for the listed interval $[a,b]$ and expand its polynomial as $q(a+(b-a)x)=\sum_{j=0}^n a_jx^j$. The binomial identity
\[
 q(a+(b-a)x)=\sum_{i=0}^n\beta_i\binom ni x^i(1-x)^{n-i},\qquad
 \beta_i=\sum_{j=0}^i a_j\frac{\binom ij}{\binom nj}
\]
gives the coefficients directly. Substitution of~\eqref{eq:reimbursement-threshold} and the two individual thresholds into this finite sum gives the positive lower bounds in the last column. Since the basis terms are nonnegative and sum to one, $q$ is at least that lower bound everywhere on the interval. For the first individual-price inequality, the factor $z$ removed in the table is positive for $z>0$; the original difference equals zero at $z=0$.

\begin{table}[htbp]
\centering
\caption{Exact polynomial sign certificates for reimbursement prices}\label{tab:price-sign-certificates}
\small
\renewcommand{\arraystretch}{1.15}
\begin{tabular}{@{}lccc@{}}
\toprule
Polynomial $q(z)$ & $[a,b]$ & Degree & Lower bound for every $\beta_i$\\
\midrule
$P_2^2P_4^2\ell_B'$ & $[0,5/7]$ & 10 & $5000000$\\
$(P_2P_4\ell_B)^2(z/\ell_B)'$ & $[0,5/7]$ & 12 & $4000000$\\
$P_2^3P_4^2(\ell_B\Phi/z)'$ & $[0,5/7]$ & 12 & $90000000$\\
$4P_2^2P_4(3/4-\ell_1\Phi/z)/z$ & $[0,4/7]$ & 7 & $2220$\\
$P_2P_4(z-\ell_1)$ & $[4/7,5/7]$ & 7 & $225$\\
$P_2P_4(\ell_2-2z)$ & $[0,1/4]$ & 7 & $1572$\\
$P_2^2P_4(\ell_2\Phi/z-1)$ & $[1/4,5/7]$ & 8 & $3598$\\
$10P_2^2P_4(\ell_1\Phi/z-7/10)$ & $[0,5/7]$ & 8 & $23694$\\
\bottomrule
\end{tabular}
\end{table}

\subsection{Do the additional commissions cover reimbursement?}\label{app:commission-budget}
At fixed prices, compensation leaves consumer choices unchanged. The envelope condition and the derivatives of total retained demand therefore give
\[
 U_b^0=-D_0(p_{1b}^0+p_{2b}^0),\qquad
 (D_1+D_2)_b^0=-\alpha(1-z)(p_{1b}^0+p_{2b}^0).
\]
For the second identity, each partial derivative $(D_1+D_2)_{p_i}$ equals $-\alpha(1-z)$ at the reference point. Substitution into~\eqref{eq:commission-accounting-main} yields
\begin{equation}
 C_b^0=t\{U_0D_0-p_0[U_0\alpha(1-z)+2D_0^2]\}
 \partial_b(p_1+p_2)\big|_0.
 \label{eq:funding-commission-slope}
\end{equation}
For $t>0$, its coefficient on the price response is negative exactly when
\[
 \ell<\frac{5z^2-20z+18}{(2z-3)(3z-8)}.
\]
Both policy regions in Proposition~\ref{prop:policy} have $\ell<2/3$. In part (a), $z<\alpha\le3/10$ and the decreasing ratio $\Phi(z)/z$ exceeds $\Phi(3/10)/(3/10)=955/649$. Thus $\ell=(1-\eta/k)/[\Phi(z)/z]\le(24/25)(649/955)<2/3$. In part (b), the construction in Appendix~\ref{app:interior} takes $\ell$ sufficiently small. The bound $2/3$ suffices because
\[
 \frac{5z^2-20z+18}{(2z-3)(3z-8)}-\frac23
 =\frac{3z^2-10z+6}{3(2z-3)(3z-8)}>0.
\]
The numerator decreases to $19/49>0$ on $[0,5/7]$, and the denominator is positive. A lower price sum therefore raises commissions in both regions. If $t=0$, commissions are identically zero.

An increase in $b$ also costs $\mathcal R_0$ on existing returns. At the reference point $b=0$, the effect of changed return volume on this bill is zero. Hence the commission-only platform's net marginal gain is $C_b^0-\mathcal R_0$, positive exactly when~\eqref{eq:commission-covers-returns} holds. Combining the price and commission responses gives the explicit test
\begin{equation}
 \frac{2tP_2(z)^2(\ell_B-\ell)[p_0U_0\alpha(1-z)+2p_0D_0^2-U_0D_0]}{\ell k(z-2)P_3(z)\mathcal R_0}>1.\label{eq:funding-budget-test}
\end{equation}
The denominator is positive, since $\mathcal R_0=2U_0(\alpha-D_0)>0$. All quantities are determined by $(\alpha,\eta,t)$ through the reference price equation. The ratio is the additional commission income divided by reimbursement on existing returns. In the two regions above, reimbursement that raises the price sum loses commissions as well as incurring the direct expense. When it lowers the price sum, profitability requires that the ratio exceed one.

\section{The common policy comparison and the range of shared payment}\label{app:policy-general}

This appendix derives the revenue comparisons used to solve the joint allocation problem. It proves the general corner classification, the cost condition excluding platform payment, and the shared-payment region. Its uniform expansion also supports the conditional sharing result proved in Appendix~\ref{app:conditional-sharing}. All use the same price equations.

\subsection{Calculating the revenue coefficients}

Fix $(\alpha,k,\eta,\xi)$ with $\alpha\in(0,1]$, $k=1-t\in(0,1]$, $0\leq\eta<k$, and finite $\xi$ satisfying~\eqref{eq:active-ratio}. The initial price equation determines $z$, $\ell=z/\alpha$, and $p_0=1-\ell$:
\[
 \Phi(z)=\alpha(1-\eta/k),\qquad
 D_0=z(1-z/2),\quad U_0=z\ell(1-z/3),\quad
 \mathcal R_0=2U_0(\alpha-D_0).
\]
All coefficients below can therefore be calculated from the original market parameters.

For completeness, the following finite differentiation gives $C_f^0$ and $C_b^0$ without requiring a separate polynomial formula. Put $r=\rho-f-b$, $p_i=1+f-w_i$, and define
\begin{align*}
 \widehat D_1&=\alpha\{w_1-\alpha w_2^2/2+\alpha f+s\},\\
 \widehat D_2&=\alpha\{w_2(1-\alpha w_1)+\alpha w_2^2/2-\alpha f-s\},\\
 \widehat U&=\alpha\{(w_1^2+w_2^2)/2-2f-s/\alpha\}\\
 &\quad+\alpha^2\{-w_1w_2^2/2+(w_1-w_2)(f+s/\alpha)+w_2^3/6\},\\
 \widehat S_1&=\alpha,\qquad
 \widehat S_2=\alpha\{1+\alpha(w_2-w_1)\},\\
 \widehat\Pi_i&=\widehat U\{(kp_i-\eta-f-b)\widehat D_i-r\widehat S_i\},\\
 \widehat C&=t\widehat U(p_1\widehat D_1+p_2\widehat D_2),\qquad
 \widehat\Gamma=\widehat\Pi_1-\widehat\Pi_2.
\end{align*}
The hats identify algebraic expressions used to calculate the leading coefficients. They omit only the terms containing $h^3$ in utility and $rh$ in firm profits, where $h=\sqrt{2(f+s/\alpha)}$. Both omitted terms are $O(\rho^{3/2})$ on the small-cost path and therefore have zero first-order policy coefficients. Define $\widehat F_i=\partial_{w_i}\widehat\Pi_i$, evaluate all following derivatives at $w_1=w_2=\ell$ and $f=b=\rho=s=0$, and set
\begin{equation}
 J=\widehat F_w,\qquad
 w_j=-J^{-1}\widehat F_j,\qquad
 C_j^0=\widehat C_j+\widehat C_w w_j,\qquad
 \Gamma_j^0=\widehat\Gamma_j+\widehat\Gamma_w w_j.
 \label{eq:policy-coefficient-recipe}
\end{equation}
Here $j\in\{f,b,\rho,s\}$ and partial derivatives hold $w$ fixed. Direct differentiation gives
\[
 \det J=\frac{k^2z^4(z-2)P_3(z)P_4(z)}{144P_2(z)^2}>0
 \qquad(0<z<5/7).
\]
The polynomial signs are those established after \eqref{eq:referencecoeff}. Thus the two-equation calculation is nonsingular under the stated conditions.

The position-value coefficients agree with~\eqref{eq:position-response-coefficients}:
\begin{align}
 \Gamma_b^0&=B,\qquad \Gamma_f^0=B+S,\qquad S>0,\nonumber\\
 \Gamma_\rho^0&=-B,\qquad \Gamma_s^0=S/\alpha.
 \label{eq:policy-position-coefficients}
\end{align}
All coefficients therefore follow from the same price system; the commission response also agrees with~\eqref{eq:funding-commission-slope}.

\subsection{Uniform revenue derivatives and exact policy optima}

Write
\[
 \varepsilon=\sqrt\rho,\qquad f=\varepsilon^2\lambda,
 \qquad b=\varepsilon^2\mu,\qquad
 \Delta=\{(\lambda,\mu):\lambda\geq0,\ \mu\geq0,\ \lambda+\mu\leq1\}.
\]
Then $h=\varepsilon\sqrt{2(\xi+\lambda)}$. In the consumer quantities for the first-product-first region, the pricing equations are smooth functions of $(\varepsilon,\lambda,\mu)$ on a neighborhood of the compact set $\{0\}\times\Delta$. The term $\sqrt{2\lambda}$ does not enter those quantities; it enters the trial-order and deviation checks in Lemma~\ref{prop:prices}. Since $\xi>0$, the remaining square root is smooth even at $\lambda=0$. All terms of order $\varepsilon$ in the pricing equations vanish. The nonsingularity of $J$ and the implicit function theorem consequently give
\begin{equation}
 p_i=p_0+\varepsilon^2 q_i(\lambda,\mu)+O(\varepsilon^3),
 \label{eq:policy-uniform-prices}
\end{equation}
uniformly on $\Delta$, where $q_i$ is affine in the policy shares. Lemma~\ref{prop:prices} establishes that these roots are the unique pricing equilibria for all sufficiently small positive $\varepsilon$. The smooth extension also gives the stated remainder in the $C^1$ norm with respect to $(\lambda,\mu)$: the remainder and both first derivatives are bounded by a constant times $\varepsilon^3$, uniformly on $\Delta$. Indeed, Taylor expansion in $\varepsilon$ can be differentiated with respect to the policy shares on a common neighborhood of this compact set. The same argument applies to the smooth revenue expressions below.

Let $\overline C_\rho$ denote the exact commission income at $f=b=0$, with $c=\eta+\rho$ and $s=\alpha\xi\rho$. Taylor expansion using \eqref{eq:policy-coefficient-recipe} gives, uniformly over $\Delta$,
\begin{align}
 C(\rho\lambda,\rho\mu)
 &=\overline C_\rho+\rho(C_f^0\lambda+C_b^0\mu)+O(\rho^{3/2}),\nonumber\\
 b\mathcal R(\rho\lambda,\rho\mu)
 &=\rho\mu\mathcal R_0+O(\rho^{3/2}),\nonumber\\
 \Gamma(\rho\lambda,\rho\mu)
 &=\rho\{-B(1-\lambda-\mu)+(\lambda+\xi)S\}
     +O(\rho^{3/2}).
 \label{eq:policy-uniform-expansion}
\end{align}
Each remainder in~\eqref{eq:policy-uniform-expansion} has order $\rho^{3/2}$ also in the $C^1$ norm with respect to the policy shares. For the second line, market returns equal $\mathcal R_0+O(\sqrt\rho)$ because shipments contain a term linear in $h$. For the third line, use~\eqref{eq:policy-position-coefficients}. Both revenue functions $P^0$ and $P^{\mathrm{ad}}=P^0+\Gamma$ therefore have this derivative control.

Define
\begin{align}
 \mathcal V^0(\lambda,\mu)
 &=C_f^0\lambda+(C_b^0-\mathcal R_0)\mu,\nonumber\\
 \mathcal V^{\mathrm{ad}}(\lambda,\mu)
 &=\mathcal V^0(\lambda,\mu)
 -B+(B+S)\lambda+B\mu+\xi S.
 \label{eq:policy-limit-objectives}
\end{align}
For either revenue model $a\in\{0,\mathrm{ad}\}$,
\begin{equation}
 P^a(\rho\lambda,\rho\mu)
 =\overline C_\rho+\rho \mathcal V^a(\lambda,\mu)+O(\rho^{3/2})
 \label{eq:policy-objective-expansion}
\end{equation}
uniformly on $\Delta$. Both functions are affine. Their values at the three vertices are exactly the entries in Table~\ref{tab:policy-gains}.

\paragraph{Proof of Proposition~\ref{prop:policy-general}.}\label{app:policy-corners}
For each revenue model, subtract $\overline C_\rho$ and divide platform revenue by $\rho$. Equation~\eqref{eq:policy-objective-expansion} gives convergence in $C^1(\Delta)$ to the corresponding affine function $\mathcal V^a$. Let $v_*$ be its strictly best vertex, and let $\delta>0$ be the smallest revenue advantage over the other two vertices. The directional derivative of $\mathcal V^a$ from $v_*$ toward either of them is at most $-\delta$. Uniform convergence of derivatives makes both directions strictly decreasing everywhere on $\Delta$ for sufficiently small positive $\rho$. Every direction from $v_*$ to another policy is a nonnegative combination of these two directions. Integrating along that segment proves that $v_*$ uniquely maximizes the exact objective. If vertex values tie, these strict directional comparisons need not hold, and the next-order terms can determine sharing.\hfill$\square$

\paragraph{Consumer funding without advertising.}\label{app:consumer-funding-example}
The first column need not be maximized by a free-return policy. An exact example takes $\alpha=k=1/2$, $\ell=1/3$, and $z=1/6$. The initial price equation gives $\eta=584/2313$. Substitution into \eqref{eq:policy-coefficient-recipe} gives
\[
 C_f^0=\frac{2584682827325}{668229583711536}>0,
 \qquad
 C_b^0-\mathcal R_0=-\frac{269548440215}{167057395927884}<0.
\]
Thus the commission-only platform uniquely chooses consumer funding at this parameter point for every fixed finite $\xi\geq9/4$ and all sufficiently small positive losses. The strict inequalities also hold in a neighborhood. These comparisons clarify why the commission-only choices in Proposition~\ref{prop:policy} require their stated market conditions.

\subsection{Proof of Proposition~\ref{prop:no-platform-payment}}\label{app:obstruction}

Let $\zeta=D_0/\alpha=\ell(1-z/2)$ be initial order retention. The initial price equation gives $\ell\leq z/\Phi(z)$, and hence
\[
 \zeta\leq\frac{z(1-z/2)}{\Phi(z)}<\frac23.
\]
The last inequality follows from
\[
 \frac23-\frac{z(1-z/2)}{\Phi(z)}
 =\frac{z(9z^2-46z+64)}{6(5z^2-26z+36)}>0.
\]
For $2/5\leq\zeta\leq2/3$, substitute $\ell=2\zeta/(2-z)$ in $T+\mathcal R_0$, where $T=(C_f^0-C_b^0)/t$ is the commission-response difference per unit commission, continued to $t=0$. This expression depends only on $(z,\ell)$. Clearing its positive denominator gives
\[
 T+\mathcal R_0=
 \frac{z^2 H(z,\zeta)}{-3\zeta(z-2)^2P_3(z)P_4(z)}>0.
\]
Here $H$ is the polynomial obtained by multiplying $T+\mathcal R_0$ by the displayed denominator and dividing by $z^2$, after the substitution for $\ell$. Under $z=5x/7$ and $\zeta=2/5+4y/15$, its Bernstein representation on $[0,1]^2$ has degrees $(12,2)$ and 39 strictly positive coefficients, with exact minimum $71859992824/41597325$. Thus $H>0$ on the whole rectangle; the displayed denominator is positive because $P_3<0<P_4$. Appendix~\ref{app:verification} explains why these coefficient bounds prove positivity.

It follows that the commission-only gain from replacing reimbursement by an equal consumer payment is strictly positive at the limit:
\[
 C_f^0-C_b^0+\mathcal R_0
 =k\mathcal R_0+t(T+\mathcal R_0)>0,
 \qquad k=1-t>0.
\]
The $C^1$ expansion above then gives, uniformly over all feasible policies,
\[
 P_f^0-P_b^0=C_f^0-C_b^0+\mathcal R_0+O(\sqrt\rho)>0.
\]
Along the same direction, $\Gamma_f-\Gamma_b=S+O(\sqrt\rho)>0$, so advertising receipts increase along the same change of payer. Starting at any policy with $b>0$, increasing $f$ and decreasing $b$ by the same amount is feasible and strictly increases profit under either revenue model. Consequently every optimum has $b=0$. This excludes reimbursement throughout the policy triangle, including shared-payment policies.

\paragraph{A condition on the original parameters.}
At fixed $\alpha$, retention $\zeta=(z/\alpha)(1-z/2)$ increases strictly in $z<1$, while $z$ decreases strictly in $\eta/k$ through $\Phi(z)=\alpha(1-\eta/k)$. The retention threshold $\zeta=2/5$ is reached at
\[
 z_R=1-\sqrt{1-4\alpha/5}.
\]
Thus the precise cost condition corresponding to this retention bound is
\begin{equation}
 \frac{\eta}{1-t}\leq1-\frac{\Phi(z_R)}{\alpha}.
 \label{eq:no-platform-primitive-sharp}
\end{equation}
The right side decreases in $\alpha$. To see this without a numerical calculation, substitute $\alpha=5z_R(2-z_R)/4$. As a function of $z_R$ it becomes
\[
 \frac{15z_R^3-90z_R^2+176z_R-96}
 {5(z_R-2)(3z_R^2-16z_R+24)}.
\]
Its derivative is $-4Q(z_R)/[5(z_R-2)^2P_2(z_R)^2]$, where
\[
\begin{aligned}
 Q(z)&=15z^4-156z^3+616z^2-1104z+768,\\
 Q(1-u)&=15u^4+96u^3+238u^2+280u+139>0
\end{aligned}
\]
for $u=1-z_R>0$. Since $z_R$ increases in $\alpha$, the claimed monotonicity follows. The threshold tends to $2/5$ as $\alpha$ falls to zero and has minimum
\[
 \frac{40\sqrt5-53}{179}>\frac15
\]
at $\alpha=1$. Therefore $\eta/(1-t)\leq1/5$ is sufficient for no reimbursement at each fixed suitability. All conclusions concern sufficiently small positive losses along~\eqref{eq:searchratio}.\hfill$\square$

The same result explains the retention restriction on the shared-payment region. When initial retention is at least $2/5$, replacing platform payment by consumer payment strictly raises profit, so their leading gains cannot coincide. Sharing in Proposition~\ref{prop:policy}(b) requires initial retention below $2/5$; its construction uses the stronger sufficient bound $\ell_0<1/3$, and hence $\zeta<1/3$.

\subsection{Firm-paid free returns: Proposition~\ref{prop:policy}(a)}\label{app:higher}
We establish the policy comparison by determining the sign of each instrument's effect throughout the policy triangle. Consider the following conditions on the original parameters
\begin{equation}
 \alpha\in[1/10,3/10],\quad t\in[1/10,1/5],\quad
 \frac\eta{1-t}\in[1/25,1/10],\quad \xi\ge9/4.\label{eq:highretprimitive}
\end{equation}
Fix a finite $\xi$, with $c=\eta+\rho$ and $s=\alpha\xi\rho$. The initial price equation implies
\[
 \frac35<\ell<\frac{13}{20},\qquad \frac3{50}<z<\frac{39}{200},\qquad \frac45\le k\le\frac9{10}.
\]
Indeed, $\eta/k=1-\Phi(\alpha\ell)/\alpha$ decreases in $\ell$ and increases in $\alpha$. At the two relevant corners its values lie above $1/10$ and below $1/25$, respectively, placing the root inside these bounds. The exact corner values are recorded in the supplement.

On this entire box, the common derivative formulas give
\begin{equation}
 C_f^0<0,\quad C_b^0-\mathcal R_0<0,\quad C_f^0+\Gamma_f^0>0,\quad C_b^0-\mathcal R_0+\Gamma_b^0<0.\label{eq:firm-region-signs}
\end{equation}
The first two inequalities show that either assigning costs to consumers or compensating firms reduces the commission-only platform's profit. The last two show that advertising reverses the incentive to assign costs to consumers while leaving reimbursement unprofitable. After positive denominators are cleared, each required sign has a Bernstein representation with strictly positive coefficients on the box. Appendix~\ref{app:verification} describes this exact sign argument; the supplement records the expressions and coefficients.

Since $k\geq4/5$ and $\xi\geq9/4$, condition~\eqref{eq:active-ratio} holds. Lemma~\ref{prop:prices} therefore gives positive bids at every feasible policy for sufficiently small positive losses, so the advertising derivatives apply throughout the triangle.

All four strict signs in~\eqref{eq:firm-region-signs} persist uniformly at these positive losses. The commission-only platform reduces both instruments to zero and uniquely chooses $(0,0)$. The advertising platform reduces $b$ to zero and raises $f$ to $\rho$, uniquely choosing $(\rho,0)$. Its extra advertising receipts exceed the lost commissions.

Finally, at the initial prices each firm retains the fraction
\[
 \frac{D_0}{\alpha}=\ell(1-z/2)>\frac35\left(1-\frac{39}{400}\right)>\frac12.
\]
The uniform price continuation preserves both majority retention and positive operating profits. After auction payments, either firm receives the second-position operating profit: the winner pays $\Pi_1-\Pi_2$, and the loser pays zero. Net firm profits are therefore positive as well. Appendix~\ref{app:finite} establishes the finite-loss comparison; Appendix~\ref{app:welfareproof} proves the associated welfare ranking.

\subsection{Proof of Proposition~\ref{prop:policy}(b): platform and consumer sharing}\label{app:interior}

On $f+b=\rho$, the consumer and platform entries differ by
\[
 E=S+C_f^0-C_b^0+\mathcal R_0.
\]
Thus the leading advertising objective is indifferent along this edge when $E=0$. We first construct market parameters at which this equality holds, then use the next profit term to locate the two commission thresholds.

Write $\mathcal S=S/k$ and use $T$ from Appendix~\ref{app:obstruction}. Both are coefficient functions determined by $(z,\ell)$. Then
\[
 E=k\mathcal S+(1-k)T+\mathcal R_0,
 \qquad K(z,\ell)=-\frac{T+\mathcal R_0}{\mathcal S-T}.
\]
The function $K$ is the value of the firm's revenue share $k$ that solves $E=0$ at a given $(z,\ell)$. On $0<z\leq\ell\leq3/8$, the exact sign calculations give, at $k=K$,
\begin{gather}
 T+\mathcal R_0<0,\quad 0<K<1,\quad
 P_b^{\mathrm{ad},0}>0,\quad C_b^0-\mathcal R_0>0,\nonumber\\
 K_\ell+(z/\ell)K_z<0,\qquad
 3a_3-2U_0\alpha^2<0.
 \label{eq:policy-sharing-signs}
\end{gather}
Here $P_b^{\mathrm{ad},0}=C_b^0-\mathcal R_0+B$ is the leading advertising-platform reimbursement derivative, and $a_3$ is derived below. For the sign verification, clear positive denominators and substitute
\[
 z=\frac{3Z}{8(Z+V+H)},\qquad
 \ell=\frac{3(Z+V)}{8(Z+V+H)},\qquad Z,V,H\geq0.
\]
The variables $Z,V,H$ are only nonnegative coordinates for this triangular domain, with $Z>0$ and $Z+V+H>0$; they parameterize only this sign calculation. Multiplying by a sufficient power of $Z+V+H$ turns each numerator into a polynomial whose terms have the same total degree. The resulting degrees and numbers of strictly positive coefficients are as follows; all other coefficients are zero.
\begin{center}\small
\begin{tabular}{lrr}
\toprule
Positive expression & Degree & Positive coefficients\\
\midrule
$-(T+\mathcal R_0)$ &12&90\\
$P_b^{\mathrm{ad},0}$ &12&90\\
$2U_0\alpha^2-3a_3$ &14&119\\
$-[K_\ell+(z/\ell)K_z]$ &27&405\\
$C_b^0-\mathcal R_0$ &20&189\\
\bottomrule
\end{tabular}
\end{center}
Each polynomial has a strictly positive pure-$Z$ coefficient, giving a strict sign for $z>0$. Every rational expression in this calculation is determined by \eqref{eq:policy-coefficient-recipe} and the displayed formula for $a_3$ below. Appendix~\ref{app:verification} explains why the coefficient signs establish the required inequalities. The inequalities $\mathcal S>0$ and $T+\mathcal R_0<0$ also give $0<K<1$ directly.

On $0<z\leq\ell\leq1/3$, one also has $K>1/4$. Since $\mathcal S-T>0$, this is equivalent to
\[
 -3T-4\mathcal R_0-\mathcal S>0.
\]
After clearing the denominator, substitute $z=Z/[3(Z+V+H)]$ and $\ell=(Z+V)/[3(Z+V+H)]$. The same polynomial construction gives 119 positive numerator coefficients and 77 positive denominator coefficients; both have a strictly positive pure-$Z$ coefficient. This proves the sign for $z>0$. Hence $t_0=1-K<3/4$, and $\xi\geq9/4$ implies~\eqref{eq:active-ratio} in a neighborhood of $t_0$.

At fixed $\alpha$, both $K(\alpha\ell,\ell)$ and $1-\Phi(\alpha\ell)/\alpha$ are positive and strictly decreasing on $0<\ell\leq3/8$. We use the subset $0<\ell\leq1/3$ to ensure positive bids throughout the policy set. For the second factor, $\Phi(z)/z<3/2$ gives $\Phi(\alpha\ell)/\alpha<9/16$, and $\Phi'>0$ gives monotonicity. Direct substitution in the coefficient formulas gives $K(\alpha\ell,\ell)\to1$ as $\ell\downarrow0$. Their product therefore decreases from one to
\[
 \eta_{\min}(\alpha)=K(\alpha/3,1/3)
    [1-\Phi(\alpha/3)/\alpha].
\]
Since $0<\eta_{\min}(\alpha)<1$, this is a nonempty interval. Every $\eta\in(\eta_{\min}(\alpha),1)$ determines a unique $\ell_0\in(0,1/3)$ and $t_0=1-K(\alpha\ell_0,\ell_0)$ with $E(t_0)=0$. The construction gives $\eta<K(\alpha\ell_0,\ell_0)=1-t_0$, so the initial equilibrium is feasible. To see how $E$ changes with commission, hold $(\alpha,\eta)$ fixed and differentiate at $t=t_0$:
\[
 \ell_t=-\frac{1-\Phi(z)/\alpha}{k\Phi'(z)}<0,
 \quad
 E_{\ell\mid\alpha}=-(\mathcal S-T)[K_\ell+(z/\ell)K_z]>0,
 \quad
 E_t=-(\mathcal S-T)+E_{\ell\mid\alpha}\ell_t<0.
\]
For the middle identity, write $E=(\mathcal S-T)(k-K)$: at $t_0$ the factor $k-K$ vanishes. The last identity also uses $dk/dt=-1$. All relevant inequalities are strict at $t_0$, so choose a fixed closed commission interval around $t_0$ on which feasibility, positive bids, $E_t<0$, and the two positive reimbursement slopes persist. The uniform expansion then preserves those reimbursement slopes over every policy for sufficiently small positive losses. Increasing $b$ at any policy with $f+b<\rho$ strictly raises profit under either revenue model. Every optimum therefore lies on $f+b=\rho$; its position value is $\rho(\lambda+\xi)S+O(\rho^{3/2})>0$, uniformly in $\lambda=f/\rho$.

\paragraph{The next term in boundary profit.}
Let $m=kp_0-\eta$ and $\omega=\alpha m-kD_0$. On $r=0$, the term $h^3$ enters firm profits through $U$ alone. Its coefficient in each pricing equation is $\alpha^2\omega/3$. The resulting coefficient in $w$ is
\[
 e_3=-J^{-1}\frac{\alpha^2\omega}{3}\binom11.
\]
At the initial prices, define the two revenue gradients
\[
 L_C=t\{D_0(2p_0D_0)+U_0[-D_0+p_0\alpha(1-z)]\},\qquad
 L_A=U_0(\omega+m\alpha^2\ell).
\]
The coefficient of $h^3$ in $C+\Gamma$, including repricing, is
\begin{align}
 a_3&=(L_C+L_A,L_C-L_A)e_3+t(2p_0D_0)\alpha^2/3\nonumber\\
 &=\frac{tz^3(2-z)
 \{\ell(21z^3-92z^2+36z+144)-15z^3+50z^2+60z-216\}}
 {3\ell^2P_3(z)}.
 \label{eq:policy-a3-derived}
\end{align}
Meanwhile $\mathcal R=\mathcal R_0-U_0\alpha^2h+O(\rho)$. Hence the terms of order $\rho^{3/2}$ in boundary profit are
\[
 a_3h^3+U_0\alpha^2(\rho-f)h.
\]
The second term reflects the reduction in compensated returns when search becomes more selective.

With $\varepsilon=\sqrt\rho$ and $f=\rho\lambda$, the pricing equations and boundary profit have a smooth extension in $(\varepsilon,\lambda,t)$ around $\varepsilon=0$. Differentiating the expansion with respect to $f$, while holding $\rho$ and $s$ fixed, gives
\begin{align}
 \mathcal M(\lambda,t,\varepsilon)
 :=\frac{dP^{\mathrm{ad}}(f,\rho-f)}{df}
 &=E(t)+\varepsilon\psi(t,\lambda)+O(\varepsilon^2),\nonumber\\
 \psi(t,\lambda)
 &=(3a_3-U_0\alpha^2)\sqrt{2(\xi+\lambda)}
   +\frac{U_0\alpha^2(1-\lambda)}{\sqrt{2(\xi+\lambda)}}.
 \label{eq:policy-smooth-edge-expansion}
\end{align}
Here the coefficients retain their dependence on $t$. The remainder and its derivatives in $(\lambda,t,\varepsilon)$ have the orders supplied by the smooth Taylor expansion, uniformly for $0\leq\lambda\leq1$ and $t$ in a sufficiently small fixed neighborhood of $t_0$. In particular,
\[
 \psi_\lambda
 =\frac{3a_3-2U_0\alpha^2}{\sqrt{2(\xi+\lambda)}}
 -\frac{U_0\alpha^2(1-\lambda)}{[2(\xi+\lambda)]^{3/2}}<0.
\]
It follows that boundary profit is strictly concave: its second derivative with respect to $f$ is $\psi_\lambda/\varepsilon+O(1)<0$. Moreover, $\mathcal M_t=E_t+O(\varepsilon)<0$.

At $\varepsilon=0$, both endpoint equations reduce to $E(t_0)=0$, with derivative $E_t(t_0)<0$. The implicit function theorem therefore gives unique smooth thresholds solving $\mathcal M(1,t_C,\varepsilon)=0$ and $\mathcal M(0,t_P,\varepsilon)=0$ near zero. Evaluating $\psi$ and $E_t$ at $t_0$ gives
\begin{align}
 t_C&=t_0-\frac{\psi(t_0,1)}{E_t(t_0)}\sqrt\rho+O(\rho),\nonumber\\
 t_P&=t_0-\frac{\psi(t_0,0)}{E_t(t_0)}\sqrt\rho+O(\rho),\nonumber\\
 t_P-t_C&=
 \frac{\psi(t_0,0)-\psi(t_0,1)}{-E_t(t_0)}\sqrt\rho+O(\rho).
 \label{eq:policy-sharing-width}
\end{align}
The last coefficient is strictly positive because $\psi(t_0,0)>\psi(t_0,1)$ and $E_t(t_0)<0$. Within the chosen neighborhood, $t\leq t_C$ gives $\mathcal M(1,t,\varepsilon)\geq0$, so strict concavity selects full consumer funding. For $t_C<t<t_P$, the derivative is positive at $\lambda=0$ and negative at $\lambda=1$, so it has a unique interior zero. For $t\geq t_P$, $\mathcal M(0,t,\varepsilon)\leq0$ selects full platform funding. At an interior optimum, implicit differentiation at fixed $\rho$ gives
\[
 \frac{df^{\mathrm{ad}}}{dt}
 =-\rho\,\frac{\mathcal M_t}{\mathcal M_\lambda}<0.
\]
The commission-only platform has a positive reimbursement slope and a boundary slope $E-S=-S<0$ with respect to $f$ at $t_0$. The same uniform continuation preserves both strict signs: first raising reimbursement reaches $f+b=\rho$, then reducing the consumer contribution along that edge reaches its unique optimum $(0,\rho)$ throughout the neighborhood.

Equation~\eqref{eq:policy-sharing-width} gives the strictly positive $\sqrt\rho$ coefficient stated in Section~\ref{sec:policy}. The width comparison follows the path $s=\alpha\xi\rho$. At each fixed positive $\rho$, the commission comparison holds $(\alpha,\eta,c,s)$ fixed.\hfill$\square$

\section{Proof of Proposition~\ref{prop:welfare}: consumer surplus and total welfare}\label{app:welfareproof}
The welfare comparison uses the same price response as the platform's funding decision. We first show what compensation does through prices, then compare compensation with assigning costs to consumers. This proves the general classification in Proposition~\ref{prop:welfare}. Appendix~\ref{app:finite} establishes the finite-loss comparisons.

\paragraph{Compensation works through prices.}
All derivatives with a superscript 0 below are evaluated at the initial prices for $s=\rho=0$, including equilibrium repricing. At that point the consumer and resource identities in~\eqref{eq:welfareidentities} give~\eqref{eq:reference-welfare-slopes}. In particular, reimbursement has no direct resource or consumer-payment effect; its leading effect operates through $\partial_b(p_1+p_2)$.

The factors multiplying the price response are positive: $U_0,D_0>0$, $1-z>0$, and $p_0>\eta/k\geq\eta$. Compensation therefore benefits both consumers and society exactly when it lowers the sum of prices. Appendix~\ref{app:welfare} established that this happens when $\ell<\ell_B(z)$; both welfare effects are negative when $\ell>\ell_B(z)$.

\paragraph{Changing the payer.}
Charging consumers adds a direct cost to their shopping decisions. Using the same price derivatives, the effect relative to compensation is
\begin{equation}
 CS_f^0-CS_b^0=-U_0\{2(\alpha-D_0)+D_0(\partial_f-\partial_b)(p_1+p_2)\}<0.\label{eq:consumer-funding-ranking}
\end{equation}
For total welfare, the same resource identities give
\[
 W_f^0=(p_0-\eta)\{2U_0\alpha(1-z)-4D_0(\alpha-D_0)
 -[U_0\alpha(1-z)+2D_0^2]\,\partial_f(p_1+p_2)|_0\}.
\]
Combining this expression with $W_b^0$ in~\eqref{eq:reference-welfare-slopes} and the price responses gives the useful bound
\begin{equation}
 W_f^0-tW_b^0<0.\label{eq:welfare-fee-bound}
\end{equation}
This bound determines the gain from changing the payer when compensation improves welfare and the consumer contribution's effect when compensation reduces welfare. Both inequalities hold throughout the feasible initial-price range $0<z<5/7$, $0<\ell\le z/\Phi(z)$, and $0<k=1-t\le1$.

To check these signs, substitute the price responses from~\eqref{eq:referencecoeff} into the displayed welfare derivatives. Dividing $-(W_f^0-tW_b^0)$ by the positive factor
\[
 \frac{2(p_0-\eta)z^3(3-z)}{3\ell(z-2)P_3(z)P_4(z)}
\]
leaves an eighth-degree polynomial in $z$. Its nine Bernstein coefficients on $[0,5/7]$ are strictly positive, with minimum $187716019847/5764801$. For consumer surplus, multiply $(CS_f^0-CS_b^0)/U_0$ by the positive factor $-\ell P_3(z)P_4(z)/z$. The result is affine in $\ell$, and is strictly negative at both $\ell=0$ and $\ell=z/\Phi(z)$ by the same coefficient test. An affine function with negative endpoint values is negative throughout that interval. Appendix~\ref{app:verification} explains why these exact coefficient bounds prove the inequalities. This proves both strict comparisons on the full stated range.

\paragraph{The welfare-maximizing policy.}
The uniform-expansion argument in Appendix~\ref{app:policy-general} also applies to $CS=U^2/2$ and $W=UV-U^2/2$: after writing $(f,b)=\rho(\lambda,\mu)$, these expressions are smooth in $(\sqrt\rho,\lambda,\mu)$. Consequently their derivatives with respect to $f$ and $b$ converge uniformly on the policy triangle to the derivatives above. Any strict sign used below therefore holds at every feasible policy for all sufficiently small positive losses.

If compensation lowers the price sum, then $CS_b^0,W_b^0>0$. Starting from any policy, increase $b$ while holding $f$ fixed until $f+b=\rho$; each positive increase raises both welfare criteria. On that edge, replace the remaining consumer contribution with reimbursement. This raises consumer surplus by~\eqref{eq:consumer-funding-ranking}, and raises total welfare because~\eqref{eq:welfare-fee-bound} implies
\[
 W_f^0-W_b^0<-(1-t)W_b^0<0.
\]
Unless the starting policy was already $(0,\rho)$, at least one of these two changes is strict. Full platform payment therefore uniquely maximizes both welfare criteria.

If compensation raises the price sum, then $CS_b^0,W_b^0<0$. The same inequalities give
\[
 CS_f^0<CS_b^0<0,\qquad W_f^0<tW_b^0\le0.
\]
Reducing each positive instrument to zero is feasible and strictly raises both welfare criteria. Firm-paid free returns therefore uniquely maximize both. Consequently,
\begin{equation}
 \begin{cases}
 (0,\rho),&\ell<\ell_B(z),\\
 (0,0),&\ell>\ell_B(z)
 \end{cases}\label{eq:welfare-funding-corner}
\end{equation}
uniquely maximizes both consumer surplus and total welfare. The proposition excludes $\ell=\ell_B(z)$, where the leading reimbursement derivative vanishes and these strict comparisons do not decide its sign.\hfill$\square$

\paragraph{Application to the two policy regions.}
In the firm-funding region~\eqref{eq:highretprimitive}, $z<1/5$ and $\ell>3/5$. The price threshold is increasing, and
\[
 \ell_B(z)<\ell_B(1/5)<\frac35<\ell.
\]
Hence $(0,0)$ uniquely maximizes both welfare criteria, which identifies the welfare optimum in the firm-funding policy region.

For the commission transition in Appendix~\ref{app:interior}, $\ell_0<1/3$, whereas $\ell_B(z)>\ell_B(0)=1/2$. The inequality $\ell<\ell_B$ persists in a sufficiently small commission neighborhood of $t_0$. Thus $(0,\rho)$ uniquely maximizes both criteria throughout that neighborhood, which identifies the welfare optimum in the sharing region. Every advertising policy with a positive consumer contribution gives strictly lower welfare. When the platform itself pays the full loss, its policy coincides with the welfare optimum.

\section{Proof of Proposition~\ref{prop:finite}}
\label{app:finite-new}\label{app:finite}\label{app:reversal}

We construct the intervals asserted in both parts of the proposition. We first give fixed market parameters and two loss intervals for part (b), verify all price deviations, and derive the policy optima. We then allow five market parameters to vary independently to establish part (a). All acceptance inequalities use rational interval bounds, as described in Appendix~\ref{app:verification}. Numerical solutions provide candidate centers and preconditioners only.

\subsection{Coverage and price equilibria}\label{app:finite-prices}
For part (b), choose the fixed market parameters
\begin{equation}
 \alpha=\frac15,\qquad t=\frac3{20},\qquad
 \eta=\frac{783}{13280},\qquad \xi=3,
 \qquad c=\eta+\rho,\quad s=\frac35\rho,
 \label{eq:finite-expanded-primitives}
\end{equation}
and take the two intervals to be
\begin{equation}
 [\underline\rho_L,\overline\rho_L]=[0.004,0.006],\qquad
 [\underline\rho_H,\overline\rho_H]=[0.01999,0.02001].
 \label{eq:finite-loss-intervals}
\end{equation}
Their endpoints satisfy the strict ordering in the proposition. Let
$p=(p_1,p_2)'$ and
$\mathcal F=(\partial_{p_1}\Pi_1,\partial_{p_2}\Pi_2)'$, with consumer
quantities from~\eqref{eq:demands}. A parameter cell $\mathcal B$ consists
of a loss interval and a rectangle of policy shares, intersected with
$f,b\geq0$ and $f+b\leq\rho$. For the low-loss region, divide
$[0.004,0.006]$ into twenty closed intervals of width $0.0001$, and use
\[
 \frac f\rho\in[i/N,(i+1)/N],\qquad
 \frac b\rho\in[j/N,(j+1)/N],\qquad
 i,j\geq0,\quad i+j\leq N-1,
\]
with $N=40$. The resulting 820 policy rectangles cover the full triangle,
including its edges. For the moderate-loss region, use the whole loss
interval $[0.01999,0.02001]$ and $N=80$, giving 3240 policy rectangles.

For each cell, let $p_c$ be a rational price center, let $X$ be its price
rectangle of half-width $0.001$, and let $Y$ be a rational preconditioning
matrix approximating $\mathcal F_p^{-1}$. Write $I$ for the identity
matrix and $\|\cdot\|_\infty$ for the maximum absolute row-sum norm.
The interval evaluations satisfy
\begin{align}
 p_c-Y\mathcal F(p_c;\mathcal B)
  +[I-Y\mathcal F_p(X;\mathcal B)](X-p_c)
 &\subset\operatorname{int}X,\label{eq:finite-new-inclusion}\\
 \|I-Y\mathcal F_p(X;\mathcal B)\|_\infty&<1.
 \label{eq:finite-new-contraction}
\end{align}
To see why these inequalities establish a root for every point in the
cell, fix its parameters and consider $p\mapsto p-Y\mathcal F(p)$.
The mean-value theorem places its image of $X$ inside the left side
of~\eqref{eq:finite-new-inclusion}; \eqref{eq:finite-new-contraction}
makes it a contraction. Thus it has exactly one fixed point in $X$.
The contraction bound also makes $Y\mathcal F_p$ nonsingular, so $Y$ and
$\mathcal F_p$ are nonsingular. The fixed point therefore solves
$\mathcal F=0$ and varies smoothly with the parameters within the cell.
Its enclosure is the left side of~\eqref{eq:finite-new-inclusion}.

For neighboring policy cells, each root enclosure lies inside the other
cell's uniqueness rectangle. The same inclusions join corresponding cells
across adjacent loss intervals. At shared parameter values, local
uniqueness makes the roots agree. Hence they form one continuous branch
over each stated loss region and its full policy triangle. This proves
uniqueness within the linked rectangles; it does not exclude disconnected
price equilibria.

On every root enclosure, the bounds give $0<U<1$, $\pi_1>\pi_2>0$, and
\[
 p_i>f,\qquad w_1>g,\qquad w_2>h,\qquad p_1-p_2<h-g.
\]
Thus Lemma~\ref{lem:consumer-choices} applies, participation is interior,
and advertising bids are positive. Each firm's profit after auction
payments is $U\pi_2>0$. Direct bounds on $D_i/S_i$ give, respectively,
more than $58.8$ and $60.2$ percent retention in the low-loss region, and
more than $59.8$ and $61.6$ percent in the moderate-loss region. In
particular, both firms retain more than half their shipments.

\subsection{All unilateral price deviations}\label{app:finite-deviations}
Fix one parameter point and the rival's price on this branch. Below
$c/(1-t)$, the retained-order margin is nonpositive, and the returned-order
margin is $-(\rho-f-b)\leq0$. Such a deviation cannot exceed the positive
equilibrium profit. Above $1+f-g$ for firm~1 or $1+f-h$ for firm~2,
the deviator has no profitable trial opportunity. It remains to check the
compact interval between these bounds.

The order switches at $p_1=p_2+h-g$ for firm~1's deviations and at
$p_2=p_1+g-h$ for firm~2's deviations. On the interval preserving the
original order, use~\eqref{eq:demands}. On the reversed interval, exchange
the product labels, use $g$ as the second trial's deduction, and subtract
$s$ for the initial inspection of product~2. Both one-sided payoffs at the
switch are checked, so the comparison includes the specified tie choice.

In the low-loss region, interval subdivision gives strictly negative
own-price curvature on each order interval. The first-order condition
therefore makes the candidate price the unique maximum on the original
order interval. On the reversed interval, the derivative at the switch
points toward the switch. Strict concavity then makes the switch the
maximum of that interval, and its payoff is strictly below equilibrium
profit.

For moderate losses, the original-order derivative is strictly positive
below a neighborhood of the candidate price and strictly negative above
it; own-price curvature is strictly negative inside that neighborhood.
The first-order condition again gives the unique maximum on the entire
original-order interval. Reversed-order profit is strictly monotone toward
the switch, whose payoff is strictly below equilibrium profit. Any part of
that interval below $c/(1-t)$ was already excluded. These comparisons
exhaust all prices and establish strict global best responses.

\subsection{Policy optima in part (b)}\label{app:finite-policy}
Nonsingularity of the price Jacobian permits implicit differentiation:
\[
 \partial_jp=-\mathcal F_p^{-1}\mathcal F_j,\qquad j\in\{f,b\}.
\]
All derivatives below include this price response and hold the market
primitives fixed. The consumer envelope relation and
\eqref{eq:welfareidentities} give
\begin{align*}
 U_j&=-\sum_iD_i\partial_jp_i-(R_1+R_2)\mathbf 1_{\{j=f\}},
 &CS_j&=UU_j,\\
 V_j&=\sum_i(p_i-c)D_{i,j}-(\rho-f)(R_{1,j}+R_{2,j}),
 &W_j&=UV_j+(V-U)U_j.
\end{align*}
Substitution into the rational enclosures establishes, throughout both
loss regions and their full policy triangles,
\[
 P^0_b<0,\qquad P^{\mathrm{ad}}_f>0,\qquad
 P^{\mathrm{ad}}_b<0,\qquad CS_f<0,\quad CS_b<0,\quad W_b<0.
\]
In the low-loss region, $P^0_f<0$ and $W_f<0$ hold throughout the triangle.
In the moderate-loss region, $W_f>0$ holds throughout the triangle, while
$P^0_f<0$ is established on the entire edge $b=0$.

These signs imply global policy comparisons along feasible paths.
For commission-only profit, lowering $b$ to zero strictly improves any
policy with $b>0$; lowering $f$ on that edge then strictly improves any
remaining policy with $f>0$. Its unique optimum is therefore $(0,0)$.
For advertising profit, lowering $b$ and then raising $f$ leads to the
unique optimum $(\rho,0)$. Lowering both payments uniquely maximizes
consumer surplus at $(0,0)$. The same path maximizes total welfare in the
low-loss region; lowering $b$ and then raising $f$ uniquely maximizes it
at $(\rho,0)$ in the moderate-loss region. This proves part (b).

The finite-loss extension of Proposition~\ref{prop:ads} also follows on
this branch. Positive bids give $A=P^{\mathrm{ad}}-P^0$. Subtracting the
policy derivatives in the low-loss cells, and differentiating
$U(\pi_1-\pi_2)$ directly in the moderate-loss cells, yields
\begin{equation}
 A_f-A_b\in
 \begin{cases}
 {[0.923,1.836]}\,10^{-3},&\rho\in[0.004,0.006],\\
 {[0.625,1.083]}\,10^{-3},&\rho\in[0.01999,0.02001].
 \end{cases}
 \label{eq:finite-equal-relief}
\end{equation}
The bounds hold over the full policy triangles. On a fixed-relief segment
$f+b>0$, advertising revenue is therefore strictly increasing in the
consumer contribution, including feasible one-sided changes at its
endpoints. At $(0,0)$ the inequality compares an initial consumer contribution
with an equally large initial reimbursement. The monotone comparison in
Appendix~\ref{app:conditional-sharing} consequently proves the finite-loss
claim in Corollary~\ref{cor:conditional-sharing}.

\subsection{Independently varying market parameters: part (a)}
\label{app:finite-box}
For part (a), take the endpoints in~\eqref{eq:finite-parameter-intervals} to be
\begin{equation}
\begin{gathered}
 {[\underline\alpha,\overline\alpha]}=[0.19,0.21],\quad
 [\underline t,\overline t]=[0.149,0.151],\\
 [\underline\eta,\overline\eta]=[0.0589,0.0590],\quad
 [\underline\rho,\overline\rho]=[0.00499,0.00501],\quad
 [\underline\xi,\overline\xi]=[2.99,3.01].
\end{gathered}
\label{eq:finite-primitive-box}
\end{equation}
All five intervals have positive length, and the loss interval has a strictly positive lower bound. Set $c=\eta+\rho$ and $s=\alpha\xi\rho$ throughout. Subdivide
$\alpha\in[0.19,0.21]$ into ten intervals of width $0.002$. In each,
the policy-share construction above with $N=20$ gives 210 triangle cells;
keep the other four parameter intervals undivided. Twenty additional
cells per suitability interval cover $b=0$. Thus the 2300 cells cover
every market and policy required in part (a).

Apply the same inclusion and contraction argument in average-price and
price-difference coordinates, $(p_1+p_2)/2$ and $p_1-p_2$. This invertible
linear change of coordinates improves the bounds on small position-profit
differences without changing the roots. Mutual inclusions link adjacent
policy and suitability cells and link the edge cells to the triangle.
They therefore define one continuous price selection over the whole
parameter region. The original-order profit is strictly concave; the
reversed-order profit is monotone toward the switch and strictly below
equilibrium profit there. Together with the price bounds in
Appendix~\ref{app:finite-deviations}, these inequalities prove global best
responses. The same enclosures establish interior participation, positive
bids and profits after the auction, and retention of at least $58.1$ and
$59.5$ percent in positions 1 and 2.

Table~\ref{tab:finite-box-bounds} reports outward-rounded rational bounds.
Except for the two rows marked $b=0$, the bounds apply throughout the
policy triangle. In particular, $W_b<0$ holds on all 2100 triangle cells.
For $W_f$ on $b=0$, refine to 100 suitability intervals and 40 intervals in $f$. Mutual root inclusions connect each of these 4000 cells to the
original branch, and every refined cell gives $W_f<0$. Thus the welfare
comparison concerns the same global price equilibria as the other policy
comparisons.

\begin{table}[htbp]
\centering
\caption{Uniform finite-loss bounds with independently varying market parameters}
\label{tab:finite-box-bounds}
\small
\renewcommand{\arraystretch}{1.15}
\begin{tabular}{lrr}
\toprule
Quantity & Lower bound & Upper bound\\
\midrule
$P^{\mathrm{ad}}_f$ & $0.001012$ & $0.002709$\\
$P^{\mathrm{ad}}_b$ & $-0.012436$ & $-0.009246$\\
$P^0_b$ & $-0.013142$ & $-0.009926$\\
$P^0_f$ on $b=0$ & $-0.000773$ & $-0.000151$\\
$CS_f$ & $-0.011951$ & $-0.007056$\\
$CS_b$ & $-0.002457$ & $-0.001130$\\
$A_f-A_b$ & $0.001086$ & $0.001730$\\
$W_b$ & $-0.003678$ & $-0.001637$\\
$W_f$ on $b=0$ & $-0.001627$ & $-0.000071$\\
\bottomrule
\end{tabular}
\end{table}

Apply the same feasible paths as in part (b): first lower $b$ to zero,
then lower $f$ for $P^0$, $CS$, and $W$, or raise $f$ for
$P^{\mathrm{ad}}$. Every nontrivial step strictly improves the relevant
objective, so the stated optima are unique. The positive bound on
$A_f-A_b$ also completes the finite-loss claims of
Proposition~\ref{prop:ads} and Corollary~\ref{cor:conditional-sharing}
for part (a).

Finally, the proportional consumer-surplus loss is exactly
\[
 \frac{CS(0,0)-CS(\rho,0)}{CS(0,0)}
 =\frac{-\int_0^\rho CS_f(f,0)\,df}{U(0,0)^2/2}.
\]
The certificate bounds utility throughout the region by
$0.0677127\leq U\leq0.0757644$. Combining this with the table and
$0.00499\leq\rho\leq0.00501$ gives a lower bound exceeding $0.0122$
and an upper bound below $0.0262$. Hence the loss lies between $1.2$ and
$2.7$ percent for every independently varying market in part (a).

\subsection{The resource comparison at \texorpdfstring{$\rho=0.02$}{rho = 0.02}}
\label{app:resource-decomposition}
This verifies the component claims following Proposition~\ref{prop:finite}.
Fix~\eqref{eq:finite-expanded-primitives} at $\rho=0.02$ and let $\Delta$
denote $(f,b)=(\rho,0)$ minus $(0,0)$. Write $D=D_1+D_2$ and
$R=R_1+R_2$. From $S_1+S_2=D+R$ and $c-\eta=\rho$,
\[
 \Delta V=\Delta M-c\Delta D-\rho\Delta R-s\Delta q.
\]
Moreover, $\Delta(UV)=\bar U\Delta V+\bar V\Delta U$ and
$\Delta(U^2/2)=\bar U\Delta U$, where bars denote endpoint means.
Subtracting proves~\eqref{eq:welfare-endpoint-decomposition}.

A fresh rational contraction at each endpoint encloses its price root
inside the corresponding uniqueness rectangle of the full-policy
certificate. Local uniqueness therefore identifies these roots with the
branch used in part (b). Evaluating the preceding identities on these
enclosures gives the following outward bounds, in units of $10^{-6}$.
\begin{center}
\begin{tabular}{lrr}
\toprule
Component & Lower bound & Upper bound\\
\midrule
Retained value per participant, $\Delta M$ & $2792.85$ & $2792.86$\\
Retained-unit cost, $-c\Delta D$ & $-643.29$ & $-643.28$\\
Return-loss saving, $-\rho\Delta R$ & $205.79$ & $205.80$\\
Inspection saving, $-s\Delta q$ & $128.57$ & $128.58$\\
\midrule
Aggregate resource contribution, $\bar U\Delta V$ & $156.10$ & $156.11$\\
Participation contribution, $(\bar V-\bar U)\Delta U$ & $-144.68$ & $-144.67$\\
Total welfare, $\Delta W$ & $11.42$ & $11.43$\\
\bottomrule
\end{tabular}
\end{center}
The lower bound on $\Delta M$ exceeds the upper bound on $c\Delta D$;
both return and inspection savings are positive. The aggregate resource
gain also exceeds the absolute participation loss, proving the stated
positive welfare effect. These component signs concern this representative
comparison; the all-policy welfare ranking on the moderate-loss interval
was proved above.

\section{Proof of Proposition~\ref{prop:entry-incidence}}\label{app:entry-extension}
The proposition has two claims: a price-response threshold when suitability is low, and persistence of the finite-loss comparisons when participation costs are close to uniform. We prove them in that order. Throughout, $\gamma>0$ is the participation elasticity introduced in Section~\ref{sec:entry}; $k=1-t$, $\ell=1-p_0$, $z=\alpha\ell$, and $m=kp_0-\eta$ retain their meanings from Appendix~\ref{app:consumer}.

\paragraph{Step 1: Participation and the price conditions.}
A consumer enters when $e\le U$. Integrating her net utility and her participation cost gives
\[
 CS=\int_0^U(U-e)\gamma e^{\gamma-1}\,de
 =\frac{U^{\gamma+1}}{\gamma+1},\qquad
 W=U^\gamma V-\int_0^Ue\gamma e^{\gamma-1}\,de
 =U^\gamma V-\frac{\gamma U^{\gamma+1}}{\gamma+1}.
\]
These are the expressions in~\eqref{eq:general-entry-welfare}. Since $U_{p_i}=-D_i$, differentiating aggregate profit yields
\[
 \partial_{p_i}(U^\gamma\pi_i)=U^\gamma\mathcal F_i^\gamma,
 \qquad
 \mathcal F_i^\gamma:=\pi_{i,p_i}-\frac{\gamma D_i\pi_i}{U}.
\]
Thus $\mathcal F_i^\gamma=0$ is firm $i$'s price condition after division by the positive participation mass. Reimbursement changes neither utility nor shopping choices at fixed prices and consumer contributions, while $\pi_{i,b}=R_i$. It follows directly that
\[
 \partial_b\partial_{p_i}(U^\gamma\pi_i)
 =U^\gamma R_{i,p_i}-\gamma U^{\gamma-1}D_iR_i,
\]
which proves~\eqref{eq:general-entry-incentive}.

\paragraph{Step 2: The reference price at low suitability.}
At zero inspection and return costs, the symmetric quantities are
\[
 D_0=\alpha\ell(1-\alpha\ell/2),\qquad
 U_0=\alpha\ell^2(1-\alpha\ell/3).
\]
The common price condition is
\[
 kD_0=m\left(\alpha+\gamma\frac{D_0^2}{U_0}\right).
\]
Substituting $m/k=1-\eta/k-\ell$ and dividing by $k\alpha$ rewrites it as
\begin{equation}
 (1-\eta/k-\ell)
 \left[1+\gamma\frac{(1-\alpha\ell/2)^2}{1-\alpha\ell/3}\right]
 =\ell(1-\alpha\ell/2).
 \label{eq:power-entry-price}
\end{equation}
The left side minus the right side is positive at $\ell=0$ and negative at $\ell=1-\eta/k$, so a root exists between them. At $\alpha=0$ the equation has the unique solution
\begin{equation}
 \ell_*:=\frac{\gamma+1}{\gamma+2}(1-\eta/k)>0.
 \label{eq:entry-limit-retention}
\end{equation}
The value $\ell_*$ is the limit of the reference quantity $\ell$. The derivative of the price equation with respect to $\ell$, after moving its right side to the left, is $-(\gamma+2)$ at $\alpha=0$. The implicit-function theorem therefore gives a unique nearby root for small $\alpha$. Any sequence of other symmetric roots would also have to converge to $\ell_*$, because the roots lie in $[0,1-\eta/k]$ and the limiting equation has no other solution. Local uniqueness excludes that possibility. Hence the symmetric reference price is uniquely determined for sufficiently low positive suitability, and $\ell\to\ell_*$.

\paragraph{Step 3: Reimbursement and both price responses.}
Policy derivatives are taken from positive costs at which product 1 is tried first, as in Appendix~\ref{app:pricing}. Let $J_\gamma$ be the Jacobian of $(\mathcal F_1^\gamma,\mathcal F_2^\gamma)$ with respect to $(p_1,p_2)$. These are price-coordinate conditions; the matrix differs from the $w$-coordinate Jacobian in~\eqref{eq:referencecoeff}.

For each $i,j$, differentiation gives
\[
 (J_\gamma)_{ij}
 =\pi_{i,p_ip_j}
 -\gamma\frac{D_{i,p_j}\pi_i+D_i\pi_{i,p_j}}{U}
 -\gamma\frac{D_i\pi_iD_j}{U^2}.
\]
The reimbursement column is especially simple. At the reference point, $R_i=\alpha-D_0$, $R_{1,p_1}=\alpha$, and $R_{2,p_2}=\alpha(1-\alpha)$, so
\[
 \partial_b\binom{\mathcal F_1^\gamma}{\mathcal F_2^\gamma}
 =\binom{\alpha-\gamma D_0(\alpha-D_0)/U_0}
 {\alpha(1-\alpha)-\gamma D_0(\alpha-D_0)/U_0}.
\]
To evaluate the low-suitability limit, use $D_0/\alpha\to\ell_*$, $U_0/\alpha\to\ell_*^2$, and $m/k\to\ell_* /(1+\gamma)$. Substitution into the displayed derivatives gives
\begin{align}
 \frac{J_\gamma}{\alpha}&\longrightarrow
 -\frac{k}{\gamma+1}
 \begin{pmatrix}\gamma^2+2\gamma+2&\gamma\\
                 \gamma&\gamma^2+2\gamma+2\end{pmatrix},
 \label{eq:entry-limit-jacobian}\\
 \frac1\alpha\partial_b\binom{\mathcal F_1^\gamma}{\mathcal F_2^\gamma}
 &\longrightarrow
 \left[1-\frac{\gamma(1-\ell_*)}{\ell_*}\right]\binom{1}{1}.
 \label{eq:entry-limit-column}
\end{align}
The limiting matrix has eigenvalues $-k(\gamma+2)$ and $-k(\gamma^2+\gamma+2)/(\gamma+1)$, both negative. Thus the price conditions are nonsingular for sufficiently small positive $\alpha$. Implicit differentiation is valid in the nearby positive-cost region. Adding the two rows of the limiting response equation yields
\begin{equation}
 \lim_{\alpha\downarrow0}\partial_b(p_1+p_2)
 =\frac{2[(\gamma+1)\ell_*-\gamma]}{k(\gamma+2)\ell_*}.
 \label{eq:power-entry-reimbursement-threshold}
\end{equation}
The derivative inside the limit is the limiting positive-cost response at each fixed $\alpha$. Its denominator is positive, and~\eqref{eq:entry-limit-retention} gives
\[
 (\gamma+1)\ell_*-\gamma
 =\frac{1-(\eta/k)(\gamma+1)^2}{\gamma+2}.
\]
Consequently, the price sum falls exactly when $\eta/k>(1+\gamma)^{-2}$, provided $\alpha$ is sufficiently small and the cost ratio is away from equality. Nonsingularity proves a local solution of the price conditions. The global price-deviation checks remain necessary to call that solution an equilibrium, as stated in Proposition~\ref{prop:entry-incidence}.

\paragraph{Step 4: Persistence of the finite-loss comparisons.}
Consider the compact parameter sets constructed in Appendix~\ref{app:finite} for Proposition~\ref{prop:finite}, together with their policy triangles. Their selected equilibrium prices lie strictly inside the same shopping region. A positive rival trial index also gives a positive lower bound for shopping utility at all relevant unilateral deviations; deviations with nonpositive margins or no profitable trials are excluded as in Appendix~\ref{app:finite-new}. We can therefore work on one compact utility interval bounded away from zero. On this interval, $U^\gamma$ and its first two derivatives converge uniformly to those of $U$ as $\gamma\to1$.

This uniform convergence preserves each part of the finite-loss argument. First, nonsingularity of the price Jacobian continues the selected price solutions, and strict own-price curvature preserves their local optimality. Second, outside a small neighborhood of each equilibrium price, the original global-deviation profit gap is strictly positive. Compactness gives a common positive gap, which remains positive for $\gamma$ close to one, including at deviations that change trial order. The continued prices are therefore global best responses.

Finally, prices and their policy derivatives vary continuously with $\gamma$. The certified strict signs of the platform and welfare derivatives, positive bids, positive post-auction profits, and majority retention consequently persist uniformly. These signs give the same unique optimal payment policies and welfare rankings. Taking the smallest of the neighborhoods needed for these comparisons proves the last claim of Proposition~\ref{prop:entry-incidence}.\hfill$\square$

\section{Exact sign and interval verification}\label{app:verification}

The preceding proofs use two elementary sign certificates. We state the
rules here to make explicit why a finite list of exact inequalities proves
a claim on a continuous parameter domain.

\paragraph{Polynomial signs.}
For a rational expression, first establish the denominator's sign and
remove only factors with a known strict sign on the stated domain.
An exact substitution may express the remaining numerator as a polynomial
in nonnegative variables with nonnegative coefficients. It is then
nonnegative; it is strictly positive wherever at least one positive
coefficient multiplies a strictly positive monomial.

Alternatively, map the relevant interval to $x\in[0,1]$ and write the
polynomial in its Bernstein basis,
\[
 Q(x)=\sum_{i=0}^n q_i\binom ni x^i(1-x)^{n-i}.
\]
The basis weights are nonnegative and sum to one by the binomial theorem.
Consequently $\min_i q_i\leq Q(x)\leq\max_i q_i$ on the whole interval;
in particular, $q_i>0$ for every $i$ proves $Q>0$. On a rectangle, products
of the one-dimensional basis weights again are nonnegative and sum to one,
so the same argument applies to all coefficients of the product basis.
The sign proofs specify the substitution, domain, and strictly signed
coefficients. This is an exact polynomial identity, not an evaluation at
selected parameter values.

\paragraph{Rational interval signs.}
Every interval operation encloses the result for all arguments in its
input intervals. The calculations use exact fractions and round the lower
endpoint down and the upper endpoint up to the rational grid $10^{-24}$.
For a square root, nonnegative rational endpoints are obtained by integer
square roots and checked by squaring: if $a^2\leq x\leq b^2$, then
$a\leq\sqrt{x}\leq b$. These operations preserve enclosure under
composition; a strictly positive lower bound or strictly negative upper
bound therefore proves the required sign throughout the cell.

For the finite-loss results, cells cover the full stated parameter and
policy domains. The inclusions
\eqref{eq:finite-new-inclusion}--\eqref{eq:finite-new-contraction}
then establish a root at every parameter point and uniqueness within its
price rectangle, by the contraction argument in
Appendix~\ref{app:finite-prices}; see also \citet{rump2010verification}.
The separate comparisons in Appendix~\ref{app:finite-deviations} establish
global best responses. Thus root certification, equilibrium verification,
and policy comparisons are distinct steps of the proof.


\begin{thebibliography}{}

\bibitem[Anderson et~al., 2009]{anderson2009option}
Anderson, E.~T., Hansen, K., and Simester, D. (2009).
\newblock The option value of returns: Theory and empirical evidence.
\newblock {\em Marketing Science}, 28(3):405--423.

\bibitem[Anderson and Renault, 2026]{anderson2026direction}
Anderson, S.~P. and Renault, R. (2026).
\newblock Search direction: Position externalities and position auction bias.
\newblock {\em The Review of Economic Studies}, 93(2):763--797.

\bibitem[Armstrong et~al., 2009]{armstrong2009prominence}
Armstrong, M., Vickers, J., and Zhou, J. (2009).
\newblock Prominence and consumer search.
\newblock {\em The RAND Journal of Economics}, 40(2):209--233.

\bibitem[Armstrong and Zhou, 2011]{armstrong2011paying}
Armstrong, M. and Zhou, J. (2011).
\newblock Paying for prominence.
\newblock {\em The Economic Journal}, 121(556):F368--F395.

\bibitem[Athey and Ellison, 2011]{athey2011position}
Athey, S. and Ellison, G. (2011).
\newblock Position auctions with consumer search.
\newblock {\em The Quarterly Journal of Economics}, 126(3):1213--1270.

\bibitem[Bower and Maxham, 2012]{bower2012return}
Bower, A.~B. and Maxham, J.~G., III (2012).
\newblock Return shipping policies of online retailers: Normative assumptions and the long-term consequences of fee and free returns.
\newblock {\em Journal of Marketing}, 76(5):110--124.

\bibitem[Chen and He, 2011]{chen2011paid}
Chen, Y. and He, C. (2011).
\newblock Paid placement: Advertising and search on the internet.
\newblock {\em The Economic Journal}, 121(556):F309--F328.

\bibitem[Compiani et~al., 2024]{compiani2024rankings}
Compiani, G., Lewis, G., Peng, S., and Wang, P. (2024).
\newblock Online search and optimal product rankings: An empirical framework.
\newblock {\em Marketing Science}, 43(3):615--636.

\bibitem[El Kihal and Shehu, 2022]{elkihal2022keep}
El Kihal, S. and Shehu, E. (2022).
\newblock It's not only what they buy, it's also what they keep: Linking marketing instruments to product returns.
\newblock {\em Journal of Retailing}, 98(3):558--571.

\bibitem[Fishman and Lubensky, 2018]{fishman2018prominence}
Fishman, A. and Lubensky, D. (2018).
\newblock Search prominence and return costs.
\newblock {\em International Journal of Industrial Organization}, 58:136--161.

\bibitem[Guide et~al., 2006]{guide2006time}
Guide, V.~D.~R., Jr., Souza, G.~C., Van Wassenhove, L.~N., and Blackburn, J.~D. (2006).
\newblock Time value of commercial product returns.
\newblock {\em Management Science}, 52(8):1200--1214.


\bibitem[Janssen et~al., 2026]{janssen2026platforms}
Janssen, M., Jungbauer, T., Preuss, M., and Williams, C. (2026).
\newblock Search platforms: Big data and sponsored positions.
\newblock {\em The Economic Journal}.
\newblock Advance article, ueag002.

\bibitem[Janssen and Williams, 2024]{janssen2024returns}
Janssen, M. and Williams, C. (2024).
\newblock Consumer search and product returns in e-commerce.
\newblock {\em American Economic Journal: Microeconomics}, 16(2):387--419.

\bibitem[Jerath and Ren, 2025]{jerath2025returns}
Jerath, K. and Ren, Q. (2025).
\newblock Consumer search and product returns.
\newblock {\em Marketing Science}, 44(3):691--710.

\bibitem[Long et~al., 2022]{long2022designing}
Long, F., Jerath, K., and Sarvary, M. (2022).
\newblock Designing an online retail marketplace: Leveraging information from
  sponsored advertising.
\newblock {\em Marketing Science}, 41(1):115--138.

\bibitem[Milgrom and Segal, 2002]{milgrom2002envelope}
Milgrom, P. and Segal, I. (2002).
\newblock Envelope theorems for arbitrary choice sets.
\newblock {\em Econometrica}, 70(2):583--601.

\bibitem[Petrikait{\.e}, 2018]{petrikaite2018returns}
Petrikait{\.e}, V. (2018).
\newblock A search model of costly product returns.
\newblock {\em International Journal of Industrial Organization}, 58:236--251.

\bibitem[Rump, 2010]{rump2010verification}
Rump, S.~M. (2010).
\newblock Verification methods: Rigorous results using floating-point arithmetic.
\newblock {\em Acta Numerica}, 19:287--449.

\bibitem[Saremi and Subramanian, 2026]{saremi2026bidding}
Saremi, E. and Subramanian, U. (2026).
\newblock Automated targeted bidding for sponsored ads on e-commerce platforms.
\newblock {\em Marketing Science}, 45(3):576--595.

\bibitem[Shulman et~al., 2011]{shulman2011competitive}
Shulman, J.~D., Coughlan, A.~T., and Savaskan, R.~C. (2011).
\newblock Managing consumer returns in a competitive environment.
\newblock {\em Management Science}, 57(2):347--362.

\bibitem[Ursu, 2018]{ursu2018rankings}
Ursu, R.~M. (2018).
\newblock The power of rankings: Quantifying the effect of rankings on online
  consumer search and purchase decisions.
\newblock {\em Marketing Science}, 37(4):530--552.

\bibitem[Vickrey, 1961]{vickrey1961counterspeculation}
Vickrey, W. (1961).
\newblock Counterspeculation, auctions, and competitive sealed tenders.
\newblock {\em The Journal of Finance}, 16(1):8--37.

\bibitem[Wang et~al., 2024]{wang2024freight}
Wang, X., Xu, Y., Choi, T.-M., and Zhou, Q. (2024).
\newblock Who should pay for the return freight in e-commerce? Platforms, retailers or consumers.
\newblock {\em International Journal of Production Economics}, 277:109375.

\bibitem[Weitzman, 1979]{weitzman1979optimal}
Weitzman, M.~L. (1979).
\newblock Optimal search for the best alternative.
\newblock {\em Econometrica}, 47(3):641--654.

\bibitem[Zhang et~al., 2025]{zhang2025browse}
Zhang, Q., Liu, Z., Kumar, S., and Yang, S. (2025).
\newblock Browse more, return less: Managing online returns through interactions with browsings and purchases.
\newblock {\em Decision Sciences}, 56(6):578--597.

\end{thebibliography}
\end{document}